\documentclass[10pt]{article}

\usepackage{amssymb}
\usepackage{graphicx}\usepackage{amsmath,amssymb}
\usepackage{color}
\usepackage{xcolor}
\usepackage{srcltx}
\usepackage{greekbf}
\usepackage{nicefrac}
\usepackage{empheq}
\usepackage{enumerate}

\usepackage{mathrsfs}

\DeclareMathAlphabet{\mathbf}{OT1}{cmr}{bx}{it}
\DeclareMathAlphabet{\mathssb}{OT1}{cmss}{bx}{n}
\DeclareMathAlphabet{\mathssn}{OT1}{cmss}{m}{n}
\DeclareMathAlphabet{\mathub}{OT1}{cmr}{b}{n}

\DeclareMathAlphabet{\mathpzc}{OT1}{pzc}%
                                 {m}{it}

\DeclareMathOperator{\Lyn}{Lin} \DeclareMathOperator{\Sim}{Sym}

\DeclareMathOperator{\Ash}{^s\!\!\Ab}
\DeclareMathOperator{\Bsh}{^s\!\Bb}
\DeclareMathOperator{\Fbs}{^s\!\Fb}
\DeclareMathOperator{\Mbs}{^s\!\Mb}
\DeclareMathOperator{\Gs}{^s\!\Gb}
\DeclareMathOperator{\Pbs}{^s\!\Pb}
\DeclareMathOperator{\nablas}{^s\nabla\!}
\DeclareMathOperator{\Divs}{^s\!\Div}
\newcommand{\<}[1]{\scriptstyle<#1>}

\newcommand\remarkname{Remark} %
\newcounter {remarkn}[section]%
\renewcommand \theremarkn {\arabic{section}.\arabic{remarkn}}%
\newenvironment{remark}{
\noindent{\emph{\remarkname}.} \rm%
}

\newcommand\remarksname{Remark} %
\newcounter {remarksn}[section]%
\renewcommand \theremarksn {\thesection.\arabic{remarksn}}%
\begin{document}

\begin{center}

\end{center}
%
% Macro per LaTeX
%
%    aggiornati al 20-set-99
%
%====  CARATTERI SPECIALI ==========================
%
\newcommand{\bydef}{\,\raise.050ex\hbox{\rm:}\kern-.025em\hbox{\rm=}\,}
\newcommand{\defby}{=\raise.075ex\hbox{\kern-.325em\hbox{\rm:}}\,}
\newcommand{\mtrp}  {{-\!\top}} % -trasposto alla "adc"
\newcommand{\bdot}  {{\scriptscriptstyle\bullet}}
\def\qed{\relax\ifmmode\hskip2em \Box\else\unskip\nobreak\hskip1em $\Box$\fi}
%
%==== VARIE =====================================
%
%\newcommand {\dim}[1] {\mathtt{dim}(#1)}
\newcommand {\eps} {\varepsilon} % espilon
\newcommand {\vp} {\varphi}      % phi
\newcommand {\0} {\textbf{0}}    % 0 bold
\newcommand {\1} {\textbf{1}}    % 0 bold
%
%==== CALLIGRAFICI ==================
\newcommand {\Ac}  {\mathcal{A}}
\newcommand {\Bc}  {\mathcal{B}}
\newcommand {\Cc}  {\mathcal{C}}
\newcommand {\Dc}  {\mathcal{D}}
\newcommand {\Ec}  {\mathcal{E}}
\newcommand {\Fc}  {\mathcal{F}}
\newcommand {\Gc}  {\mathcal{G}}
\newcommand {\Hc}  {\mathcal{H}}
\newcommand {\Kc}  {\mathcal{K}}
\newcommand {\Ic}  {\mathcal{I}}
\newcommand {\Jc}  {\mathcal{J}}
\newcommand {\Lc}  {\mathcal{L}}
\newcommand {\Mc}  {\mathcal{M}}
\newcommand {\Nc}  {\mathcal{N}}
\newcommand {\Oc}  {\mathcal{O}}
\newcommand {\Pc}  {\mathcal{P}}
\newcommand {\Rc}  {\mathcal{R}}
\newcommand {\Sc}  {\mathcal{S}}
\newcommand {\Tc}  {\mathcal{T}}
\newcommand {\Uc}  {\mathcal{U}}
\newcommand {\Vc}  {\mathcal{V}}
\newcommand {\Wc}  {\mathcal{W}}
\newcommand {\Zc}  {\mathcal{Z}}
%
%==== BOLD ====
%
\newcommand {\ab} {\mathbf{a}}
\newcommand {\bb} {\mathbf{b}}
\newcommand {\cb} {\mathbf{c}}
\newcommand {\db} {\mathbf{d}}
\newcommand {\eb} {\mathbf{e}}
\newcommand {\fb} {\mathbf{f}}
\newcommand {\gb} {\mathbf{g}}
\newcommand {\hb} {\mathbf{h}}
\newcommand {\ib} {\mathbf{i}}
\newcommand {\kb} {\mathbf{k}}
\newcommand {\lb} {\mathbf{l}}
\newcommand {\mb} {\mathbf{m}}
\newcommand {\nb} {\mathbf{n}}
\newcommand {\pb} {\mathbf{p}}
\newcommand {\qb} {\mathbf{q}}
\newcommand {\rb} {\mathbf{r}}
\renewcommand {\sb} {\mathbf{s}}
\newcommand {\tb} {\mathbf{t}}
\newcommand {\xb} {\mathbf{x}}
\newcommand {\ub} {\mathbf{u}}
\newcommand {\vb} {\mathbf{v}}
\newcommand {\wb} {\mathbf{w}}
\newcommand {\zb} {\mathbf{z}}
\newcommand {\Ab} {\mathbf{A}}
\newcommand {\Bb} {\mathbf{B}}
\newcommand {\Cb} {\mathbf{C}}
\newcommand {\Db} {\mathbf{D}}
\newcommand {\Eb} {\mathbf{E}}
\newcommand {\Fb} {\mathbf{F}}
\newcommand {\Gb} {\mathbf{G}}
\newcommand {\Hb} {\mathbf{H}}
\newcommand {\Kb} {\mathbf{K}}
\newcommand {\Jb} {\mathbf{J}}
\newcommand {\Ib} {\mathbf{I}}
\newcommand {\Lb} {\mathbf{L}}
\newcommand {\Mb} {\mathbf{M}}
\newcommand {\Nb} {\mathbf{N}}
\newcommand {\Ob} {\mathbf{O}}
\newcommand {\Pb} {\mathbf{P}}
\newcommand {\Qb} {\mathbf{Q}}
\newcommand {\Rb} {\mathbf{R}}
\newcommand {\Sb} {\mathbf{S}}
\newcommand {\Tb} {\mathbf{T}}
\newcommand {\Vb} {\mathbf{V}}
\newcommand {\Wb} {\mathbf{W}}
\newcommand {\Xb} {\mathbf{X}}
\newcommand {\Zb} {\mathbf{Z}}
%

%
%==== ASSI (minuscole) e PUNTI (maiuscole) dello spazio euclideo ====
%
\newcommand {\ax} {\mathrm{a}}
\newcommand {\bx} {\mathrm{b}}
\newcommand {\cx} {\mathrm{c}}
\newcommand {\dx} {\mathrm{d}}
\newcommand {\ex} {\mathrm{e}}
\newcommand {\fx} {\mathrm{f}}
\newcommand {\gx} {\mathrm{g}}
\newcommand {\hx} {\mathrm{h}}
\newcommand {\kx} {\mathrm{k}}
\newcommand {\lx} {\mathrm{l}}
\newcommand {\mx} {\mathrm{m}}
\newcommand {\nx} {\mathrm{n}}
\newcommand {\ox} {\mathrm{o}}
\newcommand {\px} {\mathrm{p}}
\newcommand {\qx} {\mathrm{q}}
\newcommand {\rx} {\mathrm{r}}
\newcommand {\sx} {\mathrm{s}}
\newcommand {\tx} {\mathrm{t}}
\newcommand {\xx} {\mathrm{x}}
\newcommand {\yx} {\mathrm{y}}
\newcommand {\wx} {\mathrm{w}}
\newcommand {\ux} {\mathrm{u}}
\newcommand {\vx} {\mathrm{v}}
\newcommand {\zx} {\mathrm{z}}
\newcommand {\Ax} {\mathrm{A}}
\newcommand {\Bx} {\mathrm{B}}
\newcommand {\Cx} {\mathrm{C}}
\newcommand {\Dx} {\mathrm{D}}
\newcommand {\Ex} {\mathrm{E}}
\newcommand {\Fx} {\mathrm{F}}
\newcommand {\Gx} {\mathrm{G}}
\newcommand {\Hx} {\mathrm{H}}
\newcommand {\Kx} {\mathrm{K}}
\newcommand {\Jx} {\mathrm{J}}
\newcommand {\Ix} {\mathrm{I}}
\newcommand {\Lx} {\mathrm{L}}
\newcommand {\Mx} {\mathrm{M}}
\newcommand {\Nx} {\mathrm{N}}
\newcommand {\Ox} {\mathrm{O}}
\newcommand {\Px} {\mathrm{P}}
\newcommand {\Qx} {\mathrm{Q}}
\newcommand {\Rx} {\mathrm{R}}
\newcommand {\Sx} {\mathrm{S}}
\newcommand {\Tx} {\mathrm{T}}
\newcommand {\Vx} {\mathrm{V}}
\newcommand {\Wx} {\mathrm{W}}
\newcommand {\Xx} {\mathrm{X}}
\newcommand {\Yx} {\mathrm{Y}}
\newcommand {\Zx} {\mathrm{Z}}
%

%
%==== ORLATI =============
%
\newcommand {\Real} {\mathbb{R}}
\newcommand {\Aos} {\mbox{$\scriptstyle\mathbb{A}$}}
\newcommand {\Ao} {\mathbb{A}}
\newcommand {\Bo} {\mathbb{B}}
\newcommand {\Co} {\mathbb{C}}
\newcommand {\Cop} {\mbox{$\scriptstyle\mathbb{C}$}}
\newcommand {\Do} {\mathbb{D}}
\newcommand {\Fo} {\mathbb{F}}
\newcommand {\Go} {\mathbb{G}}
\newcommand {\Io} {\mathbb{I}}
\newcommand {\Mo} {\mathbb{M}}
\newcommand {\Ko} {\mathbb{K}}
\newcommand {\No} {\mathbb{N}}
\newcommand {\Po} {\mathbb{P}}
\newcommand {\Qo} {\mathbb{Q}}
\newcommand {\Ro} {\mathbb{R}}
\newcommand {\So} {\mathbb{S}}
\newcommand {\To} {\mathbb{T}}
\newcommand {\Vo} {\mathbb{V}}
\newcommand {\Zo} {\mathbb{Z}}
%
%==== EULER SCRIPT =============
% non funzionano! perche?
%\usepackage{euler}
%\newcommand {\Gs} {\mathscr{G}}
\newcommand {\Ds} {\mathscr{D}}
\newcommand {\Ms} {\mathscr{M}}
\newcommand {\Ns} {\mathscr{N}}
\newcommand {\Vs} {\mathscr{V}}
\newcommand {\Xes} {\mathscr{X}}
\newcommand {\ges} {\mathscr{g}}
\newcommand {\wes} {\mathscr{W}}
%
%
%==== EULER FRACTUR =============
%
\newcommand {\aef} {\mathfrak{a}}
\newcommand {\fef} {\mathfrak{f}}
\newcommand {\gef} {\mathfrak{g}}
\newcommand {\hef} {\mathfrak{h}}
\newcommand {\mef} {\mathfrak{m}}
\newcommand {\nef} {\mathfrak{n}}
\newcommand {\kef} {\mathfrak{k}}
\newcommand {\wef} {\mathfrak{w}}
\newcommand {\sef} {\mathfrak{s}}
\newcommand {\zef} {\mathfrak{z}}
\newcommand {\Def} {\mathfrak{D}}
\newcommand {\Fef} {\mathfrak{F}}
\newcommand {\Mef} {\mathfrak{M}}
\newcommand {\Nef} {\mathfrak{N}}
\newcommand {\Ref} {\mathfrak{R}}
\newcommand {\Sef} {\mathfrak{S}}
\newcommand {\Xef} {\mathfrak{X}}
%
%==== TIPOGRAFICI =============
%
\newcommand {\att} {\mathtt{a}}
\newcommand {\btt} {\mathtt{b}}
\newcommand {\ctt} {\mathtt{c}}
\newcommand {\dtt} {\mathtt{d}}
\newcommand {\ftt} {\mathtt{f}}
\newcommand {\gtt} {\mathtt{g}}
\newcommand {\mtt} {\mathtt{m}}
\newcommand {\ntt} {\mathtt{n}}
\newcommand {\htt} {\mathtt{h}}
\newcommand {\ptt} {\mathtt{p}}
\newcommand {\qtt} {\mathtt{q}}
\newcommand {\rtt} {\mathtt{r}}
\newcommand {\stt} {\mathtt{s}}
\newcommand {\ttt} {\mathtt{t}}
\newcommand {\vtt} {\mathtt{v}}
\newcommand {\wtt} {\mathtt{w}}
\newcommand {\ztt} {\mathtt{z}}
\newcommand {\Ftt} {\mathtt{F}}
\newcommand {\Stt} {\mathtt{P}}
\newcommand {\Wtt} {\mathtt{W}}
%
%
%======Lowercase greek letters BOLD============
%
%
%\def\nub{\hbox{\mbo {\char 23}}}%="0?17
%\def\xib{\hbox{\mbo {\char 24}}}%="0?18
%\def\pib{\hbox{\mbo {\char 25}}}%="0?19
%\def\rhob{\hbox{\mbo {\char 26}}}%="0?1A
%\def\sigmab{\hbox{\mbo {\char 27}}}%="0?1B
\newcommand {\alfab}     {\mathbf{\alpha}}
\newcommand {\betab}     {\mathbf{\beta}}
\newcommand {\gammab}    {\mathbf{\gamma}}
\newcommand {\deltab}    {\mathbf{\delta}}
\newcommand {\epsilonb}  {\mathbf{\epsilon}}
\newcommand {\epsb}      {\mathbf{\varepsilon}}
\newcommand {\zetab}     {\mathbf{\zeta}}
\newcommand {\etab}      {\mathbf{\eta}}
\newcommand {\tetab}     {\mathbf{\teta}}
\newcommand {\vtetab}    {\mathbf{\vartheta}}
\newcommand {\iotab}     {\mathbf{\iota}}
\newcommand {\kappab}    {\mathbf{\kappa}}
\newcommand {\lambdab}   {\mathbf{\lambda}}
\newcommand {\mub}       {\mathbf{\mu}}
\font\mbo=cmmib10 scaled \magstephalf
\newcommand{\nub}   {\hbox{\mbo {\char 23}}}%="0?17
\newcommand {\csib}      {\mathbf{\xi}}
\newcommand {\xib}      {\mathbf{\xi}}
\newcommand {\pib}       {\mathbf{\pi}}
\newcommand {\varrhob}   {\mathbf{\varrho}}
\newcommand {\sigmab}    {\hbox{\mbo {\char 27}}}
\newcommand{\taub}   {\hbox{\mbo {\char 28}}}%="0?17
\newcommand {\upsilonb}  {\mathbf{\upsilon}}
\newcommand {\phib}      {\mathbf{\phi}}
\newcommand {\varphib}   {\mathbf{\varphi}}
\newcommand {\chib}      {\mathbf{\chi}}
\newcommand {\psib}       {\mathbf{\psi}}
\newcommand {\omegab}    {\mathbf{\omega}}
%
%
%======Uppercase greek letters BOLD============
%
%
\newcommand {\Gammab}    {\mathbf{\Gamma}}
\newcommand {\Deltab}    {\mathbf{\Delta}}
\newcommand {\Tetab}     {\mathbf{\Theta}}
\newcommand {\Lambdab}   {\mathbf{\Lambda}}
\newcommand {\Csib}      {\mathbf{\Xi}}
\newcommand {\Pib}       {\mathbf{\Pi}}
\newcommand {\Sigmab}    {\mathbf{\Sigma}}
\newcommand {\Phib}      {\mathbf{\Phi}}
\newcommand {\Psib}      {\mathbf{\Psi}}
\newcommand {\Omegab}    {\mathbf{\Omega}}
%
%==== LIN & COMPANY ================
%
\newcommand {\Lin} {\mathbb{L}\mathtt{in}}
\newcommand {\Sym} {\mathbb{S}\mathtt{ym}}
\newcommand {\Psym} {\mathbb{PS}\mathtt{ym}}
\newcommand {\Skw} {\mathbb{S}\mathtt{kw}}
\newcommand {\SO} {\mathcal{SO}}
\newcommand {\GL} {\mathcal{G}l}
\newcommand {\Rot} {\mathbb{R}\mathtt{ot}}
%
%==== OPERATORI ================
%
\newcommand {\tr}[1]{\mbox{tr}\, #1}
\newcommand {\psym} {\mbox{sym}}
\newcommand {\pskw} {\mbox{skw}}
\newcommand {\win}[2] {( #1 \cdot #2 )_\wedge }
\newcommand {\modulo}[1] {\left|#1\right|}
\newcommand {\sph} {\mbox{sph}}
\newcommand {\dev} {\mbox{dev}}
\newcommand {\sgn} {\mbox{sgn}}
\newcommand {\lin} {\mbox{Lin}}
%
%==== OPERATORI DIFFERENZIALI ================
%
\newcommand{\dvg} {\mathrm{div}\,}
\newcommand{\dvgt} {\mathrm{d{\widetilde i}v}\,}    % divergenza
\newcommand{\grd} {\mathrm{grad}\,}    % gradiente
\newcommand{\Grd} {\mathrm{Grad}\,}    % Gradiente
\newcommand{\dl}  {\delta}             % delta
\def\gradtwo{\mathord{\nabla^{\scriptscriptstyle(2)}}}
\def\Div{\mathop{\hbox{Div}}}
\def\div{\mathop{\hbox{div}}}
\def\mis{\mathop{\hbox{mis}}}
\def\eps{\varepsilon}
%
%==== MISCELLANEA ================
%

\newcommand\ph{\varphi}
\newcommand{\adj} {\att\dtt}     % aggiunto piccolo

\newcommand{\va}{\mathbf{a}}
\newcommand{\vn}{\mathbf{\nu}}
\newcommand{\vt}{\mathbf{\tau}}
\newcommand{\dn}{\partial_{\mathbf{\nu}}}
\newcommand{\dt}{\partial_{\mathbf{\tau}}}
\newcommand{\ord}{\scriptscriptstyle}
\newcommand{\tD}{\mathbf{E}}  %  tensore della deformazione
\newcommand{\tS}{\mathbf{S}}  %  tensore di sforzo
\newcommand{\tE}{\mathbb{C}}
\newcommand{\tPf}{\mathbb{P}}
\newcommand{\tPr}{\mathbf{P}}
\newcommand{\tC}{\mathbf{C}}
\newcommand{\tP}{{\scriptstyle\mathbb{C}}}   %  tensore piezoelettrico
\newcommand{\tDl}{\mathbf{C}}      %  tensore dielettrico

\newcommand{\Cuno}{\mathbf{c}_1}
\newcommand{\Cdue}{\mathbf{c}_2}
\newcommand{\veralf}{\mathbf{c}_\alpha} %  C1 e C2 in alfa
\newcommand{\verbet}{\mathbf{c}_\beta} %  C1 e C2 in alfa
\newcommand{\vz}{\mathbf{z}}
\newcommand{\sym}{\mathop{\mathrm{sym}}}

\newcommand{\f}{f}
\newcommand{\g}{g}
\newcommand{\h}{h}
\newcommand{\w}{w}
\renewcommand{\l}{l}

\renewcommand{\r}{r}
\newcommand{\s}{s}

\newcommand{\autof}{\mathrm{\skew 0\overline{w}}}
\newcommand{\W}{W}

\newcommand{\df}{f'}
\newcommand{\dg}{g'}
\renewcommand{\dh}{h'}
\newcommand{\ddf}{f''}
\newcommand{\ddg}{g''}
\newcommand{\ddh}{h''}

\newcommand\modv[1]{|{#1}|}

\newcommand\arr[1]{\overrightarrow{#1}}
\newcommand{\cartref}{\{O;x_1,x_2,x_3\}}
\newcommand{\orthframe}{({\bf e}_1, {\bf e}_2, {\bf e}_3)}


\newcommand\email[1]{\texttt{#1}}
\newcommand\at{:}

\begin{center}
 {\bf \Large
A New CNT-Oriented Shell Theory}
\end{center}
\medskip

\begin{center}
{\large Antonino Favata \quad
                Paolo Podio-Guidugli
}\end{center}

\begin{center}
 \noindent Dipartimento di Ingegneria Civile, Universit\`a di Roma Tor Vergata\footnote{Via Politecnico 1, 00133 Rome, Italy. \\
{\null} \quad \ Email:
\begin{minipage}[t]{30em}
\email{favata@ing.uniroma2.it} (A. Favata)\\
\email{ppg@uniroma2.it} (P. Podio-Guidugli)
\end{minipage}}
\small
\end{center}
\medskip

\begin{abstract}
\noindent {\footnotesize A theory of linearly elastic orthotropic
shells is presented, with potential application to the continuous
modeling of Carbon NanoTubes. Two relevant features are: the
selected type of \emph{orthotropic response}, which should be
suitable to capture differences in chirality; the possibility of
accounting for \emph{thickness changes} due to changes in
inter-wall separation to be expected in multi-wall CNTs. A simpler
version of the theory is also proposed, in which orthotropy is
preserved but thickness changes are excluded, intended for
possible application to single-wall CNTs. Another feature of both
versions of the present theory is boundary-value problems of
torsion,
 axial traction, uniform inner pressure, and rim flexure, can be solved explicitly in closed form. Various directions of ongoing further
 research are indicated.
\medskip

}
\end{abstract}

\section{Introduction}
\begin{figure}[h]
\centering
\includegraphics[scale=0.8]{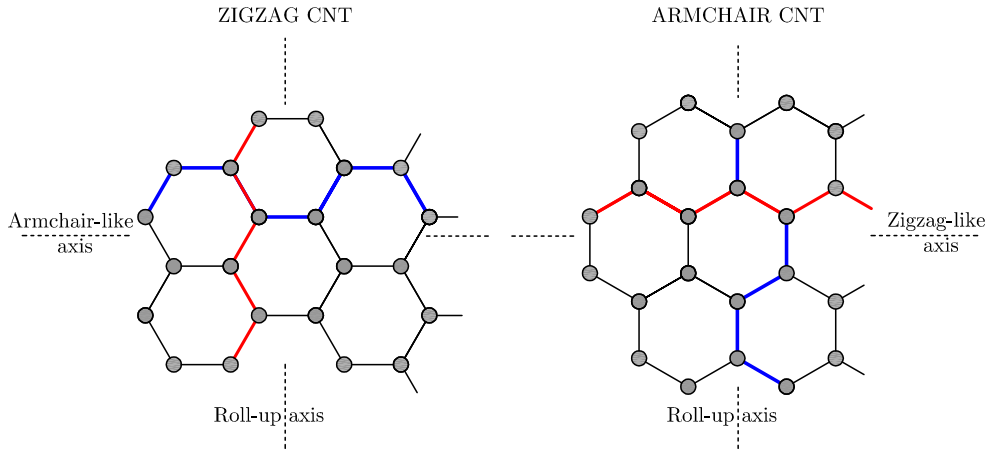}
\caption{Roll-up axes of zig-zag and armchair carbon nanotubes.}
\label{ort}
\end{figure}

\begin{figure}[h]
\centering
\includegraphics[scale=0.8]{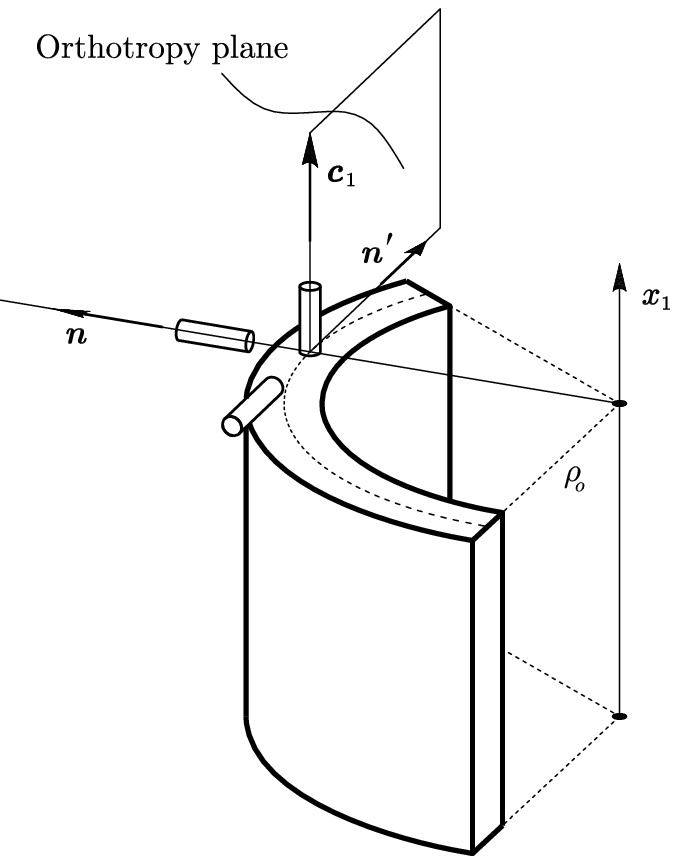}
\caption{Shell axis is chosen parallel to roll-up axis.}
\label{ortho}
\end{figure}

The application that motivated this work is the modeling of carbon
nanotubes (CNTs). When CNTs are employed as nanodevice components,
they are regarded as elastic beam-like or shell-like objects and
their mechanical response is characterized in terms of an
as-small-as-possible number of stiffness and inertia parameters.
To define and evaluate these parameters
%in terms of the relevant physical interaction forces and the resultant geometry
is the common goal of all modelers; a way to achieve it is to try
and bridge between the microscopic scale of \emph{molecular
mechanics} and the macroscopic scale of \emph{continuous structure
mechanics}, by way of a mesoscopic scale, at which concepts from
\emph{discrete structure mechanics} apply. At the onset of putting
together a bottom-up model of this sort \cite{CFPG}, we realized
that ordinary shell theories, which presume an \emph{isotropic}
three-dimensional response of the material comprising the shell,
could not possibly guarantee an accurate macroscopic account of
the mesoscopic texture of single-wall carbon nanotubes (SWCNTs): a
glance to armchair and zigzag CNTs (Figure \ref{ort})
suggests instead an \emph{orthotropic} response in planes
orthogonal to radial directions (see Figure \ref{ortho},
where the three little cylinders suggest what probes one should
cut out of a cylindrical shell-like body in order to determine its
material moduli). Moreover, when modeling multi-wall carbon
nanotubes (MWCNTs), it seems to us important to allow for
thickness distension: we conjecture that thickness changes are
essentially due to changes in the inter-wall distances, as a
consequence of the interplay of the applied loads with the van der
Waals interactions between adjacent walls. A search of the
literature convinced us that we better  produced such a shell
theory ourselves. This paper describes the results of our efforts,
results that have been partly anticipated in \cite{FPG} and that
-- so we believe -- may find application also in contexts
different from the mechanics of nanotubes.

In the next section, we expound the general lines of a theory of
linearly elastic orthotropic shells of constant referential
thickness $2\varepsilon$, whose \emph{geometry} (Section
\ref{geo}) is dictated by a piecewise smooth referential model
surface $\Sc$. Imitating the classic approach of Kirchhoff, we
specify the admissible \emph{kinematics} (Section \ref{kin}) by
choosing for the displacement field in the tubular region
$\Gc(\Sc,\varepsilon)$  a representation parameterized by a few
fields defined over $\Sc$. In the deformations we envisage, (i)
thickness may change, if the applied loads require and the
boundary conditions permit; (ii) material fibers orthogonal to
$\Sc$ must remain orthogonal to the deformed shape of $\Sc$
itself, an internal constraint we refer to as
\emph{unshearability}. Both the \emph{balance} and the
\emph{constitutive equations} of our shell theory (Sections
\ref{Equilibrium} and \ref{costa}, resp.) are inherently
consistent with the corresponding equations of three-dimensional
linear elasticity:

\noindent  - the balance equations follow from a two-dimensional
Principle of Virtual Powers that is a direct consequence of
stating the corresponding three-dimensional Principle for all
virtual velocity fields  in the linear space to which the
admissible displacement belong; they are expressed in terms of a
pair of \emph{two-dimensional stress measures} that are defined as
weighted thickness averages of the three-dimensional stress field
in $\Gc(\Sc,\varepsilon)$ (Section \ref{fmv});

\noindent - the constitutive equations are arrived at when the
three-dimensional constitutive equations for unshearable
orthotropic materials are inserted in the definitions of the
two-dimensional stress measures.

%derived in Section \ref{costa}.
% depending on seven material moduli.
%
%able to distend their thickness, and affected to the internal constraint to preserve the orthogonality
%between the transversal fibers and the middle surface. The presence of the internal constraint allow to handle with a constitutive
%law characterized by seven material moduli (in spite of nine of the unconstrained material) and to reduce the number of the equations, since
%reactive stress appears; in this way explicit analytic solutions are reasonably expected. The theory proposed is consistent with
%three-dimensional elasticity,
%in the sense that a rational deduction of from this latter is furnished, as close
%as possible in clarity and rigor to Gurtin's presentation of that theory in his
%well-known Handbuch article. We use a restricted version of the three-dimensional
% to deduce a two-dimensional version of this principle
% and the associated balance laws for shells in terms of two-dimensional stress measures and applied loads; the kinematics is based on
% a reasonable  Ansatz deducted from the prevailing internal constraint. For bigger clarity, we use direct notation, getting rid as much
%as possible of the usual esoteric paraphernalia from cartesian tensor algebra,
%surface geometry and elementary differential calculus on manifolds, which sink
%the reader in a scaring indicial sea and obstruct the way to a thorough
%understanding of the mathematical and mechanical structures underlying the
%subject.

The remaining part of the paper is dedicated to \emph{cylindrical
shells}. We begin with shells whose thickness can change. In
sections from 3.1 to 3.4, we parallel and specify the developments
of Section 2 as to, respectively, geometry, kinematics, balance
laws, and constitutive equations. Then, we confine attention to
\emph{axisymmetric equilibrium problems}, and solve explicitly and
exactly those of \emph{torsion} and \emph{axial traction}
(Sections 4.1 and 4.2) -- the cases for which experimental tests
and numerical simulations seem to be especially easy to set up for
CNTs -- as well as the problems of \emph{pressure} and \emph{rim
flexure} (Sections 4.3 and 4.4). Finally, we lay down natural
geometrical notions of  \emph{thinness} and \emph{slenderness},
and we show how remarkably the analytical solutions derived in the
previous section simplify for slender shells (Sections 5.1 and
5.2), and how \emph{effective contraction moduli} and
\emph{effective stiffnesses} can be defined for a cylindrical
shell, regarded as a traction or torsion probe (Sections 5.3 and
5.4).

Next, in Section 6, we take up orthotropic shells whose thickness
is constitutively immutable, a class that we designate by the
names of Kirchhoff and Love by analogy with the corresponding
classic plate theory. We adapt to the simpler case of
Kirchhoff-Love cylindrical shells all the formulas derived in
Sections 4 and 5, both for whatever thinness and in the small
thickness limit;  in the latter case, we show how the four
constitutive moduli characterizing the mechanical response could
be determined on the basis of simple real or computer experiments.

In our final Section 7, we briefly recapitulate our main findings,
and we indicate the directions of our future research, with
special attention to the application to CNTs of the concepts and
methods developed in the present paper.

\section{General Theory}
\subsection{Geometry}\label{geo}
In this opening section we recapitulate some well known notions,
with the main purpose of introducing our notation and terminology.

\begin{figure}[h]
\centering
\includegraphics[scale=.93]{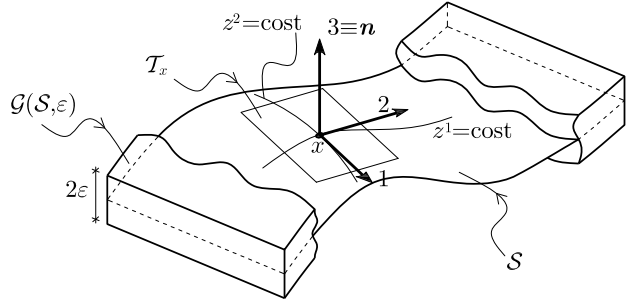}
\caption{Geometrical equipment of a typical shell-shaped region.}
\label{she}
\end{figure}

Following the approach to construct a shell theory proposed in
\cite{PPG1}, we let $\Sc$ denote a compact, regular, orientable
and oriented surface embedded in the three-dimensional Euclidean
space $\Ec$, and we let $x$ denote its typical point and $\nb(x)$
the value of its normal vector field at $x$, with $|\nb(x)|\equiv
1$. We choose an origin $o\in\Ec$, and denote by $\xb:=x-o$ the
position vector of $x$ with respect to $o$. We assume that $\Sc$
admits a \emph{tubular $\varepsilon-$neighborhood}
$\Gc(\Sc,\varepsilon)$ (see Section 2.2 of \cite{DoC}) and a
\emph{global parametrization}
$$
\Ro^2\supset\mathrm{S}\ni(z^1,z^2)\mapsto
x(z^1,z^2)\in\Sc\subset\Ec
$$
(here $\mathrm{S}$ is an open set). A point $p\in
\Gc(\Sc,\varepsilon)$ has position vector
\[
\pb:=p-o=x-o+\zeta\nb(x),\quad x\in\Sc,\;\zeta\in
I:=(-\varepsilon,+\varepsilon),
\]
with respect to $o$; $|\zeta|$ is the distance of $p$ from $x$,
the point where the straight line through $p$ perpendicular to
$\Sc$ intersects $\Sc$ itself. The mapping
\[
(z^1,z^2,\zeta)\mapsto
p(z^1,z^2,\zeta):=x(z^1,z^2,\zeta)+\zeta\nb(x(z^1,z^2))
\]
is a global parametrization of $\Gc(\Sc,\varepsilon)$, with
$(z^1,z^2,\zeta)$ the triplet of \emph{normal curvilinear
coordinates} of $p$ (Figure \ref{she}).
We term the region $\Gc(\Sc,\varepsilon)$ of $\Ec$ a
\emph{shell-shaped region},
%$-$ in short, a \emph{shell} $-$
of \emph{model surface} $\Sc$ and constant \emph{thickness}
$2\varepsilon$. The thickness of $\Gc(\Sc,\varepsilon)$ can be
visualized as the length, whatever $x\in\Sc$ one picks, of the
material fiber ${\mathcal F}(x)$ through $x$ perpendicular to
$\Sc$; clearly, ${\mathcal F}(x):=\{p\in \Gc(\Sc,\varepsilon)\,|\;
p=x+\zeta\nb(x)$.
%
%\subsubsection{Model surfaces}
%

The chosen parametrization induces a system of \emph{coordinate
curves} on $\Sc$, described by the mappings $z^\alpha\mapsto
x(z^1,z^2) \;\,(\alpha=1,2)$. The \emph{tangent space}
$\Tc_x:={\rm span}\,\{\eb_\alpha(x)\}$ to $\Sc$ at $x\equiv
(z^1,z^2)$ is spanned by the tangent vectors to the coordinate
curves at that point:
$$
\eb_\alpha(z^1,z^2):=x_{,\alpha}(z^1,z^2).
$$
On taking the normal field to be $\nb(x) =
\mathrm{vers}\big(\eb_1(x)\times \eb_2(x)\big)$, and on setting
$$
\varsigma(x):=\eb_1(x)\times\eb_2(x)\cdot\nb(x)>0,
$$
 the \emph{covariant} and \emph{contravariant bases} at $x$  are, respectively, $\{\eb_1(x),\eb_2(x),\nb(x)\}$ and
$\{\eb^1(x),\eb^2(x),\nb(x)\}$, where
$$
\varsigma(x)\eb^\alpha(x):=(-1)^\alpha\nb(x)\times\eb_{\alpha+1}(x)
\quad (\mathrm{modulo\; 2},\; \alpha \;\mathrm{not \;summed}).
$$
For the rest of this subsection we leave the indication of the
typical point $x\in\Sc$ tacit. Accordingly, we write
$$
\Pb:=\eb_i\otimes\eb^i=\eb^i\otimes\eb_i
$$
for  the \emph{metric tensor} and
$$
\Pbs:=\eb_\alpha\otimes\eb^\alpha=\eb^\alpha\otimes\eb_\alpha
$$
for the \emph{surface metric tensor}.

A vector field $\vb$ defined over $\Sc$ can be represented both in
the covariant basis and in the controvariant basis:
$$
\vb=v_i\eb^i=v^j\eb_j,\quad\textrm{with}\quad v_i:=\vb\cdot\eb_i,
\quad v^j:=\vb\cdot\eb^j,  \quad (i,j=1,2,3).
$$
 Analogously, a second-order tensor field $\Tb$ can be represented as
 $$
\Tb=T^{ij}\eb_i\otimes\eb_j=T_{ij}\eb^i\otimes\eb^j=T^{i}_j\eb_i\otimes\eb^j=T_i^j\eb^i\otimes\eb_j,
 $$
in terms of its covariant, contravariant, or mixed components
$T^{ij}=\Tb\cdot\eb^i\otimes\eb^j$,
$T_{ij}=\Tb\cdot\eb_i\otimes\eb_j$, or
$T^{i}_j=\Tb\cdot\eb^i\otimes\eb_j$ and
$T_i^j=\Tb\cdot\eb_i\otimes\eb^j$ $(i,j=1,2,3)$. \vskip 6pt
\remark \label{rem1} The physical dimensions of covariant and
contravariant basis vectors, and hence of the corresponding
components of vectors and tensors, may differ. To circumvent this
difficulty is easy, whenever it so happens that
\[
\eb_1\cdot\eb_2=0.
\]
Simply, one introduces the so called \emph{physical basis} at $x$,
that is to say, the orthonormal basis
$$\{\eb{\<{1}}(x), \eb{\<2}(x), \nb(x)\}, \quad
\eb{\<\alpha}:=\frac{\eb_\alpha}{|\eb_{\alpha}|}=\frac{\eb^\alpha}{|\eb^{\alpha}|}\,.
$$
We shall be making use of physical bases and components (e.g.,
$v{\<h}:=\vb\cdot\eb{\<h}$) in Section \ref{cilscel}, where we
deal with shells whose model surface is a right circular cylinder.
\vskip 6pt
%
%\subsubsection{Shell-shaped regions}\label{ssb}
%

At a point of $\Gc(\Sc,\varepsilon)$, the covariant and
contravariant basis vectors are, respectively,
\begin{equation}\label{basv}
\gb_\alpha:=p_{,\alpha}=\eb_\alpha+\zeta\nb_{,\alpha}, \quad
\gb_3:=p_{,3}=\nb;
\end{equation}
and
$$
\varsigma\gb^\alpha:=(-1)^\alpha\nb\times\gb_{\alpha+1} \quad
(\mathrm{modulo\; 2},\; \alpha \;\mathrm{not \;summed}),\quad
\gb^3=\nb,
$$
where
$$
\varsigma:=\gb_1\times\gb_2\cdot\nb.
$$
The \emph{metric tensor} $\Gb$ is defined to be:
$$
\Gb:=\gb^i\otimes\gb_i=\gb_i\otimes\gb^i,
$$
with
$$
\Gs:=\gb^\alpha\otimes\gb_\alpha
$$
its surface part.

The covariant bases $\{\eb_i\}$ at $x\in\Sc$ and $\{\gb_i\}$ at
$x+\zeta\nb(x)=p\in \Gc(\Sc,\varepsilon)$ are related by the
\emph{shift tensor} $-$ briefly, the \emph{shifter} $-$ $\Ab$:
$$
\Ab(x,\zeta)=\gb_i\otimes\eb^i\quad\Leftrightarrow\quad
\Ab\eb_k=\gb_k;
$$
it can be shown \cite{PPG1} that the ratio of the volume measures
at $p$ and $x$  is equal to the determinant of $\Ab$:
\begin{equation}\label{detalfa}
dvol(p)=\alpha(x,\zeta)\,dvol(x),\quad \alpha:=\det\Ab.
\end{equation}
The shifter
$$
\Bb(x,\zeta):=\gb^i\otimes\eb_i=\Bsh+\nb\otimes\nb,\quad
\Bsh:=\gb^\alpha\otimes\eb_\alpha,
$$
maps the controvariant basis at $x$ into the controvariant basis
at $p$. We have that
$$
\Bb^T\Ab=\Pb, \quad \Ab\Bb^T=\Gb.
$$
Note that
\begin{equation}\label{APW}
\Ab(x,\zeta)=\Ash(x,\zeta)+\nb(x)\otimes\nb(x),\quad
\Ash(x,\zeta):=\gb_\alpha\otimes\eb^\alpha=\Pbs(x)-\zeta \Wb(x),
\end{equation}
where
\[
\Wb:=-\nablas\nb=-\nb,_\alpha\otimes\eb^\alpha
\]
is the \emph{curvature tensor} of the oriented surface $\Sc$. We
have here denoted by $\nablas$ the operation of taking the
\emph{surface gradient} of  a smooth vector field $\vb$ over
$\Sc$: $\nablas\vb=\vb,_\alpha\otimes\eb^\alpha$. Likewise, we
denote by $\nabla$ the \emph{gradient} of a vector field $\vb$
defined over $\Gc(\Sc,\varepsilon)$:
$\nabla\vb=\vb,_i\otimes\gb^i$. Two \emph{divergence operators}
are associated with the gradient operators,
$\Div\vb:=\nabla\vb\cdot\Gb$ and $\Divs\vb:=\nablas\vb\cdot\Pbs$,
where the field $\vb$ is defined, respectively, over
$\Gc(\Sc,\varepsilon)$ and over $\Sc$. The \emph{surface
divergence} of a tensor field $\Tb$ over $\Sc$ is defined as
follows:  $\Divs(\Tb^T\vb)=:\Divs\!\Tb\cdot\vb$, for all constant
vectors $\vb$.

Let $\widetilde\ab$ the fiber-wise constant extension  to
$\Gc(\Sc,\varepsilon)$ of a vector field $\ab$ defined over $\Sc$.
When taking the gradient of $\widetilde\ab$, we have that
\[
\nabla\widetilde\ab=\widetilde\ab_{,\alpha}\otimes\gb^\alpha=\widetilde\ab_{,\alpha}\otimes(\Bsh\eb^{\alpha})=(\nablas\ab)\Bsh^T.
\]
In the following, we will not make any notational distinction
between a field defined over $\Sc$ and its  fiber-wise constant
extension  to  $\Gc(\Sc,\varepsilon)$: e.g., we shall write:
\begin{equation}\label{sugra}
\nabla\ab=(\nablas\ab)\Bsh^T.
\end{equation}
%

%In what follows, it will be clear from the context whether we regard a given vector field
%
\subsection{Kinematics}\label{kin}
The three-dimensional strain measure we use is the standard
\emph{symmetrized gradient} of the displacement field:
\[
\Eb(\ub)=\sym\nabla\ub:=\frac{1}{2}\,(\nabla\ub+\nabla\ub^T).
\]
It follows from this definition that the covariant components of
$\Eb$ are:
\[
\begin{aligned}
2E_{ij}=2\Eb\cdot\gb_i\otimes\gb_j&=(\ub_{,k}\otimes\gb^k+\gb^k\otimes\ub_{,k})\cdot(\gb_i\otimes\gb_j)\\&=(\ub\cdot\gb_i),_j+(\ub\cdot\gb_j),_i-\ub\cdot(\gb_i,_j+\gb_{j,i}),
\end{aligned}
\]
whence
\begin{equation}\label{compE}
E_{ij}=\frac{1}{2}\Big(u_{i,j}+u_{j,i}-\ub\cdot(\gb_i,_j+\gb_j,_i)\Big).
\end{equation}
%
%where $\Gamma_{ij}^h$ are the \emph{Christoffel symbols}.

We restrict attention to shell-shaped bodies
$\Gc(\Sc,\varepsilon)$ whose admissible deformations may induce
thickness changes but must keep the material fibers orthogonal to
the model surface, in the sense that, whatever $x\in\Sc$, the
deformed material fiber $\ub({\mathcal F}(x))$ must be found
orthogonal to the deformed model surface $\ub(\Sc)$. This
pointwise \emph{internal constraint} $-$ a restriction on
admissible displacement fields that, as anticipated, we refer to
as \emph{unshearability} $-$ can be expressed in terms of the
linear strain measure $\Eb$ in the following form:
\begin{equation}\label{previn}
\Eb(\ub)\nb\cdot\vb=0\quad\textrm{for all}\;\,\vb\;\textrm{such
that}\;\, \vb\cdot\nb=0,
\end{equation}
or rather, equivalently, as
\begin{equation}\label{vin}
\Eb(\ub)\nb\cdot\gb_\alpha=0\quad\textrm{in}\;\,\Gc(\Sc,\varepsilon);
\end{equation}
with the use of \eqref{basv} and \eqref{compE}, \eqref{vin}
becomes:
\begin{equation}\label{vinc}
(\ub\cdot\gb_\alpha),_3+(\ub\cdot\nb),_\alpha -
2\,\ub\cdot\nb,_\alpha=0,
%\footnote{Alternatively, these two relations may be written in the form of a system of PDEs:
%%
%\[
%u_{\alpha,3}+u_{3,\alpha}-2\,\ub\cdot\nb,_\alpha.
%\]
%%
%}
\end{equation}
with the same quantification. We look for solutions of this system
of two PDEs having the following form:
\begin{equation}\label{displ}
\ub(x,\zeta)=\overset{(0)}{\ub}(x)+\zeta\overset{(1)}{\ub}(x);
\end{equation}
%
%Since
%%
%\[
%\nabla\ub=\big(\overset{(0)}{\ub}+\zeta\overset{(1)}{\ub}),_\alpha\otimes\gb^\alpha+\overset{(1)}{\ub}\otimes\nb,
%\]
%%
%
note, in particular, that
$$
\overset{(1)}{\ub}(x)\cdot\nb(x)=\Eb(\ub(x))\cdot\nb(x)\otimes\nb(x)
$$
is the stretch of the material fiber ${\mathcal F}(x)$, uniform
all along it. Substituting \eqref{displ} into \eqref{vinc}, and
exploiting the quantification with respect to the variable
$\zeta$, we find the following restrictions on the choice of the
fields $\overset{(0)}{\ub},\overset{(1)}{\ub}$ over $\Sc$:
\begin{equation}\label{restr}
\overset{(1)}{\ub}\cdot\eb_\alpha= -
\overset{(0)}{\ub},_\alpha\cdot\nb,\qquad
\big(\overset{(1)}{\ub}\cdot\nb\big),_\alpha=0.
\end{equation}
The second of \eqref{restr} tells us that, for \eqref{vin}
\emph{to have a solution of type \eqref{displ}, the fiber stretch
must in fact be uniform all over} $\Gc(\Sc,\varepsilon)$;
moreover, the tangential part of $\overset{(1)}{\ub}$ must be
expressed as follows in terms of $\overset{(1)}{\ub}$  and the
 curvature tensor of $\Sc$:
\[
\overset{(1)}{\ub}-\big(\overset{(1)}{\ub}\cdot\nb\big)\nb=-\nablas\big(\overset{(0)}{\ub}\cdot\nb\big)-\Wb\overset{(0)}{\ub}.
\]

With the use of \eqref{APW}, we conclude that we should pick:
\begin{equation}\label{KLdispl}
{\ub}_S(x,\zeta):=\Ash(x,\zeta){\ab}(x)+w(x)\nb(x)+\zeta\big(-\nablas
w(x)+\gamma\nb(x)\big),
\end{equation}
where we have set:
\begin{equation}\label{displadd}
\ab:=\overset{(0)}{\ub}-\big(\overset{(0)}{\ub}\cdot\nb\big)\nb,\quad
w:=\overset{(0)}{\ub}\cdot\nb,\quad
\gamma:=\overset{(1)}{\ub}\cdot\nb.
\end{equation}
Note that the displacement field \eqref{KLdispl} is parameterized
by two fields over $\Sc$, the tangential vector field $\ab$ and
the scalar field $w$, and by the constant $\gamma$.\footnote{A
vector field $\ab$ (a tensor field $\Tb$) defined over $\Sc$ is
termed \emph{tangential} if it so happens that
$\ab(x)\cdot\nb(x)\equiv 0$ ($\Tb\nb\equiv \0$).} Note also that,
by taking $\gamma=0$, we recover the kinematic Ansatz typical of
the linear theory of \emph{Kirchhoff-Love shells}  \cite{PPG1};
interestingly, the displacement field:
\begin{equation}\label{kappal}
{\ub}_{K\!L}(x,\zeta)=\Ash(x,\zeta){\ab}(x)+w(x)\nb(x)-\zeta\nablas
w(x),\quad \ab(x)\cdot\nb(x)=0,
\end{equation}
can be shown to be the general solution of the system of three
PDEs that expresses the Kirchhoff-Love constraint, namely,
\begin{equation}\label{kl}
\Eb(\ub)\nb= \0\quad\textrm{in}\;\,\Gc(\Sc,\varepsilon)
\end{equation}
(cf. \eqref{vin}).
\remark That the fiber stretch $\gamma$ defined by
$\eqref{displadd}_3$ be constant at any point of
$\Gc(\Sc,\varepsilon)$ to satisfy the second of \eqref{restr} is a
counterintuitive consequence of expressing the unshearability
constraint in terms of the linear measure of deformation
$\Eb(\ub)$. In fact, on adopting one of the standard exact
deformation measures, namely,
$$
\Db:=\frac{1}{2}\big((\nabla f)^T(\nabla f)-\Ib\big),
\quad\textrm{where}\quad f(p):=p+\ub(p),
$$
one finds that
$$
\Db=\Eb+\frac{1}{2}\Hb^T\Hb,\quad \Hb:=\nabla\ub.
$$
Thus, the exact conterpart of \eqref{vin} is:
\[
0=\Db\nb\cdot\gb_\alpha=\Eb\nb\cdot\gb_\alpha+\frac{1}{2}\,\Hb\nb\cdot\Hb\gb_\alpha,
%\footnote{In fact
%$2\Db\nb\cdot\gb_\alpha=\Fb^T\Fb\nb\cdot\gb_\alpha=\Fb\nb\cdot\Fb\gb_\alpha$.}.
\]
where the term quadratic in $|\Hb|$ does not vanish in general. In
particular, for $\ub$ of the form \eqref{displ}, we find that
\[
\Hb\nb\cdot\Hb\gb_\alpha=\overset{(1)}{\ub}\cdot{\overset{(0)}{\ub}},_\alpha+\zeta\overset{(1)}{\ub}\cdot{\overset{(1)}{\ub}},_\alpha;
\]
hence, the second of \eqref{restr} is replaced by
\[
\big(\overset{(1)}{\ub}\cdot\nb+|\overset{(1)}{\ub}|^2\big),_\alpha=0,
\]
and intuition rings no bell.

\subsection{Balance Assumptions}\label{Equilibrium}
Given a material body  occupying a three-dimensional open and
bounded region $\Omega$ and a velocity field $\vb$ over $\Omega$,
the \emph{internal power expenditure} over a part $\Pi$ of
$\Omega$ associated with $\vb$  is:
\[
\int_\Pi\Sb\cdot\nabla\vb
\]
where  $\Sb$ denotes the stress field in $\Omega$; and,  the
\emph{external virtual power expenditure} over $\Pi$ is:
\[
\int_\Pi\db_o\cdot \vb+\int_{\partial\Pi}\cb_o\cdot\vb,
\]
where $(\db_o,\cb_o)$ denote, respectively, the \emph{distance
force} for unit volume and the \emph{contact force} per unit area
exerted on $\Pi$ by its own complement with respect to $\Omega$
and by the environment of the latter. These representations of
power expenditures are those typical in the theory of the
so-called \emph{simple} material bodies. In that theory, a part is
customarily a subset of non-null volume of $\Omega$ (which makes
it for a part collection deemed sufficiently rich), and a
\emph{virtual velocity field} is a smooth vector field whose
support is a part; the \emph{Principle of Virtual Powers} is the
stipulation that
\begin{equation}\label{PVPa}
\int_\Omega\Sb\cdot\nabla\vb=\int_\Omega\db_o\cdot\vb+\int_{\partial\Omega}\cb_o\cdot\vb
\end{equation}
for all virtual velocities in a collection modeled after a chosen
collection of admissible motions, an alternative stipulation being
\begin{equation}\label{PVP}
\int_\Pi\Sb\cdot\nabla\vb=\int_\Pi\db_o\cdot\vb+\int_{\partial\Pi}\cb_o\cdot\vb
\end{equation}
for all parts $\Pi$ of $\Omega$ and for all velocity fields $\vb$
consistent with the admissible motions.

The virtual-power principle is interpreted as a \emph{balance
statement} for the internal and external fields  $\Sb$ and
$\db_o,\cb_o$.
%
%The Principle of Virtual
%Powers is the stipulation that
%%
%\begin{equation}\label{PVP}
%\int_\Pi\Sb\cdot\nabla\vb=\int_\Pi\db_o\cdot\vb+\int_{\partial\Pi}\cb_o\cdot\vb
%\end{equation}
%%
%for all parts $\Pi$ of $\Omega$ and for all virtual velocities in
%a suitably rich collection.
As exemplified in \cite{PPG} for beams and plates, we use here
below a restricted version of the three-dimensional statement
\eqref{PVP} to deduce a two-dimensional Principle of Virtual
Powers and the associated balance laws for shells in terms of
two-dimensional stress measures and applied loads. The first
restriction we introduce has to do with the \emph{special shape}
of the three-dimensional bodies we consider: they must be
shell-shaped in the sense of Section \ref{geo}. The second
restriction stems from the
 \emph{special class of admissible body parts} we
choose: they all must have the same thickness of the shell-like
body. Thirdly and lastly, we pick a \emph{special class of virtual
velocities}, consistent with the representations of admissible
displacements discussed in the previous section.
%
%with
%%
%\[
%\overset{(0)}\vb=\ab+ v\nb,\quad \overset{(1)}\vb=\bb+\varphi\nb,\quad \ab\cdot\nb=\bb\cdot\nb=0,%\quad \widetilde\bb:=-\Wb\widetilde\ab-\nablas\widetilde w
%\]
%%
%the fields $\ab,v,\bb$ being defined over $\Sc$, and $\varphi$ being a constant. %,  in order to comply with \eqref{KLdispl}.
%
%
\subsubsection{Internal power expenditure. Stress measures}
The first two restrictions we listed imply that a typical part of
$\Omega\equiv\Gc(\Sc,\varepsilon)$ can be identified with the
Cartesian product $\Pi=\Pc\times I$ of $\Pc$, an open subset of
$\Sc$, and the interval $I$. As to  virtual velocities, we choose
them of the form \eqref{displ}:
\begin{equation}\label{varia}
\vb(x,\zeta)=\overset{(0)}\vb(x)+\zeta \overset{(1)}\vb(x),
\end{equation}
with $\Pc$ the intersection of the supports of the vector fields
$\overset{(0)}\vb$ and $\overset{(1)}\vb$.\footnote{Choosing
instead virtual velocities of the less general, constrained form
\eqref{KLdispl} would preclude the appearance of reactive terms in
the balance equations following from \eqref{PVP}. We shall return
on this important issue after completion of the generating
procedure we are now about to start, in Remark \ref{rmk4} at the
end of Section \ref{fmv}. } On recalling \eqref{sugra}, we have
that:
\[
\nabla\vb=\big(\nablas\overset{(0)}\vb+\zeta
\nablas\overset{(1)}\vb\big)\Bsh^T+\overset{(1)}\vb\otimes\nb.
\]
Hence,
$$
\Sb\cdot\nabla\vb=(\Sb\Bsh)\cdot\nablas\overset{(0)}\vb+\zeta(\Sb\Bsh)\cdot\nablas\overset{(1)}\vb+\Sb\nb\cdot\overset{(1)}\vb,
$$
so that, having recourse to Fubini theorem and recalling
\eqref{detalfa}, the internal power expenditure reads:
$$
\int_\Pi\Sb\cdot\nabla\vb=\int_\Pc\big(
\Fbs\cdot\nablas\overset{(0)}\vb+\Mbs\cdot\nablas\overset{(1)}\vb
+\fb^{(3)}\cdot\overset{(1)}\vb\big).
$$
Here we have made use of the following definitions:
\begin{equation}\label{2Ds}
\begin{aligned}
\Fbs(x)&:=\int_I\alpha(x,\zeta)\Sb(x,\zeta)\Bsh(x,\zeta) d\zeta=\Big(\int_I\alpha(x,\zeta)\Sb(x,\zeta)\gb^\alpha(x,\zeta) d\zeta\Big)\otimes\eb_\alpha,\\
\Mbs(x)&:=\int_I\alpha(x,\zeta)\zeta\Sb(x,\zeta)\Bsh(x,\zeta)
d\zeta=\Big(\int_I\alpha(x,\zeta)\zeta\Sb(x,\zeta)\gb^\alpha(x,\zeta)
d\zeta\Big)\otimes\eb_\alpha,
\end{aligned}
\end{equation}
and
\begin{equation}\label{2Dsh}
\fb^{(3)}(x):=\int_I\alpha(x,\zeta)\Sb(x,\zeta)\nb(x) d\zeta,
\end{equation}
for the two-dimensional stress measures that we call,
respectively,  \emph{force tensor}, \emph{moment tensor}, and
\emph{shear vector}.

It is important to observe that the force and moment tensors are
tangential. Now, for a tangential tensor field $\Tb$ over $\Sc$,
it follows from the general identity
$$
\Tb\cdot\nablas\vb=\Divs(\Tb^T\vb)-\Divs\!\Tb\cdot\vb
$$
and the standard divergence theorem that
$$
\int_\Pc\Tb\cdot\nablas\vb=\int_{\partial\Pc}\Tb\mb\cdot\vb-\int_\Pc\Divs\!\Tb\cdot\vb,
$$
where, if $\tb$ is the tangential vector to the curve
$\partial\Pc$ in the tangent plane to the surface $\Sc$,
$\mb=\tb\times\nb$ is the normal to $\partial\Pc$. Thus,
$$
\int_\Pi\Sb\cdot\nabla\vb=\int_\Pc\Big(
-\Divs\Fbs\cdot\overset{(0)}\vb+(-\Divs\Mbs+\fb^{(3)})\cdot\overset{(1)}\vb
\Big)+\int_{\partial\Pc}\Big(
\Fbs\!\mb\cdot\overset{(0)}\vb+\Mbs\!\mb\cdot\overset{(1)}\vb
\Big).
$$
% and the We will use the following nomenclature:
%\begin{equation}
%\left\{ \begin{array}{ll}
%F{\<{\alpha\alpha}}:=\Fb\cdot\eb{\<\alpha}\otimes\eb{\<\alpha} & \quad \textrm{normal membrane forces},\\
%F{\<{\alpha\beta}}:=\Fb\cdot\eb{\<\alpha}\otimes\eb{\<\beta} &
%\quad \textrm{shear membrane forces},\\
%f_{3}{\<\alpha}:=\fb^{(3)}\cdot\eb{\<\alpha}& \quad
%\textrm{transversal shear forces},
%\end{array} \right.
%\end{equation}

%\begin{equation}
%\left\{ \begin{array}{ll}
%M{\<{\alpha\alpha}}:=\Mb\cdot\eb{\<\alpha}\otimes\eb{\<\alpha} & \quad \textrm{bending moments},\\
%M{\<{\alpha\beta>}}:=\Mb\cdot\eb{\<\alpha}\otimes\eb{\<\beta} &
%\quad \textrm{twisting moments},\\
%\end{array} \right.
%\end{equation}

%
%\begin{equation}
%\left\{ \begin{array}{ll}
%f_3^3:=\fb^{(3)}\cdot\nb & \quad \textrm{thickness force},\\
%m_{3}{\<\alpha}:=\mb_3\cdot\eb{\<\alpha} &
%\quad \textrm{thickness moments}.\\
%\end{array} \right.
%\end{equation}
%

\subsubsection{External power expenditure. Applied loads}
As to power expenditure of the applied forces, we find:
$$
\int_\Pi\db_o\cdot\vb=\int_\Pc\Big( \big( \int_I\alpha\db_o
\big)\cdot\overset{(0)}\vb+\big( \int_I\alpha\zeta\db_o
\big)\cdot\overset{(1)}\vb \Big)
$$
for the distance force, and
\begin{equation}
\begin{aligned}
\int_{\partial\Pi}\cb_o\cdot\vb&=\int_\Pc\Big((\alpha^+\cb_o^++\alpha^-\cb_o^-)\cdot\overset{(0)}\vb+\varepsilon(\alpha^+\cb_o^+-\alpha^-\cb_o^-)\cdot\overset{(1)}\vb
\Big)+\\
&+\int_{\partial\Pc}\Big(
\big(\int_I\cb_o\big)\cdot\overset{(0)}\vb
+\big(\int_I\zeta\cb_o\big)\cdot\overset{(1)}\vb\Big)
\end{aligned}
\end{equation}
for the contact force,\footnote{Here we have made use of the fact
that
$\partial\Pi=\{\Pc\times\{\pm\varepsilon\}\}\cup\{\partial\Pc\times
I \}$.} where we have set:
$$
\alpha^\pm(x):=\alpha(x,\pm\varepsilon), \quad
\cb_o^\pm(x):=\cb_o(x,\pm\varepsilon), \quad x\in\Pc.
$$
We are now in a position to define the load fields over $\Sc$
induced by the three-dimensional fields $(\db_o,\cb_o)$. These
are:
\begin{itemize}
\item the \emph{distance force} and \emph{distance couple}
per unit area
\begin{equation}\label{forceload}
\begin{aligned}
\qb_o(x)&:=\int_I\alpha(x,\zeta)\db_o(x,\zeta)\,d\zeta+\alpha^+(x)\cb_o^+(x)+\alpha^-(x)\cb_o^-(x),\\
\rb_o(x)&:=\int_I\alpha(x,\zeta)\zeta\db_o(x,\zeta)\,d\zeta+\varepsilon\big(\alpha^+(x)\cb_o^+(x)-\alpha^-(x)\cb_o^-(x)\big);
\end{aligned}
\end{equation}
\item the \emph{contact force} and  \emph{contact couple} per unit length
$$
\lb_o(x):=\int_I\cb_o(x,\zeta)d\,\zeta,\quad
\mb_o(x):=\int_I\zeta\cb_o(x,\zeta)\,d\zeta.
$$
\end{itemize}
All in all, the external virtual power expenditure takes the
following form:
$$
\int_\Pi\db_o\cdot\vb+\int_{\partial\Pi}\cb_o\cdot\vb=\int_\Pc(\qb_o\cdot\overset{(0)}\vb+\rb_o\cdot\overset{(1)}\vb)+\int_{\partial\Pc}(\lb_o\cdot\overset{(0)}\vb+\mb_o\cdot\overset{(1)}\vb).
$$
\subsubsection{Principle of Virtual Powers. Field equations. Boundary equations}\label{bc}
The two-dimensional Principle of Virtual Powers we arrive at is:
\begin{equation}\label{2DPVP}
\begin{aligned}
0&=\int_\Pc\Big((-\Divs\Fbs-\qb_o)\cdot\overset{(0)}\vb
+(-\Divs\Mbs+\,\fb^{(3)}-\rb_o)\cdot\overset{(1)}\vb \Big)+\\
&+\int_{\partial\Pc}\Big((\Fbs\!\mb-\lb_o)\cdot\overset{(0)}\vb+(\Mbs\!\mb-\mb_o)\cdot\overset{(1)}\vb
\Big),
\end{aligned}
\end{equation}
a statement to hold for every pair of vector fields
$\overset{(0)}\vb$ and $\overset{(1)}\vb$ on $\Sc$ and for every
part $\Pc$ of $\Sc$. Under the standard blanket assumptions of
smoothness, and with the use of a standard localization lemma,
\eqref{2DPVP} yields the \emph{(two-dimensional) field equations}
to hold at any interior point of $\Sc$:
\begin{equation}\label{bil}
\begin{aligned}
&\Divs\Fbs+\qb_o=\mathbf{0},\\
&\Divs\Mbs-\fb^{(3)}+\rb_o=\mathbf{0}.
\end{aligned}
\end{equation}
%
%
%Component-wise, the general balance equations \eqref{bil} can be written in the following form:
%%
%\begin{equation}\label{bildue}
%\boxed{
%\begin{aligned}
%F^{\delta\alpha}|_\alpha-W_\alpha^\delta\,F^{3\alpha}+q_o^\delta&=0
%\quad
%(\delta=1,2),\\
%F^{3\alpha}|_\alpha+W_{\beta\alpha} {F}^{\beta\alpha}+q_o^3&=0;\\
%M^{\delta\alpha}|_\alpha- {F}^{3\delta}+r_o^\delta&=0 \quad
%(\delta=1,2),\\
%M^{3\alpha}|_\alpha+W_{\beta\alpha}{M}^{\beta\alpha}-\fb^{(3)}\cdot\nb+r_o^3&=0,
%\end{aligned}
%}
%\end{equation}
%%
%where a vertical bar denotes covariant differentiation (given a tensor field $\Tb$ over $\Sc$, $T^{\delta\alpha}|_\alpha:=T^{\delta\alpha}_{,\alpha}+\gamma_{\beta\alpha}^{\,\delta}T^{\beta\alpha}+\gamma_{\alpha\beta}^{\beta}T^{\delta\alpha}$, with $\Tb^{\delta\alpha}:=\Tb\cdot\eb^\delta\otimes\eb^\alpha$) and where $\gamma_{\beta\alpha}^{\,\delta}:=\eb^\delta\cdot\eb_\beta,_\alpha$ are the surface Christoffel symbols.

Granted \eqref{bil}, what remains of \eqref{2DPVP} is:
\begin{equation}\label{issa}
\int_{\partial\Pc}\Big((\Fbs\mb-\lb_o)\cdot\overset{(0)}\vb+(\Mbs\mb-\mb_o)\cdot\overset{(1)}\vb
\Big)=0
\end{equation}
for all admissible variations $\overset{(0)}\vb$ and
$\overset{(1)}\vb$. Localization of \eqref{issa} yields different
results according to where it is performed. At an interior point
of $\Sc$, where it can be combined with arbitrariness in the
choice of $\Pc$, we have that:
\begin{equation}\label{issalocD}
\Fbs\!\mb=\lb_o,\quad \Mbs\!\mb=\mb_o,
\end{equation}
two relations that parallel the classic relation between stress
tensor and contact-force vector for three-dimensional Cauchy
bodies.\footnote{See \cite{PGV2} for a discussion of this issue
that covers second-gradient materials as well.} At a point of
$\partial\Sc$ where a Dirichlet boundary condition prevails --
that is to say, where one or more components of the boundary trace
of the displacement field are prescribed -- the corresponding
components of the admissible variations must vanish; accordingly,
the complementing components of both vectors $(\Fbs\!\mb-\lb_o)$
and $(\Mbs\!\mb-\mb_o)$ must also vanish (yielding boundary
equations of Neumann type), because localization of \eqref{issa}
leads to:
\begin{equation}\label{issalocN}
(\Fbs\!\mb-\lb_o)\cdot\overset{(0)}\vb=0\quad\textrm{and}\quad
(\Mbs\!\mb-\mb_o)\cdot\overset{(1)}\vb=0,
\end{equation}
for all admissible choices of $\overset{(0)}\vb$ and
$\overset{(1)}\vb$.

\vskip 6pt \remark The shear vector $\fb^{(3)}$, defined by
\eqref{2Dsh} and entering the last of equations \eqref{bildue}, is
one of the \emph{force vectors}:
\begin{equation}\label{2forze}
\fb^{(i)}(x):=\int_I\alpha(x,\zeta)\Sb(x,\zeta)\gb^i(x,\zeta)
d\zeta;
\end{equation}
the \emph{moment vectors} are:
\begin{equation}\label{2momenti}
\mb^{(i)}(x):=\int_I\alpha(x,\zeta)\zeta\Sb(x,\zeta)\gb^i(x,\zeta)
d\zeta.
\end{equation}
With these definitions, we may further set:
\[
\Fb:=\fb^{(i)}\otimes\eb_i=\Fbs+\fb^{(3)}\otimes\nb,\quad
\Mb:=\mb^{(i)}\otimes\eb_i,\
\]
with
\begin{equation}\label{2sforzi}
\Fbs=\fb^{(\alpha)}\otimes\eb_\alpha,\quad
\Mbs=\mb^{(\alpha)}\otimes\eb_\alpha,
\end{equation}
and
\begin{equation}\label{2forzem}
\fb^{(\alpha)}=\Fbs\eb^\alpha,\quad \mb^{(\alpha)}=\Mbs\eb^\alpha.
\end{equation}
In terms of force and moment vectors, equations \eqref{bil}
become:
\begin{equation}\label{biltre}
\begin{aligned}
&{\fb^{(\alpha)}},_{\alpha}+\gamma_{\alpha\beta}^{\,\beta}\,\fb^{(\alpha)}+\qb_o=\mathbf{0},\\
&{\mb^{(\alpha)}},_{\alpha}+\gamma_{\alpha\beta}^{\,\beta}\,\mb^{(\alpha)}-\fb^{(3)}+\rb_o=\mathbf{0}.
\end{aligned}
\end{equation}
\vskip 6pt We call  \emph{membrane forces} the components
$F^{\alpha\beta}:=\Fb\cdot\eb^\alpha\otimes\eb^\beta=
\fb^{(\beta)}\cdot\eb^\alpha= \Fbs\cdot\eb^\alpha\otimes\eb^\beta$
-- respectively, \emph{normal m. f.} for $\alpha=\beta$ and
\emph{shear m.f.} for $\alpha\neq\beta$; and we call
$M^{\alpha\alpha}:=\Mb\cdot\eb^\alpha\otimes\eb^\alpha=
\mb^{(\alpha)}\cdot\eb^\alpha=\Mbs\cdot\eb^\alpha\otimes\eb^\alpha$
$(\alpha=1,2)$ the \emph{bending moments} and
$M^{\alpha\beta}:=\Mb\cdot\eb^\alpha\otimes\eb^\beta=
\mb^{(\alpha)}\cdot\eb^\beta=\Mbs\cdot\eb^\alpha\otimes\eb^\beta$,
$(\alpha,\beta=1,2, \alpha\neq\beta)$ the \emph{twisting moments}.
Finally, we call $F^{3\alpha}=\fb^{(3)}\cdot \eb^\alpha$ the
\emph{transverse shears},  $F^{33}=f^{(3)3}=\fb^{(3)}\cdot \nb$
the \emph{thickness shear}, and
$M^{3\alpha}=\mb^{(3)}\cdot{\eb^\alpha}$ the \emph{thickness
moments}. Component-wise, the general field equations \eqref{bil}
can be written in the following form:
\begin{equation}\label{bildue}
\begin{aligned}
F^{\delta\alpha}|_\alpha-W_\alpha^\delta\,F^{3\alpha}+q_o^\delta&=0
\quad
(\delta=1,2),\\
F^{3\alpha}|_\alpha+W_{\beta\alpha} {F}^{\beta\alpha}+q_o^3&=0;\\
M^{\delta\alpha}|_\alpha- {F}^{3\delta}+r_o^\delta&=0 \quad
(\delta=1,2),\\
M^{3\alpha}|_\alpha+W_{\beta\alpha}{M}^{\beta\alpha}-F^{33}+r_o^3&=0,
\end{aligned}
\end{equation}
where a vertical bar denotes covariant differentiation (given a
tensor field $\Tb$ over $\Sc$,
$T^{\delta\alpha}|_\alpha:=T^{\delta\alpha}_{,\alpha}+\gamma_{\beta\alpha}^{\,\delta}T^{\beta\alpha}+\gamma_{\alpha\beta}^{\beta}T^{\delta\alpha}$,
with $\Tb^{\delta\alpha}:=\Tb\cdot\eb^\delta\otimes\eb^\alpha$)
and where
$\gamma_{\beta\alpha}^{\,\delta}:=\eb^\delta\cdot\eb_\beta,_\alpha$
are the surface Christoffel symbols.

 \remark
A shell-shaped body $\Gc(\Sc,\varepsilon)$ is in a \emph{membrane
regime} if it so happens that the following conditions:
\begin{equation}\label{membr}
\fb^{(\alpha)}\cdot \nb=0, \quad \fb^{(3)}=\mathbf{0},\quad
\mb^{(\alpha)}=\mathbf{0}
\end{equation}
hold identically in $\Gc(\Sc,\varepsilon)$. The first two of
$\eqref{membr}$ imply that $ F^{3\alpha}=0$, the fourth that $
M^{i\alpha}=0$. Consequently,  equations \eqref{bildue} reduce to:
\begin{equation}\label{membreq}
\begin{aligned}
&F^{\delta\alpha}|_{\alpha}+q_o^\delta=0,\\
&W_{\beta\alpha}{F}^{\beta\alpha}+q_o^3=0,
\end{aligned}
\end{equation}
together with the following compatibility condition on the data:
$$
\rb_o=\mathbf{0}.
$$

%We can observe that the contact-force vector $\fb^{(\alpha)}$ can
%be written in terms of the force tensor
%$\Fb=\int_I\alpha(x,\zeta)\Sb(x,\zeta)\Bb(x,\zeta) d\zeta$ like
%$\fb^{(\alpha)}=F^{i\alpha}\eb_i$ and similarly the contact-moment
%vector is written in terms of the moment tensor
%$\Mb=\int_I\alpha(x,\zeta)\zeta\Sb(x,\zeta)\Bb(x,\zeta) d\zeta$ as
%$\mb^{(\alpha)}=M^{i\alpha}\eb_i$.
% }

%It is easy to show that an alternative form for the balance
%equations is given by the following relations:
%\begin{equation}
%\begin{aligned}
%&F^{\delta\alpha}|_{\alpha}-W_\alpha^\delta
%F^{3\alpha}+q_{o}^\delta=0,\\
%&F^{3\alpha}|_{\alpha}-W_{\beta\alpha}
%F^{\beta\alpha}+q_{o}^3=0,\\
%&M^{\delta\alpha}|_{\alpha}-W_\alpha^\delta
%M^{t\alpha}-f_3^\delta+r_{o}^\delta=0,\\
%&M^{3\alpha}|_\alpha+W_{\beta\alpha}M^{\beta\alpha}-F_3+r_o^3=0,
%\end{aligned}
%\end{equation}
%where, for a tensor field $\Tb$,
%$T^{ih}|_h:=T^{ih}_{,h}+T^{hk}\Gamma_{hk}^i+T^{ih}\Gamma_{hk}^k$
%is the covariant derivative of the component $T^{im}$ in the
%direction $\gb_m$; for a scalar field $h$,
%$h^\alpha|_\beta:=h^\alpha_{,\beta}+h^\delta\Gamma^{\alpha}_{\delta\beta}$
%is the covariant derivative of $h^\alpha$ in the direction
%$\gb_\beta$; $\Wb:=-\nablas \nb$ is the \emph{Weingarten tensor}.

\subsection{Constitutive Assumptions}\label{costa}
The penultimate step in our construction of a shell theory
consists in specifying how the two-dimensional stress measures
\eqref{2Ds} depend on the admissible deformations \eqref{KLdispl}.
We wish to come up with the simplest theory accommodating both an
orthotropic response and the unshearability constraint discussed
in Section \ref{kin}. To this end, we confine attention to cases
when the material response is uniform all over
$\Gc(\Sc,\varepsilon)$. Moreover, all along each fixed material
fiber ${\mathcal F}(x)$ we take the orthotropy plane orthogonal to
that fiber, so that the orthotropy and the tangent planes coincide
at $x\in\Sc$; in particular, whenever use is made of orthogonal
curvilinear coordinates, as is the case for the cylindrical shells
treated in Section \ref{cilscel}, we take the orthotropy axes
tangent to both the coordinate lines $z^\alpha=\textrm{const.}$

\subsubsection{Orthotropic elasticity tensors}
Let $\Co$ denote the fourth-order \emph{tensor of elasticity}, a
linear transformation of the space of symmetric tensors, that
specifies the stress response to deformations in the parent
three-dimensional theory we are going to select. An orthogonal
tensor $\Qb$ is a \emph{symmetry transformation} for the linearly
elastic material described by $\Co$   if it so happens that
$$
\Qb\Co[\Eb]\Qb^T=\Co[\Qb\Eb\Qb^T],\quad\textrm{for all symmetric
tensors}\;\,\Eb,
$$
i.e., that
\[
\Qo\Co=\Co\Qo,
\]
where the fourth-order tensor $\Qo$ is such that
$\Qo[\Ab]:=\Qb\Ab\Qb^T$ for every $\Ab$ in the space the second
order tensors $\Lyn$; the \emph{symmetry group} of $\Co$ is the
set
$$
\mathscr{G}_\Co:=\{\Qb\in \mathrm{Orth}\,|\,\Qo\Co=\Co\Qo  \}.
$$
Given a subgroup $\mathscr{G}$ of the orthogonal group (or, for
what it matters here, of the group of all rotations), one seeks a
representation formula for all elasticity tensors $\Co$ such that
$\mathscr{G}_\Co\supset\mathscr{G}$, i.e., for all elasticity
tensors sharing a given symmetry group. A linearly elastic
material is called \emph{orthotropic} when its stress response is
insensitive to a rotation of $\pi$ about a given axis $\cb$, i.e.,
when the symmetry group of its elasticity tensor includes that
rotation. We give here below a general representation formula for
the elasticity tensors in question.

Let $\{\cb_i\; (i=1,2,3)\}$ be an orthonormal basis of vectors.
Consider the following orthonormal  basis for the linear space Sym
of all symmetric tensors:
\begin{equation}\label{base}
\begin{aligned}
\Vb_\alpha&=\frac{1}{\sqrt{2}}(\cb_\alpha\otimes\cb_3+\cb_3\otimes\cb_\alpha)\;\,(\alpha=1,2),\quad \Vb_3=\cb_3\otimes\cb_3,\\
\Wb_\alpha&=\cb_\alpha\otimes\cb_\alpha \;\,
(\alpha\,\;\textrm{not summed}),\quad
\Wb_3=\frac{1}{\sqrt{2}}\,(\cb_1\otimes\cb_2+\cb_2\otimes\cb_1).
\end{aligned}
\end{equation}
With the use of this basis, any orthotropic elasticity tensor can
be written in the following form:
\begin{equation}\label{othotr}
\begin{aligned}
\Co&=\Co_{1111}\Wb_1\otimes\Wb_1+\Co_{2222}\Wb_2\otimes\Wb_2+\Co_{3333}\Vb_3\otimes\Vb_3+\\
&+\Co_{1212}\Wb_3\otimes\Wb_3+\Co_{3131}\Vb_1\otimes\Vb_1+\Co_{2323}\Vb_2\otimes\Vb_2+\\
&+\Co_{1122}(\Wb_1\otimes\Wb_2+\Wb_2\otimes\Wb_1)+\Co_{1133}(\Wb_1\otimes\Vb_3+\Vb_3\otimes\Wb_1)+\\
&+\Co_{2233}(\Wb_2\otimes\Vb_3+\Vb_3\otimes\Wb_2)
\end{aligned}
\end{equation}
(cf. \cite{Gu}); the orthotropic material class is then
parameterized by the 9 elastic moduli
$\Co_{1111},\ldots,\Co_{2233}$.
\subsubsection{The elasticity tensor of unshearable orthotropic materials}\label{2.5.2}
As a direct consequence of the fact that the shell model we are
after incorporates a kinematic constraint, the appropriate
three-dimensional response is captured by an elasticity tensor
somewhat simpler than \eqref{othotr}. To see why, and to derive
such an elasticity tensor, we apply to our present case a general
representation result in constrained linear elasticity
\cite{PGV,PPG}.

We make the shell geometry agree with the geometry intrinsic to
the material response, in the sense that, at any fixed point
$x\in\Sc$, we identify $\cb_3$ with $\nb(x)$. With this
identification, the internal constraint \eqref{previn} can be read
as the requirement that all admissible strains be orthogonal to
the following subspace of $\Sim$:
\begin{equation}\label{a}
\mathscr{D}^\perp:=\textrm{span}(\!\Vb_\alpha,\, \alpha=1,2).
\end{equation}
Accordingly, the space $\Sim$ is split into the direct sum of two
orthogonal subspaces:
\begin{equation}\label{b}
\Sim=\mathscr{D}\oplus\mathscr{D}^\perp,
\end{equation}
and the stress is split into \emph{reactive} and \emph{active}
parts:
\begin{equation}\label{RA1}
\Sb=\Sb^{(R)}+\Sb^{(A)},
\end{equation}
with
\begin{equation}\label{RA2}
\begin{aligned}
\Sb^{(R)}&=\psi_\alpha^{(R)}\Vb_\alpha, \quad \psi_\alpha^{(R)}\in\Ro,\\
\Sb^{(A)}&=\widetilde\Co[\Eb],\quad
\widetilde\Co:\mathscr{D}\rightarrow\mathscr{D};
\end{aligned}
\end{equation}
here the coefficients $\psi_\alpha^{(R)}$ are constitutively
unspecified, and the \emph{constraint space} $\mathscr{D}$
implicitly defined by \eqref{a} and \eqref{b} can be identified as
\begin{equation}\label{constrsp}
\mathscr{D}=\mathrm{span}(\!\Vb_3; \Wb_i,\, i=1,2,3).
\end{equation}
On applying a general result proved in \cite{PGV}, a
representation for the desired elasticity tensor $\widetilde\Co$
can be deduced from the one given for $\Co$ in \eqref{othotr}:
\[
\widetilde\Co=\Po_{\mathscr{D}}\,\Co|_{\mathscr{D}},\quad\Po_{\Ds}:=\Io-\Vb_\alpha\otimes\Vb_\alpha.
\]
where $\Po_{\mathscr{D}}$ denotes the orthogonal projector of
$\Sim$ on $\mathscr{D}$. One finds the 7-parameter representation
\begin{equation}\label{ctilde}
\begin{aligned}
\widetilde\Co&=\,\Co_{1111}\Wb_1\otimes\Wb_1+\Co_{2222}\Wb_2\otimes\Wb_2+\Co_{3333}\Vb_3\otimes\Vb_3+\Co_{1212}\Wb_3\otimes\Wb_3+\\
&+\Co_{1122}(\Wb_1\otimes\Wb_2+\Wb_2\otimes\Wb_1)+\Co_{1133}(\Wb_1\otimes\Vb_3+\Vb_3\otimes\Wb_1)+\\
&+\Co_{2233}(\Wb_2\otimes\Vb_3+\Vb_3\otimes\Wb_2).
\end{aligned}
\end{equation}
As is customary, we assume that $\widetilde\Co$ is
positive-definite, i.e., that
\begin{equation}\label{invert}
\Eb\cdot\widetilde\Co[\Eb]>0\quad \textrm{for
all}\;\,\Eb\in{\mathscr{D}}\setminus\{\0\}.
\end{equation}
%

%
%Projecting on the constraint manifold, we obtain:
%\begin{equation}
%\begin{aligned}
%\widetilde\Co=\Po_\Ds\Co|_{{\Ds_g}}=&\,\Co_{1111}\Wb_1\otimes\Wb_1+\Co_{2222}\Wb_2\otimes\Wb_2+\Co_{3333}\Vb_3\otimes\Vb_3+\Co_{1212}\Wb_3\otimes\Wb_3+\\
%&\Co_{1122}(\Wb_1\otimes\Wb_2+\Wb_2\otimes\Wb_1)+\Co_{1133}(\Wb_1\otimes\Vb_3+\Vb_3\otimes\Wb_1)+\\
%&\Co_{2233}(\Wb_2\otimes\Vb_3+\Vb_3\otimes\Wb_2).
%\end{aligned}
%\end{equation}
%
\vskip 4pt \remark In case the internal constraint \eqref{previn}
is reinforced \emph{\`a la} Kirchhoff-Love as specified by
\eqref{kl}, so as to exclude thickness changes, the above
procedure yields the  elasticity tensor
\begin{equation}\label{chat}
\widehat\Co=\Co_{1111}\Wb_1\otimes\Wb_1+\Co_{2222}\Wb_2\otimes\Wb_2+\Co_{1212}\Wb_3\otimes\Wb_3+\Co_{1122}(\Wb_1\otimes\Wb_2+\Wb_2\otimes\Wb_1),
\end{equation}
a linear transformation of the constraint space $\,\widehat{\mathscr{D}}=\mathrm{span}(\!\Wb_i,\, i=1,2,3)\,$ into itself. %We shall make use of this 4-parameter representation when we deal with cylindrical shells.
\subsubsection{The technical moduli}
In technical applications of classic isotropic elasticity, the two
Lam\'e constants are replaced by the technical moduli $E$ and
$\nu$ of Young and Poisson, plus the shear modulus $G$, under the
condition that $E=2(1+\nu)G$. Likewise, we here replace the seven
Lam\'e-like constants $\Co_{1111},\ldots,\Co_{2233}$ in
\eqref{ctilde} by an equivalent list of ten technical moduli --
three of them being Young-like, six Poisson-like, and one
shear-like -- that must satisfy three independent algebraic
conditions. The technical moduli in question are precisely those
that enter the following representation of the \emph{compliance
tensor} $\widetilde\Co^{-1}$:
\begin{equation}\label{ctildeinv}
\begin{aligned}
\widetilde\Co^{-1}&=\frac{1}{E_1}\Wb_1\otimes\Wb_1+\frac{1}{E_2}\Wb_2\otimes\Wb_2+\frac{1}{E_3}\Vb_3\otimes\Vb_3+\frac{1}{2G_{12}}\Wb_3\otimes\Wb_3+\\
&-\frac{\nu_{12}}{E_1}(\Wb_1\otimes\Wb_2+\Wb_2\otimes\Wb_1)-\frac{\nu_{13}}{E_1}(\Wb_1\otimes\Vb_3+\Vb_3\otimes\Wb_1)+\\
&-\frac{\nu_{23}}{E_2}(\Wb_2\otimes\Vb_3+\Vb_3\otimes\Wb_2)
\end{aligned}
\end{equation}
(as is well known, the positivity assumption \eqref{invert}
guarantees invertibility of  $\widetilde\Co$).

To confirm that the three moduli $E_1,E_2$ and $E_3$ are
Young-like, imagine to induce a state of uniaxial traction in the
direction, say, $\cb_1$ in a specimen made of the material under
examination, so that the stress is $\Sb=\sigma\Wb_1$ and the
corresponding strain is
$$\Eb=\sigma\,\widetilde\Co^{-1}[\Wb_1]=\frac{\sigma}{E_1}(\Wb_1-\nu_{12}\Wb_2-\nu_{13}\Vb_3).$$
Then, we find that
\begin{equation}\label{Y}
E_1=\frac{\Sb\cdot\Wb_1}{\Eb\cdot\Wb_1},
\end{equation}
the ratio between the axial stress and the corresponding axial
deformation. Moreover, we find that
\begin{equation}\label{P}
\nu_{12}=-\frac{\Eb\cdot\Wb_2}{\Eb\cdot\Wb_1},\quad
\nu_{13}=-\frac{\Eb\cdot\Vb_3}{\Eb\cdot\Wb_1},
\end{equation}
the Poisson-like negative ratios of the transverse deformations in
the directions $\cb_2$ and $\cb_3$ and the axial deformation in
the direction $\cb_1$. The defining formulae for
$E_2,\nu_{21},\nu_{23}$ and $E_3,\nu_{31},\nu_{32}$ are completely
analogous to \eqref{Y} and \eqref{P}. Due to the built-in
symmetries of $\widetilde\Co$, it turns out that
\begin{equation}\label{coeff}
\frac{E_1}{E_2}=\frac{\nu_{12}}{\nu_{21}},\quad
\frac{E_1}{E_3}=\frac{\nu_{13}}{\nu_{31}}, \quad
\frac{E_2}{E_3}=\frac{\nu_{23}}{\nu_{32}}.
\end{equation}
Finally, the formula
$$
2G=\frac{\Sb\cdot\Wb_3}{\Eb\cdot\Wb_3},
$$
establishes the nature of $G$ as a shear modulus.

It is straightforward to see that $G=\Co_{1212}$.
The other technical moduli can be written as follows in terms of
the components of $\widetilde\Co$:
\begin{equation}\label{moduli}
\begin{aligned}
E_1&=\frac{\Co_{1111}\Co_{2222}\Co_{3333}-\Co_{3333}\Co_{1122}^2-\Co_{2222}\Co_{1133}^2-\Co_{1111}\Co_{2233}^2+2\Co_{1122}\Co_{1133}\Co_{2233}}{\Co_{2222}\Co_{3333}-\Co_{2233}^2},\\
E_2&=\frac{\Co_{1111}\Co_{2222}\Co_{3333}-\Co_{3333}\Co_{1122}^2-\Co_{2222}\Co_{1133}^2-\Co_{1111}\Co_{2233}^2+2\Co_{1122}\Co_{1133}\Co_{2233}}{\Co_{1111}\Co_{3333}-\Co_{1133}^2},\\
E_3&=\frac{\Co_{1111}\Co_{2222}\Co_{3333}-\Co_{3333}\Co_{1122}^2-\Co_{2222}\Co_{1133}^2-\Co_{1111}\Co_{2233}^2+2\Co_{1122}\Co_{1133}\Co_{2233}}{\Co_{1111}\Co_{2222}-\Co_{1122}^2};\\
\nu_{12}&=\frac{\Co_{3333}\Co_{1122}-\Co_{1133}\Co_{2233}}{\Co_{2222}\Co_{3333}-\Co_{2233}^2},\quad \nu_{21}=\frac{\Co_{3333}\Co_{1122}-\Co_{1133}\Co_{2233}}{\Co_{1111}\Co_{3333}-\Co_{1133}^2},\\
\nu_{13}&=\frac{\Co_{2222}\Co_{1133}-\Co_{1122}\Co_{2233}}{\Co_{2222}\Co_{3333}-\Co_{2233}^2},
\quad
\nu_{31}=\frac{\Co_{2222}\Co_{1133}-\Co_{1122}\Co_{2233}}{\Co_{1111}\Co_{2222}-\Co_{1122}^2},\\
\nu_{23}&=\frac{\Co_{1111}\Co_{2233}-\Co_{1122}\Co_{1133}}{\Co_{1111}\Co_{3333}-\Co_{1133}^2},
\quad
\nu_{32}=\frac{\Co_{1111}\Co_{2233}-\Co_{1122}\Co_{1133}}{\Co_{1111}\Co_{2222}-\Co_{1122}^2}.
\end{aligned}
\end{equation}

To sum up, the 7-parameter representation \eqref{ctilde} can be
re-written in the form:
\begin{equation}\label{consta}
\begin{aligned}
\widetilde\Co&=\Delta^{-1}\Big(E_1(1-\nu_{23}\,\nu_{32})\Wb_1\otimes\Wb_1+
E_2(1-\nu_{13}\,\nu_{31})\Wb_2\otimes\Wb_2\\
&+E_3(1-\nu_{12}\,\nu_{21})\Vb_3\otimes\Vb_3+2\Delta\, G\,
\Wb_3\otimes\Wb_3\\
&+E_1(\nu_{21}+\nu_{23}\,\nu_{31})(\Wb_1\otimes\Wb_2+\Wb_2\otimes\Wb_1)\\
&+E_1(\nu_{31}+\nu_{21}\,\nu_{32})(\Wb_1\otimes\Vb_3+\Vb_3\otimes\Wb_1)\\
&+E_2(\nu_{32}+\nu_{31}\,\nu_{12})(\Wb_2\otimes\Vb_3+\Vb_3\otimes\Wb_2)\Big),
\end{aligned}
\end{equation}
where
\begin{equation}\label{constb}
\Delta:=1-\nu_{12}\,\nu_{21}-\nu_{13}\,\nu_{31}-\nu_{23}\,\nu_{32}-2\nu_{12}\,\nu_{23}\,\nu_{31}.
\end{equation}
\subsection{Force and Moment Vectors and Tensors}\label{fmv}
The final step in the assemblage of our shell theory -- the posing
of initial- and boundary-value
 problems -- demands that the balance equations \eqref{bildue} are written in terms of the
 parameters involved in the general representation \eqref{KLdispl} for an admissible displacement field.
 All we have to do to construct the corresponding parametric representations for those components of the
 force and moment tensors that enter the balance equations is to insert  in the definitions \eqref{2forze}
 and \eqref{2momenti} the general parametric representation for the stress
 field in $\Gc(\Sc,\varepsilon)$, and then make use of definitions \eqref{2sforzi}. Now, the required stress
 representation is obtained by combining \eqref{RA1}-\eqref{RA2} -- with $\cb_3\equiv\nb(x)$ --
 and \eqref{ctilde} -- with $\Eb=\Eb(\ub_S)$, and $\ub_S$ given by \eqref{KLdispl}; one finds:
\begin{equation}\label{Eqcost}
\Sb(x,\zeta)=\sb^{(R)}(x,\zeta)\otimes\nb(x)+\nb(x)\otimes\sb^{(R)}(x,\zeta)
+\widetilde\Co[\Eb(\ub_S(x,\zeta))],
\end{equation}
where the restriction to the fiber ${\mathcal F}(x)$ of the
reactive field $\sb^{(R)}$ is a vector field perpendicular to
$\nb(x)$.\footnote{It follows from the first of \eqref{RA2} and
the first two of
 \eqref{base} that $\Sb^{(R)}=\sb^{(R)}\otimes \cb_3+\cb_3\otimes\sb^{(R)},$ with
  $\sb^{(R)}=\frac{1}{\sqrt 2}\,\psi_\alpha^{(R)}\cb_\alpha$.} Consequently, the force
   and moment vectors have both reactive and active parts, namely,
\begin{equation}\label{2forzedef}
\begin{aligned}
& \overset{R}{\fb}{}^{(\alpha)}(x)\!=\!\Big(\!\int_I\alpha(x,\zeta)\big(\sb^{(R)}(x,\zeta)\cdot\gb^{\alpha}(x,\zeta)\big)d\zeta\Big)\nb(x),\\
&\overset{R}{\fb}{}^{(3)}(x)\!=\!\int_I\alpha(x,\zeta)\sb^{(R)}(x,\zeta)d\zeta,\quad\overset{A}{\fb}{}^{(i)}(x)=\int_I\alpha(x,\zeta)\,\widetilde\Co[\Eb(\ub_S(x,\zeta))]\gb^{i}(x,\zeta)
d\zeta,
%\\
%&\overset{A}{\fb}{}^{(i)}(x)=\int_I\alpha(x,\zeta)\,\widetilde\Co[\Eb(\ub_S(x,\zeta))]\gb^{i}(x,\zeta) d\zeta,
\end{aligned}
\end{equation}
and
\begin{equation}\label{2momdef}
\begin{aligned}
&
\overset{R}{\mb}{}^{(\alpha)}(x)\!=\!\Big(\!\int_I\alpha(x,\zeta)\zeta\big(\sb^{(R)}(x,\zeta)\cdot\gb^{\alpha}(x,\zeta)\big)d\zeta\Big)\nb(x),\\&\overset{R}{\mb}{}^{(3)}(x)\!=\!\int_I\alpha(x,\zeta)\zeta\sb^{(R)}(x,\zeta)d\zeta,\quad\overset{A}{\mb}{}^{(i)}(x)=\int_I\alpha(x,\zeta)\zeta\,\widetilde\Co[\Eb(\ub_S(x,\zeta))]\gb^{i}(x,\zeta)
d\zeta.
\end{aligned}
\end{equation}
It is not difficult to check that
\begin{equation}
 \overset{R}{\fb}{}^{(i)}(x)\cdot \overset{A}{\fb}{}^{(i)}(x)=0,\quad\overset{R}{\mb}{}^{(i)}(x)\cdot \overset{A}{\mb}{}^{(i)}(x)=0.
\end{equation}
It is also easy to see, in the light of \eqref{2sforzi}, that the
force and moment tensors $\Fbs$ and $\Mbs$ have active and
reactive parts as well. Thus, the balance equations one arrives at
are not \emph{pure}, in the sense that, in addition to the
parameter fields, they also include reactive terms. In fact, the
transverse shears are reactive, and equations $\eqref{bildue}_3$
relate them to the active bending moments:
\begin{equation}\label{react}
F^{3\delta}=\,M^{\delta\alpha}|_\alpha+r_o^\delta.\footnote{In
fact, $F^{3\delta}=\fb^{(3)}\cdot
\eb^\delta=\overset{R}{\fb}{}^{(3)}\cdot \eb^\delta$; and,
$M^{\delta\alpha}= \mb^{(\delta)}\cdot\eb^\alpha=
\overset{A}{\mb}{}^{(\delta)}\cdot\eb^\alpha$. }
\end{equation}
The thickness moments are also reactive, and $\eqref{bildue}_5$
relates their divergence to the active bending moments and the
active thickness shear:
\begin{equation}\label{reactm}
M^{3\alpha}|_\alpha=-W_{\beta\alpha}{M}^{\beta\alpha}+F^{33}-r_o^3.
\end{equation}

These observations suggest a \emph{sequential strategy} to solve a
shell problem within our present theory, where
 the unknowns are (the fields $\ab, w$ and the constant $\gamma$ that parameterize) the displacement
  field $\ub_S$ \emph{and} the reaction force and moment fields: firstly, by the use of the projection
  operator $\Po_{\mathscr{D}}$ defined in the Subsection \ref{2.5.2}, one derives a set of  reaction-free
  consequences of the balance equations; secondly, one solves such `purified' system of equations for $\ub_S$;
  thirdly, one returns to the full balance equations, where the active terms can now be computed explicitly, and solves them for the
  reactive fields.\footnote{The annihilation procedure of the reactive terms occurring in the three-dimensional balance
  equations of constrained linear elasticity is discussed in \cite{PPG}, Section 17.2.}
\vskip 6pt \remark \label{rmk4}As anticipated in footnote 2,
reactive forces and moments are found in the balance equations
\eqref{bildue}, because the class of variations \eqref{varia} we
used to derive those equations from the Principle of Virtual
Powers \eqref{PVP} is visibly larger than the class
\eqref{KLdispl} of admissible displacements (this will not be the
case in the next section). Although we here do not
 pursue this issue any further, we recall that having at one's disposal the reactive dynamical
  descriptors associated with the kinematical Ansatz adopted to construct a lower-dimensional
  structure theory can be proved beneficial to  improve  the pointwise approximation of the relative
  three-dimensional stress field \cite{PPG69,LPG1,LPG2,PPGU}.
\vfill \pagebreak

\section{Cylindrical Shells: Generalities}\label{cilscel}
\subsection{Geometry}
\begin{figure}[h]
\centering
\includegraphics[scale=.93]{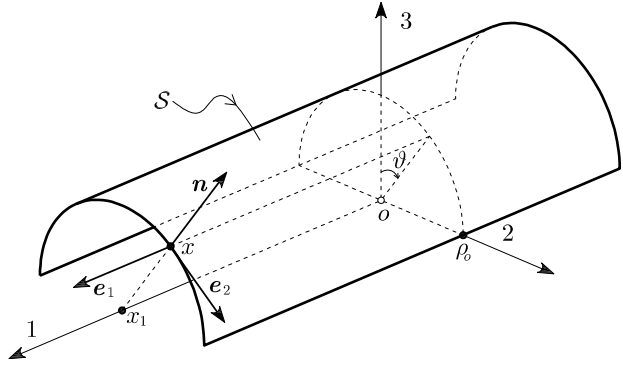}
\caption{Geometrical equipment of the model surface of a
cylindrical shell.} \label{cyl}
\end{figure}

We now restrict our attention to shells whose model surface $\Sc$
is a portion of a right circular cylinder, that we parameterize as
usual by means of cylindrical coordinates:
$$
z^1=x_1, \quad z^2=\vartheta
$$
(Figure \ref{cyl}).
 For $\rho_o$ the radius of the directrix of $\Sc$ and $\nb$ the
normal to $\Sc$, and for $\varepsilon<\rho_o$ the thickness of the
shell-shaped region $\Gc(\Sc,\varepsilon)$,  the position vectors
with respect to the origin $o$ of two typical points
$p\in\Gc(\Sc,\varepsilon)$ and $x\in\Sc$ are, respectively,
$$
p-o=\pb(x_1,\vartheta,\zeta)=\xb(x_1,\vartheta)+\zeta\nb(\vartheta)=x_1\cb_1+\rho_o\left(1+\frac{\zeta}{\rho_o}
\right)\nb(\vartheta).
$$
and
$$
x-o=\xb(x_1,\vartheta)=x_1\cb_2+\rho_o\nb(\vartheta), \quad
\nb(\vartheta)=\sin\vartheta\cb_1+\cos\vartheta\cb_3.
$$
The relative covariant and contravariant bases can both be
represented in terms of the orthonormal basis
$\{\cb_1,\nb^\prime(\vartheta),\nb(\vartheta)\}$, that serves as
physical basis (recall Remark \ref{rem1}):
\begin{equation}
\begin{aligned}
&\gb_1(p)=\eb_1(x), \quad \eb_1(x)=\cb_1,\\
&\gb_2(p)=\left(1+\frac{\zeta}{\rho_o}\right)\eb_2(x), \quad
\eb_2(x)=\rho_o\nb'(\vartheta),\\
&\gb_3(p)=\eb_3(x)=\nb(\vartheta).
\end{aligned}
\end{equation}
and
\begin{equation}
\begin{aligned}
&\gb^1(p)=\eb^1(x), \quad \eb^1(x)=\cb_1,\\
&\gb^2(p)=\left(1+\frac{\zeta}{\rho_o}\right)^{-1}\eb^2(x), \quad
\eb^2(x)=\rho_o^{-1}\nb'(\vartheta),\\
&\gb^3(p)=\eb^3(x)=\nb(\vartheta).
\end{aligned}
\end{equation}
%It is easy to show that the normal coordinates
%$(x_1,\vartheta,\zeta)$ are also principal, and that the physical
%bases in $p$ and $x$ are the same:
%\begin{equation}
%\begin{aligned}
%&\gb{\<1}(p)=\eb{\<1}(x)=\cb_1,\\
%&\gb{\<2}(p)=\eb{\<2}(x)=\nb'(\vartheta),\\
%&\gb{\<3}(p)=\eb{\<3}(x)=\nb(\vartheta).
%\end{aligned}
%\end{equation}
%
The surface shifter defined by $\eqref{APW}_2$ turns out to be:
$$
\Ash(\vartheta,\zeta)=\cb_1\otimes\cb_1+\left(
1+\frac{\zeta}{\rho_o}
\right)\nb'(\vartheta)\otimes\nb'(\vartheta),
$$
whence
$$
\alpha(\zeta)=1+\frac{\zeta}{\rho_o}\,.
$$
Finally, the only nonnull Christoffel symbols on $\Sc$ are:
$$
\gamma_{22}^{\;3}=-\rho_o, \qquad
\gamma_{23}^{\;2}=\gamma_{32}^{\;2}=\rho_o^{-1}.
$$
\subsection{Kinematics}
The displacement field \eqref{KLdispl} now
 reads:
\begin{equation}\label{discil}
\begin{aligned}
\ub_C(x_1,\vartheta, \zeta)=&\left(
\cb_1\otimes\cb_1+\left(1+\frac{\zeta}{\rho_o}
\right)\nb'(\vartheta)\otimes\nb'(\vartheta)\right)\,\ab(x_1,\vartheta)+\\
&+\big(w(x_1,\vartheta)+\zeta\gamma\big)\nb(\vartheta)-\zeta\bigg(w_{,1}(x_1,\vartheta)\cb_1+\rho_o^{-1}w_{,2}(x_1,\vartheta)\nb'(\vartheta)
\bigg);
\end{aligned}
\end{equation}
its representation in the physical basis is:
\begin{equation}\label{displl}
\ub_C(x_1,\vartheta,
\zeta)=u{\<1}(x_1,\vartheta,\zeta)\cb_1+u{\<2}(x_1,\vartheta,\zeta)\nb'(\vartheta)+u{\<3}(x_1,\vartheta,\zeta)\nb(\vartheta),
\end{equation}
with components:
\begin{equation}\label{spost}
\begin{aligned}
&u{\<1}=a{\<1}-\zeta w_{,1}\,,\\
&u{\<2}=\left(1+\frac{\zeta}{\rho_o}\right)\,a{\<2}-\frac{\zeta}{\rho_o}\,w_{,2}\,,\\
&u{\<3}=w+\zeta\gamma\,.
\end{aligned}
\end{equation}
Since
%the displacement gradient is generally defined to be
%$\nabla\ub=\ub_{,i}\otimes\gb^i$, in cylindrical coordinates we
%have that
$$
\nabla\ub_C=\ub_{,i}\otimes\gb^i=\ub_{,1}\otimes\cb_1+
\frac{1}{\rho_0}\left( 1+\frac{\zeta}{\rho_o} \right)
^{-1}\ub_{,2}\otimes\nb'+\ub_{,3}\otimes\nb,
$$
the nonnull physical components of the strain tensor are:
\begin{equation}\label{straincil}
\begin{aligned}
E{\<{11}}&=\nabla\ub_C\cdot\cb_1\otimes\cb_1=u{\<1}_{,1}=a{\<1}_{,1}-\zeta w_{,11}\,,\\
2E{\<{12}}&=2E{\<{21}}=\nabla\ub_C\cdot(\cb_1\otimes\nb'+\nb'\otimes\cb_1)=u{\<2}_{,1}+\rho_0\left(
1+\frac{\zeta}{\rho_o} \right) ^{-1} u{\<1}_{,2}\\&=\left(
1+\frac{\zeta}{\rho_o}
\right)a{\<2}_{,1}-\frac{\zeta}{\rho_o}w_{,21}+\frac{1}{\rho_0}\left(
1+\frac{\zeta}{\rho_o} \right) ^{-1}(a{\<1}_{,2}-\zeta
w_{,12})\,,\\
E{\<{22}}&=\nabla\ub_C\cdot\nb'\otimes\nb'=\rho_0\left(
1+\frac{\zeta}{\rho_o} \right)
^{-1}(u_{<2>,2}+u_{<3>})\\
&=\frac{1}{\rho_0}\left( 1+\frac{\zeta}{\rho_o} \right)
^{-1}\left( \left( 1+\frac{\zeta}{\rho_o}
\right)\,a{\<2}_{,2}-\frac{\zeta}{\rho_o} w_{,22}+w +\zeta\gamma\right),\\
E{\<{33}}&=\nabla\ub_C\cdot\nb\otimes\nb=u{\<3}_{,3}=\gamma\,.
\end{aligned}
\end{equation}
%Finally, combining these relations with \eqref{spost}, we obtain:
%\begin{equation}
%\begin{aligned}
%E{\<{11}}&=a{\<1}_{,1}-\zeta w_{,11}\,,\\
%2E{\<{12}}&=\left( 1+\frac{\zeta}{\rho_o}
%\right)a{\<2}_{,1}-\frac{\zeta}{\rho_o}w_{,21}+\Big(\rho_o\left(
%1+\frac{\zeta}{\rho_o} \right)  \Big)^{-1}(a{\<1}_{,2}-\zeta
%w_{,12})\,,\\
%E{\<{22}}&=\Big(\rho_o\left( 1+\frac{\zeta}{\rho_o} \right)
%\Big)^{-1}\Bigg( \left( 1+\frac{\zeta}{\rho_o}
%\right)\,a{\<2}_{,2}-\frac{\zeta}{\rho_o} w_{,22}+w \Bigg),\\
%E{\<{33}}&=\gamma.
%\end{aligned}
%\end{equation}
\subsection{Balance Assumptions}
We exploit the virtual-power procedure detailed in Section
\ref{Equilibrium}, with the difference anticipated in Remark
\ref{rmk4}: the variations we now employ have the same structure
as the admissible displacements \eqref{displl}-\eqref{spost};
hence, no reactive contributions are going to enter  the field and
boundary equations that the procedure delivers. Precisely, the
variations in question have the following form:
\begin{equation}\label{vars}
\begin{aligned}
&\overset{(0)}\vb=v_1\cb_1+v_2\nb'+v_3\nb,\\
&\overset{(1)}\vb=-v_3,_1\cb_1+\rho_o^{-1}(v_2-v_3,_2)\nb'+v_4\nb,
\end{aligned}
\end{equation}
with the scalar fields $v_i=v_i(x_1,\vartheta)$ $(i=1,2,3)$
compactly supported in $\Sc$ and with $v_4$ a constant.
\subsubsection{Field equations} With a view toward deriving the field equations to which the general balances \eqref{bil} reduce in the present situation, we firstly take $v_4=0$ in \eqref{vars}.
In this instance,  on setting:
\[
\overset{(0)}\fb:=-(\Divs\Fbs+\qb_o),
\quad\overset{(0)}\mb:=-(\Divs\Mbs-\fb^{(3)}+\rb_o),
\]
the formulation \eqref{2DPVP} of the Principle of Virtual Powers
reads:
\[\begin{aligned}
0&=\int_\Pc\big(\overset{(0)}\fb\cdot\overset{(0)}\vb
+\overset{(0)}\mb\cdot\overset{(1)}\vb \big)=\int_\Pc\Big(
v_1\big(\overset{(0)}\fb\cdot\cb_1\big)+v_2\big(\overset{(0)}\fb+\rho_o^{-1}\,\overset{(0)}\mb\big)\cdot\nb^\prime+\\
&+v_3\big(\overset{(0)}\fb+\big(\overset{(0)}\mb\cdot\cb_1\big),_1+\rho_o^{-1}\big(\overset{(0)}\mb\cdot\nb^\prime\big),_2\big)\cdot\nb\Big).
\end{aligned}
\]
Hence, three field equations must hold at each interior point of
$\Sc$, namely,
\begin{equation}\label{pre3}
\begin{aligned}
&\overset{(0)}\fb\cdot\cb_1=0,\\
&\big(\overset{(0)}\fb+\rho_o^{-1}\,\overset{(0)}\mb\big)\cdot\nb^\prime=0,\\
&\big(\overset{(0)}\fb+\big(\overset{(0)}\mb\cdot\cb_1\big),_1+\rho_o^{-1}\big(\overset{(0)}\mb\cdot\nb^\prime\big),_2\big)\cdot\nb=0.
\end{aligned}
\end{equation}

To find the balance law that follows from testing the equality of
internal and external powers by way of virtual velocity fields of
the form:
\begin{equation}\label{v4}
\vb=\zeta\overset{(1)}\vb,\quad \overset{(1)}\vb=v_4\,\nb,
\end{equation}
with $v_4$ an arbitrary real number, we turn to the formulation
\eqref{PVPa} of the Principle: indeed, $v_4$ must be taken
constant over the whole of $\Omega$, because it is meant to be a
variation of the constant thickness strain $E{\<{33}}=\gamma$ in
the representation \eqref{spost} of the admissible
displacement.\footnote{More generally, we point out that, whenever
the admissible motions are such as to suggest the use of  test
fields of the form $\vb=v_o\wb$, with $v_o$ an arbitrary constant
and $\wb$ a given vector field, then \eqref{PVPa} is the
appropriate formulation of the Principle of Virtual Powers, and
the associated balance information is an integral relation of the
form:
\[
\int_\Omega\Sb\cdot\nabla\wb=\int_\Omega\db_o\cdot\wb+\int_{\partial\Omega}\cb_o\cdot\wb.
\]
}
 Given that, for any $\vb$ as in \eqref{v4},
\[
\nabla\vb=v_4(\nb\otimes\nb+\frac{\zeta}{\rho_o}\,\nb^\prime\otimes\nb^\prime),
\]
and that $v_4$ can be chosen arbitrarily,  \eqref{PVPa} and
\eqref{v4} yield:
\begin{equation}\label{pre4}
\int_{-l}^{+l}\int_0^{2\pi}\left(\int_I\alpha\big(\Sb\cdot\nb\otimes\nb+\frac{\zeta}{\rho_o}\,\Sb\cdot\nb^\prime\otimes\nb^\prime-\zeta\db_o\cdot\nb\big)d\zeta\right)dx_1d\vartheta=0\,.
\end{equation}
In terms of physical components of the force and moment tensors,
the system of equations  \eqref{pre3} and \eqref{pre4} can be
written as:
\begin{equation}\label{equicil}
\begin{aligned}
&F{\<{11}}_{,1}+\rho_o^{-1}F{\<{12}}_{,2}+q_{o}{\<1}=0,\\
&(F{\<{21}}+\rho_o^{-1}M{\<{21}})_{,1}+\rho_o^{-1}(F{\<{22}}+\rho_o^{-1}M{\<{22}})_{,2}+q_{o}{\<2}+\rho_o^{-1}r_{o}{\<2}=0,\\
&M{\<{11}}_{,11}+\rho_o^{-1}(M{\<{12}}+M{\<{21}})_{,12}+\frac{1}{\rho_o^2}M{\<{22}}_{,22}+\\
&\hspace{2cm}-\rho_o^{-1}F{\<{22}}+q_{o}{\<3}+r_{o}{\<1}_{,1}+\rho_o^{-1}r_{o}{\<2}_{,2}=0,\\
&\int_{-l}^{+l}\int_0^{2\pi}\left(\rho_o^{-1}M{\<{22}}+F{\<{33}}-r_{o}{\<3}\right)dx_1d\vartheta=0;
\end{aligned}
\end{equation}
the following version in terms of contravariant components may
easy a comparison with \eqref{bildue}:
\[
\begin{aligned}
&{F^{11}},_1+{F^{12}},_2+q_o^1=0,\\
&{F^{21}},_1+{F^{22}},_1+\rho_o^{-1}({M^{21}},_1+{M^{22}},_2+r_o^2)+q_o^2=0,\\
&{M^{11}},_{11}+{M^{22}},_{22}+(M^{21}+{M}^{12})_{,12}-\rho_o{F}^{22}+q_o^3+r_{o,1}^1+r_{o,2}^2=0,\\
&\int_{-l}^{+l}\int_0^{2\pi}\left(\rho_oM^{22}+F^{33}-r_o^3\right)dx_1d\vartheta=0.
\end{aligned}
\]
\subsubsection{Boundary equations} We start from the general weak statement  \eqref{issa}, repeated here for the reader's convenience:
$$
\int_{\partial\Pc}\Big((\Fbs\mb-\lb_o)\cdot\overset{(0)}\vb+(\Mbs\mb-\mb_o)\cdot\overset{(1)}\vb
\Big)=0,
$$
where we now insert variations of type \eqref{vars}. For
simplicity, we restrict attention to parts of the model surface
whose boundary consists of the union of two identical parts
$\mathpzc{d}_0$, $\mathpzc{d}_1$ of directrices at different axial
abscissae and the relative two segments of generatrices
$\mathpzc{g}_0$, $\mathpzc{g}_1$, that is to say, with reference
to Figure \ref{bc},
\begin{figure}[h]
\centering
\includegraphics[scale=.93]{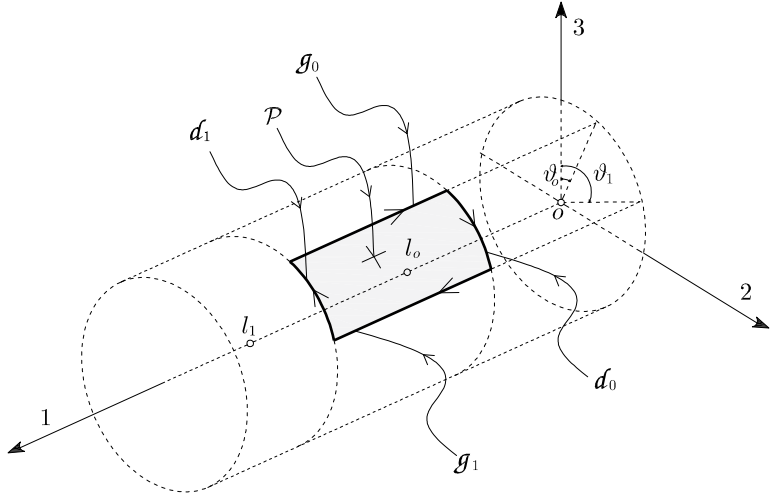}
\caption{A typical part of the model surface of a cylindrical
shell, bounded by generatrices and directrices.} \label{bc}
\end{figure}
parts $\Pc$ such that
\[
\partial\Pc=\bigcup_{a}\,\mathpzc{d}_a\cup\, \mathpzc{g}_a\;\,(a=0,1),
\]
with
\[
\mathpzc{d}_a:=\{x\in\Sc\,|\;x\equiv
(l_a,\vartheta),\;\vartheta\in(\vartheta_o,\vartheta_1)\},\;\mathpzc{g}_a:=\{x\in\Sc\,|\;x\equiv
(x_1,\vartheta_a),\;x_1\in(l_0,l_1)\}.
\]

Over $\mathpzc{d}_1$, where $\mb=\cb_1$, the integral condition
\eqref{issa} takes the form:
\begin{equation}\label{dir}
\begin{aligned}
0=&\int_{\vartheta_o}^{\vartheta_1}\Big((\Fbs\cb_1-\lb_o)\cdot(v_1\cb_1+v_2\nb'+v_3\nb)+\\
&+(\Mbs\cb_1-\mb_o)\cdot(-v_3,_1\cb_1+\rho_o^{-1}(v_2-v_3,_2)\nb'+v_4\nb)\Big)\Big|_{x_1=l_1}\,d\vartheta=\\
&=\int_{\vartheta_o}^{\vartheta_1}\Big(v_1(F{\<{11}}-l_{o}{\<1})+v_2(F{\<{21}}+\rho_o^{-1}M{\<{21}}-l_{o}{\<2}-\rho_o^{-1}m_{o}{\<2})+\\
&+v_3(F{\<{31}}-\rho_o^{-1}M{\<{21}}_{,2}-l_{o}{\<3}-\rho_o^{-1}m_{o}{\<2}_{,2})+\\
&-v_3,_1(M{\<{11}}-m_{o}{\<1})+v_4(M{\<{31}}-m_{o}{\<3})\Big)\Big|_{x_1=l_1}\,d\vartheta+\\
&+\Big(v_3(M{\<{21}}-\rho_o^{-1}m_{o}{\<2})\Big)\Big|_{(l_1,\vartheta_0)}^{(l_1,\vartheta_1)}\,;
\end{aligned}
\end{equation}
over $\mathpzc{g}_1$, where $\mb=\nb'(\vartheta)$,  \eqref{issa}
becomes:
\begin{equation}\label{gen}
\begin{aligned}
0=&\int_{l_o}^{l_1}\Big((\Fbs\nb^\prime-\lb_o)\cdot(v_1\cb_1+v_2\nb'+v_3\nb)+\\
&+(\Mbs\nb^\prime-\mb_o)\cdot(-v_3,_1\cb_1+\rho_o^{-1}(v_2-v_3,_2)\nb'+v_4\nb)\Big)\Big|_{\vartheta=\vartheta_1}\,dx_1=\\
&=\int_{l_o}^{l_1}\Big(v_1(F{\<{12}}-l_o{\<1})+v_2(F{\<{22}}+\rho_o^{-1}M{\<{22}}-l_o{\<2}-\rho_o^{-1}m_o{\<2})+\\
&+v_3(F{\<{32}}-\rho_o^{-1}M{\<{12}}_{,1}-l_o{\<3}-\rho_o^{-1}m_o{\<2}_{,1})+\\
&-v_{3,2}(M{\<{22}}-m_o{\<{1}})+v_4(M{\<{32}}-m_o{\<3})
\Big)\Big|_{\vartheta=\vartheta_1}\,dx_1+\\
&+\Big(v_3(M{\<{12}}-\rho_o^{-1}m_{o}{\<2})\Big)\Big|_{(l_0,\vartheta_1)}^{(l_1,\vartheta_1)}\,;
\end{aligned}
\end{equation}
needless to say, relations completely similar to \eqref{dir} and
\eqref{gen} hold, respectively, at  $\mathpzc{d}_0$ and
$\mathpzc{g}_0$.

As mentioned in Section \ref{bc}, if  ${\overset{(1)}{u}}\!_3$ or
anyone of the components of $\overset{(0)}\ub$ and
$\nabla{\overset{(0)}{u}}\!_3$ is assigned at a boundary point,
then $v_4$ or the corresponding component of $\overset{(0)}\vb$
and $\nabla{\overset{(0)}{v}}\!_3$   must vanish at that point; on
the other hand, the arbitrary variation of each of the remaining
parameters in \eqref{vars} induces at the same point a boundary
equation of Neumann type: for example, inspection of \eqref{dir}
shows that, if $\,\overset{(0)}\ub\cdot\cb_1=a\<1\>$ is assigned
over $\mathpzc{d}_a$, then $v_1$ must be taken there identically
null, while the Neumann boundary equations
\[
F{\<{21}}+\rho_o^{-1}M{\<{21}}=l_{o}{\<2}+\rho_o^{-1}m_{o}{\<2}\;\,\textrm{etc.}
\]
must hold; in other words, at a point of $\mathpzc{d}_a$
admissible boundary conditions must consist of a list of mutually
exclusive assignments of the one or the other element of the
following five power-conjugate pairs:
\begin{equation}\label{conto1}
\begin{aligned}
&(F{\<{11}},  a{\<1}),\;
(F{\<{21}}+\rho_o^{-1}M{\<{21}},a{\<2}),\\
&(F{\<{31}}-\rho_o^{-1}M{\<{21}}_{,2}, w),\;(M{\<{11}}, w,_1), \;
(M{\<{31}}, \gamma).
\end{aligned}
\end{equation}
Likewise, from \eqref{gen} we deduce that the boundary conditions
at a point of $\mathpzc{g}_a$ should consist of mutually exclusive
assignments of the pairs:
\begin{equation}\label{conto2}
\begin{aligned}
&(F{\<{12}},  a{\<1}),\; (F{\<{22}}+\rho_o^{-1}M{\<{22}},a{\<2},\\
&(F{\<{32}}-\rho_o^{-1}M{\<{12}}_{,1}, w),\; \;(M{\<{22}},
w,_2),\big(M{\<{32}}, \gamma).
\end{aligned}
\end{equation}
%\begin{equation}
%\begin{aligned}
%&\big(F{\<{12}},  a{\<1}\big),\quad
%\big(F{\<{22}}+\rho_o^{-1}M{\<{22}},a{\<2}\big),\\
%&\big(F{\<{32}}-\rho_o^{-1}M{\<{12}}_{,1}, w\big),\quad
%\big(M{\<{32}}, \gamma\big).
%\end{aligned}
%\end{equation}
%
\subsubsection{Inertial interactions. Evolution equations}
The time evolution of an unshearable orthotropic shell is ruled by
the partial differential equations that follow from \eqref{bil}
when the inertial force is separated from the rest of the distance
force per unit volume of $\Gc(\Sc,\varepsilon)$. To this effect,
we set:
\[
\db_o^{\rm in}:=-\delta_o\ddot\ub \quad\textrm{and}\quad
\db_o^{\rm ni}:=\db_o-\db_o^{\rm in},
\]
for, respectively, the inertial and noninertial distance forces
(here $\delta_o$ denotes the mass density per unit reference
volume and a superposed dot signifies time differentiation). We
then set:
\begin{equation}\label{inforze}
\qb_o^{\rm
in}:=-\int_I\alpha(x,\zeta)\delta_o(x,\zeta)\ddot{\ub}(x,\zeta;t)\,d\zeta\quad\textrm{and}\quad
\rb_o^{\rm
in}:=-\int_I\alpha(x,\zeta)\zeta\delta_o(x,\zeta)\ddot{\ub}(x,\zeta;t)\,d\zeta,
\end{equation}
for the \emph{inertial force} and the \emph{inertial couple} per
unit area of $\Sc$. Finally, we return to definitions
$\eqref{forceload}_1,_2$ and set:
$$
 \qb_o^{\rm ni}:=\qb_o-\qb_o^{\rm in} \quad \textrm{and}\quad  \rb_o^{\rm ni}:=\rb_o-\rb_o^{\rm in},
$$
for the relative noninertial loadings. In conclusion, the balance
equations \eqref{bil} take the evolutionary form:
\begin{equation}\label{evol}
\begin{aligned}
&\qb_o^{\rm
in}=\Divs\Fbs+\qb_o^{\rm ni},\\
&\rb_o^{\rm in}=\Divs\Mbs-\fb^{(3)}+\rb_o^{\rm ni}.
\end{aligned}
\end{equation}
In the case of cylindrical shells whose mass distribution is
uniform, definitions \eqref{inforze} yield:
\begin{equation}\label{quzero}
\begin{aligned}
q_{o}^{\rm in}{\<1}&=-\bar{\delta}_o\left(\ddot
a{\<1}-\frac{1}{3}\varepsilon^2\rho_o^{-1}\ddot w_{,1}\right),\\
q_{o}^{\rm
in}{\<2}&=-\bar\delta_o\left(\left(1+\frac{1}{3}\frac{\varepsilon^2}{\rho_o^2}
\right)\,\ddot
a{\<2}-\frac{1}{3}\frac{\varepsilon^2}{\rho_o^2}\,\ddot w_{,2}
\right),\\
q_{o}^{\rm in}{\<3}&=-\bar\delta_o\Big(\ddot
w+\frac{1}{3}\frac{\varepsilon^2}{\rho_o}\ddot\gamma \Big),\\
r_{o}^{\rm in}{\<1}&=-\bar\delta_o\Big(\rho_o^{-1}\,\ddot
a{\<1}-\ddot w_{,1}\Big),\\
r_{o}^{\rm
in}{\<2}&=-\bar\delta_o\frac{1}{3}\frac{\varepsilon^2}{\rho_o}\left(
\,\ddot a{\<2}-\,\ddot w_{,2}
\right),\\
r_{o}^{\rm
in}{\<3}&=-\bar\delta_o\frac{\varepsilon^2}{3\rho_o}\Big(\ddot
w+\rho_o\ddot\gamma \Big),
\end{aligned}
\end{equation}
where $\bar\delta_o:=(2\varepsilon)\delta_o$ is the uniform mass
density per unit area of the model surface $\Sc$. Hence, the
evolution equations corresponding to the balance equations
\eqref{equicil} are:
\begin{equation}\label{evol1}
\begin{aligned}
&\bar{\delta}_o\left(\ddot
a{\<1}-\frac{1}{3}\varepsilon^2\rho_o^{-1}\ddot
w_{,1}\right)=F{\<{11}}_{,1}+\rho_o^{-1}F{\<{12}}_{,2}+q^{\rm ni}_{o}{\<1},\\
&\bar\delta_o\left(\left(1+\frac{2}{3}\frac{\varepsilon^2}{\rho_o^2}
\right)\,\ddot
a{\<2}-\frac{2}{3}\frac{\varepsilon^2}{\rho_{o}^2}\,\ddot w_{,2}
\right)=\\
&\hspace{0.7cm}=(F{\<{21}}+\rho_o^{-1}M{\<{21}})_{,1}+\rho_o^{-1}(F{\<{22}}+\rho_o^{-1}M{\<{22}})_{,2}+q^{\rm ni}_{o}{\<2}+\rho_o^{-1}r_o^{\rm ni}{\<2},\\
&\bar\delta_o\left(\ddot w+è\frac{\varepsilon^2}{3\rho_o}\left(\ddot\gamma+\ddot a{\<1}_{,1}-\rho_o\ddot w_{,11}+\rho_o^{-1}\ddot a{\<{2}}_{,2}-\rho_o^{-1}\ddot w_{,2} \right)\right)=\\
&\hspace{0.7cm}=M{\<{11}}_{,11}+\rho_o^{-1}(M{\<{12}}+M{\<{21}})_{,12}+\frac{1}{\rho_o^2}M{\<{22}}_{,22}-\rho_o^{-1}F{\<{22}}+\\
&\hspace{0.7cm}+q^{\rm ni}_{o}{\<3}+r_o^{\rm
ni}{\<1}_{,1}+\rho_o^{-1}r_o^{\rm ni}{\<2}_{,2},\\
&\bar\delta_o\frac{\varepsilon^2}{3\rho_o}\Big(\ddot
w+\rho_o\ddot\gamma \Big)=\rho_o^{-1}M{\<{22}}+F{\<{33}}-r^{\rm
ni}_{o}{\<3}.
\end{aligned}
\end{equation}

Needless to say, these equations are to be equipped with a set of
initial conditions
 for the unknown fields $a{\<{1}},a{\<{2}},w$, and $\gamma$, and for their time rates.

\subsection{Constitutive Assumptions}
The components of force and moment tensors that appear in
\eqref{equicil} depend on the active part of the three-dimensional
stress field. The latter has the following expression in terms of
the constitutive law \eqref{consta} and the strain tensor
\eqref{straincil}:
\begin{equation}\label{Eqcostcil}
\Sb(x,\zeta)=\widetilde\Co[\Eb(\ub_C(x,\zeta))]\,;
\end{equation}
in components, this equation reads:
\begin{equation}\label{stresscomp}
\begin{aligned}
&S{\<{11}}=\frac{E_1}{\Delta}(1-\nu_{23}\nu_{32})\,E{\<{11}}+\frac{E_1}{\Delta}(\nu_{21}+\nu_{23}\nu_{31})\,E_{\<{22}}+\frac{E_1}{\Delta}(\nu_{31}+\nu_{21}\nu_{32})\,E{\<{33}},\\
&S{\<{22}}=\frac{E_2}{\Delta}(1-\nu_{13}\nu_{31})\,E{\<{22}}+\frac{E_1}{\Delta}(\nu_{21}+\nu_{23}\nu_{31})\,E{\<{11}}+\frac{E_2}{\Delta}(\nu_{32}+\nu_{31}\nu_{12})\,E{\<{33}},\\
&S{\<{33}}=\frac{E_3}{\Delta}(1-\nu_{12}\nu_{21})\,E{\<{33}}+\frac{E_1}{\Delta}(\nu_{31}+\nu_{21}\nu_{32})\,E{\<{11}}+\frac{E_2}{\Delta}(\nu_{32}+\nu_{31}\nu_{12})\,E{\<{22}},
\\
&S{\<{12}}=S{\<{21}}=2G\, E{\<{12}},\\[2.mm]
\end{aligned}
\end{equation}
where the constitutive modulus $\Delta$ is defined as in
\eqref{constb}, the components $E{\<{\alpha\beta}}$ as in
\eqref{straincil}. On inserting these relations in the appropriate
consequences of definitions \eqref{2sforzi}, we find:
\begin{equation}%\label{1bil}
\begin{aligned}
F{\<{11}}=&\int_I\left(1+\frac{\zeta}{\rho_o}
\right)S{\<{11}}=2\frac{\varepsilon}{\rho_o}\frac{E_1}{\Delta}\Bigg[(1-\nu_{23}\nu_{32})\left(\rho_oa{\<1}_{,1}-\frac{1}{3}\varepsilon^2w_{,11}\right)\\
&+(\nu_{21}+\nu_{23}\nu_{32})(a{\<2}_{,2}+w)+(\nu_{31}+\nu_{21}\nu_{32})\rho_o\gamma
\Bigg],\\
\end{aligned}
\end{equation}
\begin{equation}\label{2bil}
\begin{aligned}
F{\<{22}}=&\int_IS{\<{22}}=2\frac{\varepsilon}{\rho_o}\frac{E_2}{\Delta}\Bigg[(1-\nu_{13}\nu_{31})\left(a{\<2}_{,2}-\left(1-\frac{1}{2\frac{\varepsilon}{\rho_o}}\log\frac{1+\frac{\varepsilon}{\rho_o}}{1-\frac{\varepsilon}{\rho_o}}
\right)(w_{,22}-\rho_o\gamma)+\right.\\
&\frac{1}{2\frac{\varepsilon}{\rho_o}}\log\frac{1+\frac{\varepsilon}{\rho_o}}{1-\frac{\varepsilon}{\rho_o}}\,w\left.\right.\Bigg)+\frac{\rho_o}{\eta}(\nu_{21}+\nu_{23}\nu_{31})\,a{\<1}_{,1}+(\nu_{32}+\nu_{31}\nu_{12})\rho_o\gamma
\Bigg],
\end{aligned}
\end{equation}
\begin{equation}\label{4bil}
\begin{aligned}
F{\<{33}}=&\int_I\left(1+\frac{\zeta}{\rho_o}
\right)S{\<{33}}=2\frac{\varepsilon}{\rho_o}\frac{E_3}{\Delta}\Bigg[(1-\nu_{12}\nu_{21})\rho_o\gamma+\\
&+\frac{\nu_{31}+\nu_{21}\nu_{32}}{\lambda}\left(\rho_{o}a{\<1}_{,1}-\frac{1}{3}\varepsilon^2
w_{,11}\right)
+\frac{\nu_{32}-\nu_{31}\nu_{12}}{\mu}\left(a{\<2}_{,2}+w \right)
\Bigg],
\end{aligned}
\end{equation}
\begin{equation}\label{1bil}
\begin{aligned}
 F{\<{21}}=&\int_I\left(1+\frac{\zeta}{\rho_o}
\right)S{\<{21}}=2\varepsilon
G\left[\left(1+\frac{1}{3}\frac{\varepsilon^2}{\rho_o^2}
\right)\,a{\<2}_{,1}-\frac{1}{3}\frac{\varepsilon^2}{\rho_o^2}\,
w_{,21}+\frac{1}{\rho_o}a{\<1}_{,2} \right],
\end{aligned}
\end{equation}
and
\begin{equation}\label{5bil}
\begin{aligned}
\hspace{-1cm}M{\<{11}}=&\int_I\left(1+\frac{\zeta}{\rho_o}
\right)\zeta
S{\<{11}}=\frac{2}{3}\frac{\varepsilon^3}{\rho_o}\frac{E_1}{\Delta}\Bigg[(1-\nu_{23}\nu_{32})(a{\<1}_{,1}-\rho_o w_{,11})+\\
&+(\nu_{21}+\nu_{23}\nu_{31})(a{\<2}_{,2}-w_{,22})+(\nu_{31}+\nu_{21}\nu_{32})\gamma
\Bigg],
\end{aligned}
\end{equation}

\begin{equation}\label{6bil}
\begin{aligned}
\hspace{-3.8cm}M{\<{21}}=\int_I\left(1+\frac{\zeta}{\rho_o}
\right)\zeta
S{\<{21}}=\frac{4}{3}\frac{\varepsilon^3}{\rho_o}G(a{\<2}_{,1}-w_{,12}),
\end{aligned}
\end{equation}

\begin{equation}\label{7bil}
\begin{aligned}
M{\<{12}}=&\int_I\zeta
S{\<{12}}=\frac{2}{3}\frac{\varepsilon^3}{\rho_o}G\Bigg[a{\<2}_{,1}-w_{,21}+3\frac{\rho_o}{\varepsilon^2}\left(1-\frac{1}{\frac{\varepsilon}{\rho_o}}\log\frac{1+\frac{\varepsilon}{\rho_o}}{1-\frac{\varepsilon}{\rho_o}}
\right)a{\<1}_{,2}+\\
&-3\frac{1}{\frac{\varepsilon^2}{\rho_o^2}}\left(1-\log\frac{1+\frac{\varepsilon}{\rho_o}}{1-\frac{\varepsilon}{\rho_o}}
\right)w_{,12} \Bigg],
\end{aligned}
\end{equation}

\begin{equation}\label{8bil}
\begin{aligned}
M{\<{22}}=&\int_I\zeta
S{\<{22}}=2\varepsilon\frac{E_2}{\Delta}\Bigg[(1-\nu_{13}\nu_{31})\left(1-\frac{1}{2\frac{\varepsilon}{\rho_o}}\log\frac{1+\frac{\varepsilon}{\rho_o}}{1-\frac{\varepsilon}{\rho_o}}\right)\left(w-w_{,22}+\rho_o\gamma
 \right)+\\
 &-\frac{1}{3}\varepsilon^2\frac{\nu_{12}+\nu_{23}\nu_{31}}{\eta}w_{,11}\Bigg],
\end{aligned}
\end{equation}
where
\begin{equation}\label{moreconst}
\frac{E_1}{E_2}=\frac{\nu_{12}}{\nu_{21}}=:\frac{1}{\eta},\quad
\frac{E_1}{E_3}=\frac{\nu_{13}}{\nu_{31}}=:\frac{1}{\lambda},
\quad \frac{E_2}{E_3}=\frac{\nu_{23}}{\nu_{32}}=:\frac{1}{\mu}.
\end{equation}

The equations governing the equilibria of unshearable cylindrical
shells are arrived at when the above constitutive equations are
inserted into the balances \eqref{equicil}. In their general form,
those equations are complicated to solve analytically; we choose
not to list them here. However, certain highly symmetric problems
admit  simple and explicit solutions. We deal with such problems
in the next section, while the noticeable simplifications obtained
when the approximations judged appropriate for thin and slender
shells will be introduced and exemplified in Section 5.
%
%%
%\vfill
%\pagebreak
%
%
\section{Cylindrical Shells: Axisymmetric Boundary-Value Problems}
A boundary-value problem for a cylindrical shell is
\emph{axisymmetric} if the load and confinement data induce
solution displacement fields whose physical components $u{\<{i}}$
are all independent of the circumferential coordinate $\vartheta$,
that is to say, in view of \eqref{spost}, if
\begin{equation}\label{cilcil}
u{\<1}=a{\<1}-\zeta w', \quad u{\<2}=\left(1+\frac{\zeta}{\rho_o}
\right)a{\<2}, \quad u{\<3}=w+\zeta\gamma,
\end{equation}
where a prime denotes differentiation with respect to $x_1$, the
only space variable from which all of the parameter fields
$a{\<1},a{\<2}$, and $w$, may depend. When the displacement field
has the form \eqref{cilcil},

\noindent (i) the strain components \eqref{straincil} take the
simpler form:
\begin{equation}\label{strainax}
\begin{aligned}
E{\<{11}}&=a{\<1}'-\zeta w''\,,\\
E{\<{12}}&=E{\<{21}}=\frac{1}{2}\left( 1+\frac{\zeta}{\rho_o}
\right)a{\<2}',\\
E_{\<{22}}&=\left(\rho_o\left(
1+\frac{\zeta}{\rho_o} \right) \right)^{-1}(w+\zeta\gamma),\\
E{\<{33}}&=\gamma;
\end{aligned}
\end{equation}
(ii) the constitutive equations \eqref{1bil}-\eqref{8bil} for the
force and moment components become:
\begin{equation}\label{constF}
\begin{aligned}
F{\<{11}}&=2\frac{\varepsilon}{\rho_o}\frac{E_1}{\Delta}\Bigg[(1-\nu_{23}\nu_{32})\left(\rho_oa{\<1}'-\frac{1}{3}\varepsilon^2
w''\right)
\\
&+(\nu_{21}+\nu_{23}\nu_{31})w+(\nu_{31}+\nu_{21}\nu_{32})\rho_o\gamma
\Bigg],\\
F{\<{22}}&=2\frac{\varepsilon}{\rho_o}\frac{E_2}{\Delta}\Bigg[(1-\nu_{13}\nu_{31})\left(\left( 1-\frac{1}{2\frac{\varepsilon}{\rho_o}}\log\frac{1+\frac{\varepsilon}{\rho_o}}{1-\frac{\varepsilon}{\rho_o}}\right)\rho_o\gamma+\frac{1}{2\frac{\varepsilon}{\rho_o}}\log\frac{1+\frac{\varepsilon}{\rho_o}}{1-\frac{\varepsilon}{\rho_o}}\,w\right)\\
&+\rho_o\frac{\nu_{21}+\nu_{23}\nu_{31}}{\eta}\,a{\<1}'+(\nu_{32}+\nu_{31}\nu_{12})\rho_o\gamma \Bigg]\\
F{\<{21}}&=2\varepsilon
G\left(1+\frac{1}{3}\frac{\varepsilon^2}{\rho_o^2}
\right)\,a{\<2}',\\
F{\<{33}}&=2\frac{\varepsilon}{\rho_o}\frac{E_3}{\Delta}\Bigg[(1-\nu_{12}\nu_{21})\rho_o\gamma+\frac{\nu_{31}+\nu_{21}\nu_{32}}{\lambda}\left(\rho_{o}a{\<1}'-\frac{\varepsilon^2}{3}\,w''\right)+\frac{\nu_{32}+\nu_{31}\nu_{12}}{\mu}\,w\Bigg],
\end{aligned}
\end{equation}
and
\begin{equation}\label{constM}
\begin{aligned}
M{\<{11}}&=-\frac{2}{3}\frac{\varepsilon^3}{\rho_o}\frac{E_1}{\Delta}\Bigg[(1-\nu_{23}\nu_{32})(\rho_o
w''-a{\<1}')-(\nu_{31}+\nu_{21}\nu_{32})\gamma
\Bigg],\\
M{\<{21}}&=\frac{4}{3}\frac{\varepsilon^3}{\rho_o}G\,a{\<2}',\quad
M{\<{12}}=\frac{2}{3}\frac{\varepsilon^3}{\rho_o}G\,a{\<2}',\\
M_{\<{22}}&=2\varepsilon\frac{E_2}{\Delta}\Bigg[(1-\nu_{13}\nu_{31})\left(1-\frac{1}{2\frac{\varepsilon}{\rho_o}}\log\frac{1+\frac{\varepsilon}{\rho_o}}{1-\frac{\varepsilon}{\rho_o}}\right)(w+\rho_o\gamma)-\frac{1}{3}\varepsilon^2
\frac{\nu_{12}+\nu_{23}\nu_{31}}{\eta}w''\Bigg];
\end{aligned}
\end{equation}
(iii) the reaction-free equations $\eqref{equicil}$ reduce to the
field equations:
\begin{equation}\label{equil}
\begin{aligned}
&F{\<{11}}'+q_{o}{\<1}=0,\\
&(F{\<{21}}+\rho_o^{-1}M{\<{21}})'+q_{o}{\<2}+\rho_o^{-1}r_{o}{\<2}=0,\\
&M{\<{11}}''-\rho_{o}^{-1}F{\<{22}}+q_{o}{\<3}+r_{o}{\<1}'=0,\\
\end{aligned}
\end{equation}
holding in the interval $(-l,+l)$, plus the integral relation:
\begin{equation}\label{integ}
\int_{-l}^{+l}\left(\rho_o^{-1}M{\<{22}}+F{\<{33}}-r_{o}{\<3}\right)dx_1=0.
\end{equation}

\noindent (iv) the reaction-free boundary conditions consist in
specifications at $\pm l$ of one of the elements in each of the
pairs $\eqref{conto1}_1$, $\eqref{conto1}_2$, and
$\eqref{conto1}_4$:
\begin{equation}\label{conto2}
\begin{aligned}
&(F{\<{11}},  a{\<1}),\quad
(F{\<{21}}+\rho_o^{-1}M{\<{21}},a{\<2}),\quad(M{\<{11}},
w^\prime);
\end{aligned}
\end{equation}

Insertion of \eqref{constF} and \eqref{constM} into \eqref{equil}
and \eqref{integ} yields the system of equations ruling
axisymmetric boundary-value problems in our theory. We are going
to solve this system for assignments of Neumann data
corresponding, respectively, to problems of torsion, axial
traction, pressure, and rim flexure. Interestingly, as we shall
quickly demonstrate, this system splits into one equation for the
circumferential displacement $a{\<2}$ plus a system of three
equations for the axial and radial displacements $a{\<1}$ and $w$
and the thickness stretch $\gamma$.
\vskip 6pt \remark As mentioned in closing Section \ref{fmv}, once
the equilibrium displacement field has been found, the axial
distributions of reactive stress measures $F{\<{31}}$ and
$M{\<{31}}$ consistent with boundary conditions compatible with
$\eqref{conto1}_3$ and $\eqref{conto1}_5$ can be determined  by a
use of, respectively, the balance equation $\eqref{react}_1$ and
the balance equation \eqref{reactm}, keeping into account
\eqref{integ}. The situations of our present interest occur when
homogeneous Neumann data are prescribed  at the boundary, so that
the problems to solve are:
\begin{equation}\label{rebordo1}
F{\<{31}}=M{\<{11}}^\prime
+r_{o}{\<1}\;\;\textrm{in}\;(-l,+l),\;\, F{\<{31}}(\pm l)=0,
\end{equation}
and
\begin{equation}\label{redordo2}
M{\<{31}}^\prime=0\;\;\textrm{in}\;(-l,+l),\;\, M{\<{31}}(\pm
l)=0\quad\Leftrightarrow\quad M{\<{31}}\equiv 0
\;\;\textrm{in}\;[-l,+l].
\end{equation}
Note that equation $\eqref{rebordo1}_1$ has the familiar structure
of the moment balance in Bernoulli-Navier rod theory, with
$F{\<{31}}$ playing the role of the shear resultant, $M{\<{11}}$
of the bending moment, and $r_{o}{\<1}$ of the diffused applied
couples; both  $F{\<{31}}$ here and the shear resultant in that
classic rod theory have reactive nature, as a consequence of one
and the same unshearability constraint.

\subsection{Torsion}\label{torsio}
From $\eqref{constF}_3$ and $\eqref{constM}_2$, we have that
\begin{equation}\label{torc}
F{\<{21}}+\rho_o^{-1}M{\<{21}}=2\varepsilon
\left(1+\frac{\varepsilon^2}{\rho_o^2} \right)G\,a{\<2}';
\end{equation}
with this, equation $\eqref{equil}_2$ takes the form of a
second-order equation for $a\<2$:
\begin{equation}\label{tors}
2\varepsilon\left(1+ \frac{\varepsilon^2}{\rho_o^2}\right)G\,
a{\<2}''+q_o{\<2}+\rho_o^{-1}r_o{\<2}=0.
\end{equation}
This equation accounts for whatever twisting about its axis an
unshearable cylindrical shell may have; it can be associated with
boundary conditions specifying the values at $x_1=\pm l$ of either
$a\<2$ or
%$F{\<{21}}+\rho_o^{-1}M{\<{21}}$ (that is to say, in view of \eqref{torc}, of
$a{\<{2}}^\prime$.

When the only applied load is a distribution of end tractions
statically equivalent to two mutually balancing torques of
magnitude
\begin{equation}\label{tigrande}
T=(2\pi\rho_o^2)t, \quad\textrm{with}\;\, t=O(\varepsilon),
\end{equation}
 the thickness stretch $\gamma$ and the axial displacement  $a{\<{1}}$ vanish, and \eqref{tors}, the only relevant  equation, reduces to $a{\<{2}}'=$
a constant, whose value is determined by the boundary condition:
 \begin{equation}\label{storci}
F{\<{21}}+\rho_o^{-1}M{\<{21}}=t=r_T\,a{\<{2}}^\prime,\quad
r_T:=2\varepsilon\left(1+\frac{\varepsilon^2}{\rho_o^2}\right)\,G\,;
 \end{equation}
 it follows from \eqref{storci} that a twisted
shell of the type we study undergoes a rotation per unit length
\begin{equation}\label{rotu}
\Theta:= \rho_o^{-1}a{\<{2}}^\prime
\end{equation}
proportional to $\,(\rho_o r_T)^{-1}$. Moreover, given that the
function $a{\<{2}}$ must be odd,
\begin{equation}\label{tors1}
a{\<{2}}(x_1)=\frac{1}{1+\frac{\varepsilon^2}{\rho_o^2}}\frac{\varepsilon^{-1}t}{2G}\,x_1\,.
\end{equation}
%
%%
%\[
%\rho_o^{-1}\left(1+\frac{\varepsilon^2}{\rho_o^2}\right)^{-1}\frac{\varepsilon^{-1}t}{2G},
%\]
%
%in agreement with Saint-Venant's classic result for linearly
%elastic isotropic twisted tubes.
%It is not difficult to convince ourselves that all the other displacement fields are null, as well as
%all the stress descriptors unless $M{\<{21}}$.
\vskip 6pt
\remark %\subsection{Twist waves}\label{twist}
In line with $\eqref{quzero}_2$, set
\[
q_o{\<2}=q_o^{in}{\<2}=-\bar\delta_o\left(1+\frac{1}{3}\frac{\varepsilon^2}{\rho_o^2}
\right)\,\ddot a{\<2}
\]
in equation \eqref{tors}, so that it takes the form of the
classical wave equation:
\begin{equation}\label{torsev}
\frac{\partial^2a{\<{2}}(x_1,t)}{\partial
t^2}-c^2\frac{\partial^2a{\<{2}}(x_1,t)}{\partial
x_1^2}=0,\quad\textrm{with}\quad
c^2=\frac{G}{\delta_o}\,\frac{1+\frac{\varepsilon^2}{\rho_o^2}}{1+\frac{2}{3}\frac{\varepsilon^2}{\rho_o^2}}.
\end{equation}
Then, %whatever the slenderness of the cylindrical shell under study,
$$
a{\<{2}}(x_1,t)=\varphi(x_1+ct)+\psi(x_1-ct),
$$
and two \emph{twist waves} propagate along the axis with speed
$|c|$, the one in the positive direction the other in the negative
direction.

\vskip 6pt \remark In all boundary-value problems we shall solve
next -- we recall, axial traction, uniform pressure, and rim
flexure   -- the twisting loads are null. As a consequence, in all
three cases, the second equation of system \eqref{equil} and the
boundary equations:
\[
\big(F{\<{21}}+\rho_o^{-1}M{\<{21}}\big)(\pm l)=0
\]
together imply that the construct
$(F{\<{21}}+\rho_o^{-1}M{\<{21}})$ is identically null in
$[-l,+l]$. Hence, by the constitutive relations $\eqref{constF}_3$
and $\eqref{constM}_2$, the circumferential displacement $a{\<2}$
has to have constant value, as is the case for a rotation about
the $x_1-$axis; we note that this, because of $\eqref{constM}_2$,
implies that
\[
M{\<{21}}(\pm l)=0,
\]
and we take $a{\<2}\equiv 0$. In fact, as is customary in
elasticity with Neumann data, in all three cases we expect to
arrive at a displacement solution being unique to within an
ignorable rigid motion. Moreover, given the common built-in
symmetries, there will be no loss of generality in searching for
solutions with $a\<1$ an odd function of $x_1\in[-l,+l]$, and $w$
even.

\subsection{Traction}\label{traz}
Let us take all distance forces and couples null  and all boundary
conditions of Neumann type and homogeneous, except for a
distribution of tractions at the ends equivalent to two mutually
balancing axial forces of magnitude
\begin{equation}\label{axtraz}
P=(2\pi\rho_o)p,\quad \textrm{with}\;\,p=O(\varepsilon).
\end{equation}

Two of the balance equations \eqref{equil} are in force:
\begin{equation}\label{ecco}
\begin{aligned}
&F{\<{11}}'=0,\\
&M{\<{11}}''-\rho_{o}^{-1}F{\<{22}}=0;
\end{aligned}
\end{equation}
the accompanying boundary equations are:
\begin{equation}\label{dircili}
F{\<{11}}(\pm l)=p,\quad M{\<{11}}(\pm l)=0, \quad M{\<{11}}'(\pm
l)=0
\end{equation}
the last one following from \eqref{rebordo1}. Now,
$\eqref{ecco}_1$ and $\eqref{dircili}_1$ imply that the membrane
force $F{\<{11}}$ is constant and equal to $p$ in the closed
interval $[-l,+l]$; then, making use of $\eqref{constF}_1$, we
obtain that the differential relation:
\begin{equation}\label{ali}
\begin{aligned}
(1-\nu_{23}\nu_{32})\Big(\rho_o
a{\<1}^\prime-\frac{1}{3}\varepsilon^2 w^{\prime\prime}\Big)&+
(\nu_{21}+\nu_{23}\nu_{31})w+\\
&+(\nu_{31}+\nu_{21}\nu_{32})\rho_o\gamma=\rho_o\Delta\,
\frac{\varepsilon^{-1}p}{2E_1}
\end{aligned}
\end{equation}
must hold in $[-l,+l]$. Moreover, with the use of
$\eqref{constM}_1$ and under the parity assumptions we made for
$a\<1$ and $w$, we see that the boundary conditions
$\eqref{dircili}_2$ take the common form:
\begin{equation}\label{gam}
(1-\nu_{23}\nu_{32})\Big(a{\<1}'(l)-\rho_o w''(l)\Big)+
(\nu_{31}+\nu_{21}\nu_{32}\big)\gamma=0.
\end{equation}
Finally, with the use of $\eqref{constM}_1$ again and of
$\eqref{constF}_2$, and under the provisional assumption that the
traction problem admits a solution with $\gamma$ a constant,
$\eqref{ecco}_2$ becomes:
\begin{equation}\label{brandi}
\begin{aligned}
&\frac{1}{3}\varepsilon^2\frac{1-\nu_{23}\nu_{32}}{\eta}\Big(\rho_o^2
w''-\rho_o  a{\<1}'\Big)^{\prime\prime}+\\
&+(1-\nu_{13}\nu_{31})\left(\left( 1-\frac{1}{2\frac{\varepsilon}{\rho_o}}\log\frac{1+\frac{\varepsilon}{\rho_o}}{1-\frac{\varepsilon}{\rho_o}}\right)\rho_o\gamma+\frac{1}{2\frac{\varepsilon}{\rho_o}}\log\frac{1+\frac{\varepsilon}{\rho_o}}{1-\frac{\varepsilon}{\rho_o}}\,w\right)+\\
&+\frac{\nu_{21}+\nu_{23}\nu_{31}}{\eta}\rho_o
a{\<1}'+(\nu_{32}+\nu_{31}\nu_{12})\rho_o\gamma=0\,.
\end{aligned}
\end{equation}
%
%together with the other two boxed equations \eqref{bill} and
%\eqref{ali}, this equation composes the system of ODEs we should
%solve for $ a{\<1}$, $w$, and $\gamma$.

On eliminating $a{\<1}'$ by means of \eqref{ali}, \eqref{brandi}
yields:
\begin{equation}\label{eqdiffi}
\varepsilon^2 \rho_o^2\left( 1-
\frac{1}{3}\frac{\varepsilon^2}{\rho_o^2}\right)\,w{''''}+\varepsilon^2
a\,w''+b\,w+c\,\rho_o\Delta\,\frac{\varepsilon^{-1}p}{2E_1}+d\,\rho_o\gamma=0,
\end{equation}
where
\begin{equation}\label{ostanti}
\begin{aligned}
&a:=\frac{2(\nu_{21}+\nu_{23}\nu_{31})}{1-\nu_{23}\nu_{32}},\\
&b:=\frac{3}{1-\nu_{23}\nu_{32}}\left(\eta(1-\nu_{13}\nu_{31})\,\frac{1}{2\frac{\varepsilon}{\rho_o}}\log\frac{1+\frac{\varepsilon}{\rho_o}}{1-\frac{\varepsilon}{\rho_o}}-\frac{(\nu_{21}+\nu_{23}\nu_{31})^2}{1-\nu_{23}\nu_{32}}\right),\\
&c:=\frac{3(\nu_{21}+\nu_{23}\nu_{31})}{(1-\nu_{23}\nu_{32})^2},\\
&d:=\frac{3}{1-\nu_{23}\nu_{32}}\Bigg(\eta(1-\nu_{13}\nu_{31})\left(1-\frac{1}{2\frac{\varepsilon}{\rho_o}}\log\frac{1+\frac{\varepsilon}{\rho_o}}{1-\frac{\varepsilon}{\rho_o}} \right)-\\
&\hspace{0.6cm}-\frac{(\nu_{31}+\nu_{21}\nu_{32})(\nu_{21}+\nu_{23}\nu_{31})}{1-\nu_{23}\nu_{32}}+\eta(\nu_{32}+\nu_{31}\nu_{12})\Bigg).
\end{aligned}
\end{equation}
The general solution of the homogeneous equation associated with
\eqref{eqdiffi} has the form:
\begin{equation}\label{w}
w_h(x_1)=c_1\big(\exp(\alpha_1x_1)+\exp(-\alpha_1x_1)\big)+c_2\big(\exp(\alpha_2
x_1)+\exp(-\alpha_2 x_1)\big),
\end{equation}
where
\begin{equation}\label{alpha}
{\alpha_1}^2:=\frac{-a+\sqrt{a^2-4b\frac{\rho_o^2}{\varepsilon^2}\left(
1-
\frac{1}{3}\frac{\varepsilon^2}{\rho_o^2}\right)}}{2\rho_o^2\left(
1- \frac{1}{3}\frac{\varepsilon^2}{\rho_o^2}\right)}, \quad
{\alpha_2}^2:=\frac{-a-\sqrt{a^2-4b\frac{\rho_o^2}{\varepsilon^2}\left(
1-
\frac{1}{3}\frac{\varepsilon^2}{\rho_o^2}\right)}}{2\rho_o^2\left(
1- \frac{1}{3}\frac{\varepsilon^2}{\rho_o^2}\right)}\,.
\end{equation}
With this, we write:
\begin{equation}\label{freccia}
w(x_1)=w_h(x_1)+w_p,\quad w_p:=-\rho_o\left(
\frac{c}{b}\,\Delta\,\frac{\varepsilon^{-1}p}{2E_1}+\frac{d}{b}\,\gamma\right),
\end{equation}
with $w_p$ the constant solution of \eqref{eqdiffi}.

With a view to determining the coefficients $c_1$ and $c_2$, we
firstly return to the boundary condition $\eqref{dircili}_3$,
that, when combined with $\eqref{constM}_1$, reads:
\[
\rho_o w'''(l)-a{\<1}''(l)=0;
\]
in addition, by differentiating \eqref{ali} and invoking
continuity up to the boundary of the resultant expression, we
obtain that %%
%\[
%(1-\nu_{23}\nu_{32})\Big(\rho_o
%a{\<1}^{\prime}-\frac{1}{3}\varepsilon^2
%w^{\prime\prime}\Big)^\prime+(\nu_{21}+\nu_{23}\nu_{31})w^\prime=0,
%\]
%%
%whence
%
\[
\rho_o a{\<1}^{\prime\prime}(l)=\frac{1}{3}\varepsilon^2
w^{\prime\prime\prime}(l)-\frac{\nu_{21}+\nu_{23}\nu_{31}}{1-\nu_{23}\nu_{32}}w^\prime(l);
\]
the last two relations together imply that
\begin{equation}\label{br3}
\left(1-\frac{1}{3}\frac{\varepsilon^2}{\rho_o^2}\right)\rho_ow^{\prime\prime\prime}(l)
+\frac{\nu_{21}+\nu_{23}\nu_{31}}{1-\nu_{23}\nu_{32}}\frac{1}{\rho_o}w^\prime(l)=0.
\end{equation}
Secondly, on eliminating $a{\<{1}}'(l)$ in \eqref{gam} by means of
\eqref{ali}, we obtain:
\begin{equation}\label{br4}
\left(1-\frac{1}{3}\frac{\varepsilon^2}{\rho_o^2}\right)\,\rho_o\,w^{\prime\prime}(l)+
\frac{\nu_{21}+\nu_{23}\nu_{31}}{1-\nu_{23}\nu_{32}}\frac{1}{\rho_o}\,w(l)-\frac{1}{(1-\nu_{23}\nu_{32})}\Delta\,
\frac{\varepsilon^{-1}p}{2E_1}=0.
\end{equation}
On taking \eqref{freccia} into account, the system of equations
\eqref{br3} and \eqref{br4} determines the coefficients $c_\alpha$
in \eqref{w}:

\begin{equation}\label{consttr}
\begin{aligned}
&c_1=\kappa_1\,\frac{\exp(2\alpha_1\,l)-1}{\alpha_1\big(\exp(2\alpha_1\,l)-1\big)\big(\exp(2\alpha_2\,l)+1\big)-\alpha_2\big(\exp(2\alpha_1\,l)+1\big)\big(\exp(2\alpha_2\,l)-1\big)},\\
&c_2=\kappa_2\,\frac{\exp(2\alpha_2\,l)-1}{\alpha_2\big(\exp(2\alpha_2\,l)-1\big)\big(\exp(2\alpha_1\,l)+1\big)-\alpha_1\big(\exp(2\alpha_2\,l)+1\big)\big(\exp(2\alpha_1\,l)-1\big)},\\
\end{aligned}
\end{equation}
with
\begin{equation}
\begin{aligned}
&\kappa_1:=-\frac{\alpha_2\Big(\frac{c}{b}\left(1-\frac{\nu_{21}\nu_{23}\nu_{31}}{1-\nu_{23}\nu_{32}}
\right)\Delta\frac{\varepsilon^{-1}p}{2E_1}+\frac{\nu_{21}+\nu_{23}\nu_{21}}{1-\nu_{23}\nu_{32}}\frac{d}{b}\gamma
 \Big)}{{\alpha_1}^2\rho_o\left(1+\frac{1}{3}\frac{\varepsilon^2}{\rho_o^2}
\right)+\rho_o^{-1}\frac{\nu_{21}+\nu_{23}\nu_{31}}{1-\nu_{23}\nu_{32}}},\\
&\kappa_2:=-\frac{\alpha_1\Big(\frac{c}{b}\left(1-\frac{\nu_{21}\nu_{23}\nu_{31}}{1-\nu_{23}\nu_{32}}
\right)\Delta\frac{\varepsilon^{-1}p}{2E_1}+\frac{\nu_{21}+\nu_{23}\nu_{21}}{1-\nu_{23}\nu_{32}}\frac{d}{b}\gamma
 \Big)}{{\alpha_2}^2\rho_o\left(1+\frac{1}{3}\frac{\varepsilon^2}{\rho_o^2}
\right)+\rho_o^{-1}\frac{\nu_{21}+\nu_{23}\nu_{31}}{1-\nu_{23}\nu_{32}}};
\end{aligned}
\end{equation}
%As anticipated, it is not difficult to convince ourselves that
%$$
%\lim_{l\rightarrow\infty}\,c_1=\lim_{l\rightarrow\infty}\,c_2=0;
%$$
%hence, in particular, \eqref{wappr} holds, as well as
%\eqref{gamm}, for $r_o{\<3}=0$.
note that both the coefficients $c_\alpha$ depend on $\gamma$.

\begin{figure}[!h]
\centering
\includegraphics[scale=0.7]{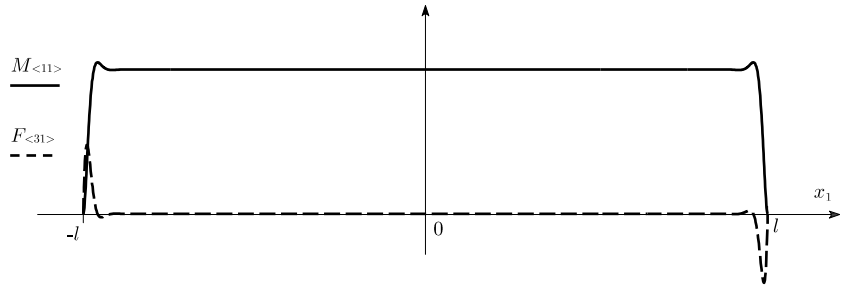}
\caption{Axial Traction Problem: qualitative diagrams of bending
moment $M{\<{11}}$ and shear force $F{\<{31}}$.} \label{d_trac}
\end{figure}

\begin{figure}[!h]
\centering
\includegraphics[scale=0.4]{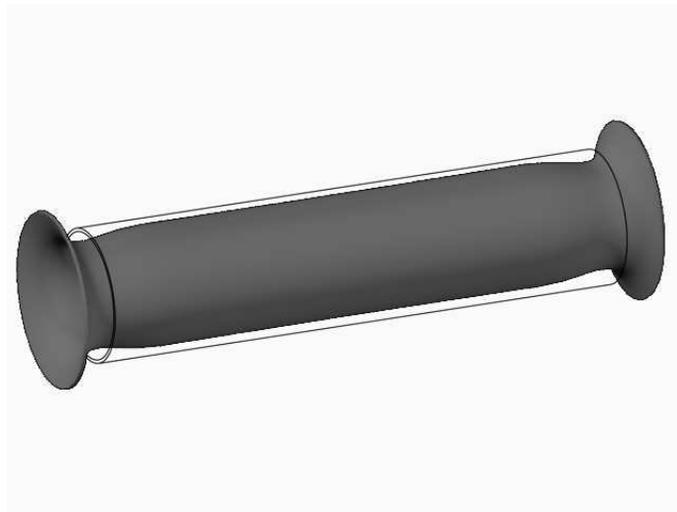}
\caption{Axial Traction Problem: cartoon visualization of deformed
and undeformed shapes.} \label{traction_ghost}
\end{figure}

Having found the form of the radial displacement $w$ in $[-l,+l]$,
we revert to equation \eqref{ali} to find the axial displacement
$a{\<{1}}$. A simply calculation yields:
$$
\begin{aligned}
a{\<{1}}(x_1)=&\left[\left(1+\frac{\nu_{21}+\nu_{23}\nu_{31}}{1-\nu_{23}\nu_{32}}\frac{c}{b}\right)\Delta\frac{\eps^{-1}p}{2E_1}+\right.\\
&\left.+(1-\nu_{23}\nu_{32})^{-1}\left(\frac{d}{b}(\nu_{21}+\nu_{23}\nu_{31})-(\nu_{31}+\nu_{21}\nu_{32})\right)\gamma\right]\,x_1\,.
+\\
&+\frac{1}{3\rho_o\alpha_1\alpha_2}\Bigg[\alpha_2c_1\left(\eps^2{\alpha_1}^2-3\,\frac{\nu_{21}+\nu_{23}\nu_{31}}{1-\nu_{23}\nu_{32}}
\right)\big(\exp(\alpha_1x_1)-\exp(-\alpha_1x_1)\big) + \\
&+\alpha_1c_2\left(\eps^2{\alpha_2}^2-3\,\frac{\nu_{21}+\nu_{23}\nu_{31}}{1-\nu_{23}\nu_{32}}
\right)\big(\exp(\alpha_2x_1)-\exp(-\alpha_2x_1)\big)\Bigg]
\end{aligned}
$$
(needless to say, the parity condition $a{\<{1}}(0)=0$ is
satisfied). The one task remaining is to find the constant
$\gamma$. This we do by a sequence of manipulations of the
integral balance equation \eqref{integ}:

\noindent (i) using $\eqref{constF}_4$ and $\eqref{constM}_3$, we
give equation \eqref{integ} the form:
\begin{equation}\label{pregammat}
\begin{aligned}
\int_{-l}^{+l}&\Bigg((1-\nu_{13}\nu_{31})\!\left(1-\frac{1}{2\frac{\varepsilon}{\rho_o}}\log\frac{1+\frac{\varepsilon}{\rho_o}}{1-\frac{\varepsilon}{\rho_o}}\right)\!(w+\rho_o\gamma)-\frac{1}{3}\varepsilon^2\frac{\nu_{12}+\nu_{23}\nu_{31}}{\eta}w^{\prime\prime}+\\
&+\mu(1-\nu_{12}\nu_{21})\rho_o\gamma+\frac{\nu_{31}+\nu_{21}\nu_{32}}{\eta}\left(\rho_{o}a{\<1}^\prime-\frac{1}{3}\varepsilon^2
w^{\prime\prime}\right)
 +(\nu_{32}-\nu_{31}\nu_{12})w\Bigg)=0;
\end{aligned}
\end{equation}
(ii) on expunging the construct  $(\rho_o
a{\<1}^\prime-\frac{1}{3}\varepsilon^2 w^{\prime\prime})$ by means
of \eqref{ali},  we have:
\[
%\int_{-l}^{+l}\Bigg(\varepsilon^2a_0 w_h^{\prime\prime}+a_1
%(w_h+w_p)-\rho_o\left(\Delta\frac{\varepsilon^{-1} p}{2E_1}
%+a_2\,\gamma\right)\Bigg)=0,\\
%
2\varepsilon^2a_0
\,w_h^{\prime}(l)+a_1\int_{-l}^{+l}w_h=2l\left(\rho_o\left(\Delta\frac{\varepsilon^{-1}
p}{2E_1} +a_2\,\gamma\right)-a_1w_p\right),
\]
where
\[
\begin{aligned}
&a_0:=\frac{1}{3}\, \frac{1-\nu_{23}\nu_{31}}{(\nu_{31}+\nu_{21}\nu_{32})(\nu_{12}+\nu_{23}\nu_{31})},\\
&a_1:=\eta\frac{1-\nu_{23}\nu_{32}}{\nu_{31}+\nu_{21}\nu_{32}}\,\Bigg(\nu_{21}+\nu_{32}\nu_{31}-\nu_{23}+\nu_{31}\nu_{12}-(1-\nu_{13}\nu_{31})\left(1-\frac{1}{2\frac{\varepsilon}{\rho_o}}\log\frac{1+\frac{\varepsilon}{\rho_o}}{1-\frac{\varepsilon}{\rho_o}}  \right)  \Bigg),\\
%&a_2:=\eta\frac{1-\nu_{23}\nu_{32}}{\nu_{31}+\nu_{21}\nu_{32}},\\
&a_2:=
\eta\frac{1-\nu_{23}\nu_{32}}{\nu_{31}+\nu_{21}\nu_{32}}\,\left((1-\nu_{13}\nu_{31})\left(1-\frac{1}{2\frac{\varepsilon}{\rho_o}}\log\frac{1+\frac{\varepsilon}{\rho_o}}{1-\frac{\varepsilon}{\rho_o}}\right)
+\mu(1-\nu_{12}\nu_{21})\right)+\\
&\hspace{0.7cm}-(\nu_{31}+\nu_{21}\nu_{32})\,;
\end{aligned}
\]
(iii) in view of \eqref{freccia}, we end up with:
\begin{equation}\label{gammat}
\begin{aligned}
\left(a_1\frac{d}{b}+a_2\right)\gamma=&-\left(1+c\,\frac{a_1}{b}
\right)\Delta
\frac{\varepsilon^{-1}p}{2E_1}+\\
&+\frac{2}{l\rho_o}\Big({\alpha_1}^{-1}c_1(a_1+\varepsilon^{2}a_o{\alpha_1}^2)\sinh(\alpha_1l)+
{\alpha_2}^{-1}c_2(a_1+\varepsilon^{2}a_o{\alpha_2}^2)\sinh(\alpha_2l)
\Big),
\end{aligned}
\end{equation}
an implicit equation for $\gamma$. Figures
\ref{d_trac}-\ref{traction_ghost} help visualizing some relevant
features of the analytic solution we just constructed.

\begin{figure}[!h]
\centering
\includegraphics[scale=0.7]{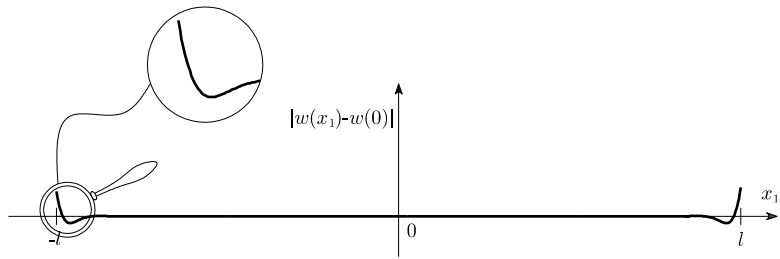}
\caption{Axial Traction Problem: radial displacement $w$.}
\label{w_trac}
\end{figure}

\newpage
\subsection{Pressure}\label{up}
We now let  the cylindrical shell we study be subject to a uniform
pressure $q_o{\<{3}}=\varpi=O(\varepsilon)$, all the other applied
loads being null. Accordingly, the field and boundary equations
\eqref{ecco} and \eqref{dircili} are replaced by, respectively,
\begin{equation}\label{eqpr}
\begin{aligned}
&F{\<{11}}'=0,\\
&M{\<{11}}''-\rho_o^{-1}F{\<{22}}+\varpi=0;
\end{aligned}
\end{equation}
and
\begin{equation}\label{bcpr}
F{\<{11}}(\pm l)=0, \quad M{\<{11}}(\pm l)=0,\quad
M{\<{11}}^\prime(\pm l)=0.
\end{equation}
Equations $\eqref{eqpr}_1$ and $\eqref{bcpr}_1$ imply that
$F{\<{11}}$ is identically null, whence, in view of
$\eqref{constF}_1$,  that
\begin{equation}\label{prali}
\begin{aligned}
\rho_o a{\<1}^\prime-\frac{1}{3}\varepsilon^2 w^{\prime\prime}=-
\frac{\nu_{21}+\nu_{23}\nu_{31}}{1-\nu_{23}\nu_{32}}w-\frac{\nu_{31}+\nu_{21}\nu_{32}}{1-\nu_{23}\nu_{32}}\rho_o\gamma\quad\textrm{over}\;\,[-l,+l].
\end{aligned}
\end{equation}
With this, $\eqref{constM}_1$  and $\eqref{constF}_2$, equation
$\eqref{eqpr}_2$ takes the form:
\begin{equation}\label{preqdiff}
\varepsilon^2 \rho_o^2\left( 1-
\frac{1}{3}\frac{\varepsilon^2}{\rho_o^2}\right)\,w{''''}+\varepsilon^2
a\,w''+b\,w-\tilde{c}\rho_o^2\Delta\frac{\varepsilon^{-1}\varpi}{2E_1}+d\,\rho_o\gamma=0,
\end{equation}
where
$$
\tilde{c}:=\frac{3}{1-\nu_{23}\nu_{32}},
$$
the other constants being defined by $\eqref{ostanti}_{1,2,4}$.
The general solution of the homogeneous equation associated to
\eqref{preqdiff} has the form \eqref{w}:
\begin{equation}\label{prw}
w_h(x_1)=\tilde
c_1\big(\exp(\alpha_1x_1)+\exp(-\alpha_1x_1)\big)+\tilde
c_2\big(\exp(\alpha_2 x_1)+\exp(-\alpha_2 x_1)\big),
\end{equation}
with the coefficients $\alpha_\delta$ given by \eqref{alpha};
then,
\begin{equation}\label{prfreccia}
w(x_1)=w_h(x_1)+w_p,\quad w_p:=\rho_o\left( \frac{\tilde
c}{b}\,\rho_o\Delta\,\frac{\varepsilon^{-1}\varpi}{2E_1}-\frac{d}{b}\,\gamma\right),
\end{equation}
with $w_p$ the constant solution of \eqref{preqdiff}. The boundary
conditions $\eqref{bcpr}_{2,3}$ become:
\begin{equation}\label{prbr4}
\left(1-\frac{1}{3}\frac{\varepsilon^2}{\rho_o^2}\right)\,\rho_o\,w^{\prime\prime}(l)+
\frac{\nu_{21}+\nu_{23}\nu_{31}}{1-\nu_{23}\nu_{32}}\frac{1}{\rho_o}\,w(l)=0,
\end{equation}
\begin{equation}\label{prbr3}
\left(1-\frac{1}{3}\frac{\varepsilon^2}{\rho_o^2}\right)\rho_ow^{\prime\prime\prime}(l)
+\frac{\nu_{21}+\nu_{23}\nu_{31}}{1-\nu_{23}\nu_{32}}\frac{1}{\rho_o}w^\prime(l)=0,
\end{equation}
and are expedient to determine the constants $\tilde c_1$ and
$\tilde c_2$:
\begin{equation}\label{cbarra}
\begin{aligned}
&\tilde c_1=\tilde\kappa_1\,\frac{\exp(2\alpha_1\,l)-1}{\alpha_1\big(\exp(2\alpha_1\,l)-1\big)\big(\exp(2\alpha_2\,l)+1\big)-\alpha_2\big(\exp(2\alpha_1\,l)+1\big)\big(\exp(2\alpha_2\,l)-1\big)},\\
&\tilde c_2=\tilde\kappa_2\,\frac{\exp(2\alpha_2\,l)-1}{\alpha_2\big(\exp(2\alpha_2\,l)-1\big)\big(\exp(2\alpha_1\,l)+1\big)-\alpha_1\big(\exp(2\alpha_2\,l)+1\big)\big(\exp(2\alpha_1\,l)-1\big)},\\
\end{aligned}
\end{equation}
with
\begin{equation}
\begin{aligned}
&\tilde\kappa_1:=\frac{\alpha_2\Big(\frac{\tilde
c}{b}\left(1-\frac{\nu_{21}\nu_{23}\nu_{31}}{1-\nu_{23}\nu_{32}}
\right)\rho_o\Delta\frac{\varepsilon^{-1}\varpi}{2E_1}-\frac{\nu_{21}+\nu_{23}\nu_{21}}{1-\nu_{23}\nu_{32}}\frac{d}{b}\gamma
 \Big)}{{\alpha_1}^2\rho_o\left(1+\frac{1}{3}\frac{\varepsilon^2}{\rho_o^2}
\right)+\rho_o^{-1}\frac{\nu_{21}+\nu_{23}\nu_{31}}{1-\nu_{23}\nu_{32}}},\\
&\tilde\kappa_2:=\frac{\alpha_1\Big(\frac{\tilde
c}{b}\left(1-\frac{\nu_{21}\nu_{23}\nu_{31}}{1-\nu_{23}\nu_{32}}
\right)\rho_o\Delta\frac{\varepsilon^{-1}\varpi}{2E_1}-\frac{\nu_{21}+\nu_{23}\nu_{21}}{1-\nu_{23}\nu_{32}}\frac{d}{b}\gamma
 \Big)}{{\alpha_2}^2\rho_o\left(1+\frac{1}{3}\frac{\varepsilon^2}{\rho_o^2}
\right)+\rho_o^{-1}\frac{\nu_{21}+\nu_{23}\nu_{31}}{1-\nu_{23}\nu_{32}}}.
\end{aligned}
\end{equation}
Finally, by a sequence of steps completely analogous to the one
leading to \eqref{gammat}, the integral balance \eqref{integ}
yields
%
%\begin{equation}\label{prorsu}
%\int_{-l}^{+l}\Bigg(\varepsilon^2a_o w_h^{\prime\prime}+a_1
%(w_h+w_p)-a_2\rho_o\gamma\Bigg) dx_1=0,
%\end{equation}
%which yields
an implicit equation for the constant $\gamma$:
\begin{equation}\label{gammap}
\begin{aligned}
\left(a_1\frac{d}{b}+a_2\right)\gamma=&a_1\frac{\tilde c}{b}\rho_o\Delta\frac{\eps^{-1}\varpi}{2E_1}+  \\
&+\frac{2}{l\rho_o}\Big({\alpha_1}^{-1}c_1(a_1+\varepsilon^{2}a_o{\alpha_1}^2)\sinh(\alpha_1l)+
{\alpha_2}^{-1}c_2(a_1+\varepsilon^{2}a_o{\alpha_2}^2)\sinh(\alpha_2l)
\Big).
\end{aligned}
\end{equation}
\begin{figure}[!h]
\centering
\includegraphics[scale=0.7]{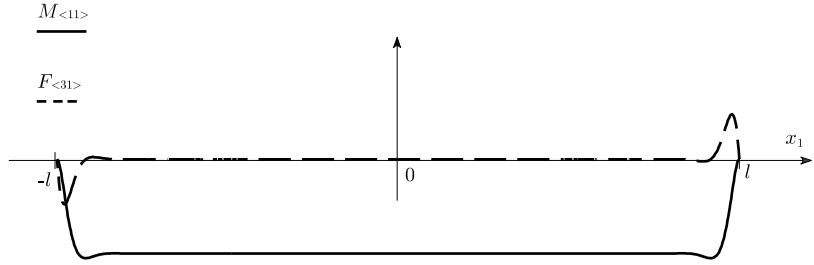}
\caption{Pressure Problem: qualitative diagrams of bending moment
$M{\<{11}}$ and shear force $F{\<{31}}$.} \label{w_trac}
\end{figure}

\begin{figure}[h]
\centering
\includegraphics[scale=0.7]{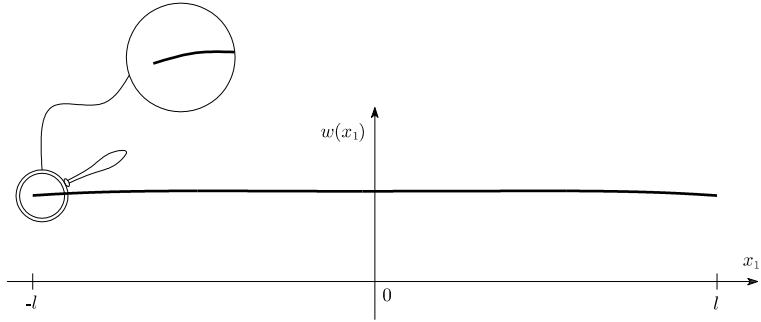}
\caption{Pressure Problem: radial displacement $w$.}
\label{w_pres}
\end{figure}

\subsection{Rim flexure}
Lastly, we consider the case when a uniform distribution of
bending couples $m=O(\varepsilon)$ per unit length is applied at
both rims of the cylinder. This time, the field equations we have
to satisfy are:
\begin{equation}\label{recco}
\begin{aligned}
&F{\<{11}}'=0,\\
&M{\<{11}}''-\rho_{o}^{-1}F{\<{22}}=0,\\
\end{aligned}
\end{equation}
with the boundary conditions:
\begin{equation}\label{rdircili}
F{\<{11}}(\pm l)=0,\quad M{\<{11}}(\pm l)=m,\quad
M^\prime{\<{11}}(\pm l)=0.
\end{equation}
Just as in the pressure case, equations $\eqref{recco}_1$ and
$\eqref{rdircili}_1$ imply that \eqref{prali} holds; moreover,
%\begin{equation}\label{rali}
%\begin{aligned}
%(1-\nu_{23}\nu_{32})\Big(\rho_o
%a{\<1}^\prime-\frac{1}{3}\varepsilon^2
%w^{\prime\prime}\Big)+(\nu_{21}+\nu_{23}\nu_{31})w+(\nu_{31}+\nu_{21}\nu_{32})\rho_o\gamma=0.
%\end{aligned}
%\end{equation}
with the use of $\eqref{constM}_1$, $\eqref{constF}_2$, and
\eqref{prali}, equation $\eqref{recco}_2$ takes the form:
\begin{equation}\label{reqdiffi}
\varepsilon^2 \rho_o^2\left( 1-
\frac{1}{3}\frac{\varepsilon^2}{\rho_o^2}\right)\,w{''''}+\varepsilon^2
a\,w''+b\,w+d\,\rho_o\gamma=0
\end{equation}
(cf. \eqref{preqdiff}), where the constants are defined in
\eqref{ostanti}. We set:
\[
w(x_1)= -\,\frac{d}{b}\rho_o\gamma+w_h(x_1),
\]
with
\begin{equation}\label{rfw}
w_h(x_1)=\hat
c_1\big(\exp(\alpha_1x_1)+\exp(-\alpha_1x_1)\big)+\hat
c_2\big(\exp(\alpha_2 x_1)+\exp(-\alpha_2 x_1)\big),
\end{equation}
the constants $\alpha_1,\alpha_2$ being given by \eqref{alpha}. On
writing the boundary conditions $\eqref{rdircili}_{2,3}$ in terms
of displacements:
\begin{equation}\label{rbr4}
\left(1-\frac{1}{3}\frac{\varepsilon^2}{\rho_o^2}\right)\,\rho_o\,w^{\prime\prime}(l)+
\frac{\nu_{21}+\nu_{23}\nu_{31}}{1-\nu_{23}\nu_{32}}\frac{1}{\rho_o}\,w(l)=-\frac{3}{2}\,\Delta\,
\frac{m\rho_o}{\varepsilon^3 E_1},
\end{equation}
\begin{equation}\label{rbr3}
\left(1-\frac{1}{3}\frac{\varepsilon^2}{\rho_o^2}\right)\rho_ow^{\prime\prime\prime}(l)
+\frac{\nu_{21}+\nu_{23}\nu_{31}}{1-\nu_{23}\nu_{32}}\frac{1}{\rho_o}w^\prime(l)=0,
\end{equation}
we determine the constants $\hat c_1$ and $\hat c_2$:
$$
\hat c_1=\hat k_1^{-1}\left(\frac{3}{2}\,\Delta\,
\frac{m\rho_o}{\varepsilon^3
E_1}-\frac{d}{b}\gamma\right)\,\frac{\alpha_2\big(\exp(2\alpha_2l)-1)
\big)\,\exp(\alpha_1l)}{\alpha_1\big(\exp(2\alpha_2l)+1
\big)\big(\exp(2\alpha_1l)-1\big)-\alpha_2\big(\exp(2\alpha_1l)+1
\big)\big(\exp(2\alpha_2l)-1\big)},
$$
$$
\hat c_2=\hat k_2^{-1}\left(\frac{3}{2}\,\Delta\,
\frac{m\rho_o}{\varepsilon^3
E_1}-\frac{d}{b}\gamma\right)\,\frac{\alpha_1\big(\exp(2\alpha_1l)-1)
\big)\,\exp(\alpha_2l)}{\alpha_2\big(\exp(2\alpha_1l)+1
\big)\big(\exp(2\alpha_2l)-1\big)-\alpha_1\big(\exp(2\alpha_2l)+1
\big)\big(\exp(2\alpha_1l)-1\big)},
$$
where
$$
\begin{aligned}
&\hat
k_1=\left(1-\frac{1}{3}\frac{\varepsilon^2}{\rho_o^2}\right)\rho_o{\alpha_1}^2+\rho_o^{-1}\,\frac{\nu_{21}+\nu_{23}\nu_{31}}{1-\nu_{23}\nu_{32}},\\
&\hat
k_2=\left(1-\frac{1}{3}\frac{\varepsilon^2}{\rho_o^2}\right)\rho_o{\alpha_2}^2+\rho_o^{-1}\,\frac{\nu_{21}+\nu_{23}\nu_{31}}{1-\nu_{23}\nu_{32}}.\\
\end{aligned}
$$
Once again, the integral condition balance \eqref{integ} yields an
implicit equation for $\gamma$:
\begin{equation}\label{gammarf}
\begin{aligned}
\left(a_1\frac{d}{b}+a_2\right)\gamma=\frac{2}{l\rho_o}\Big({\alpha_1}^{-1}c_1(a_1+\varepsilon^{2}a_o{\alpha_1}^2)\sinh(\alpha_1l)+
{\alpha_2}^{-1}c_2(a_1+\varepsilon^{2}a_o{\alpha_2}^2)\sinh(\alpha_2l)
\Big).
\end{aligned}
\end{equation}
\vfill \pagebreak
\begin{figure}[!h]
\centering
\includegraphics[scale=0.7]{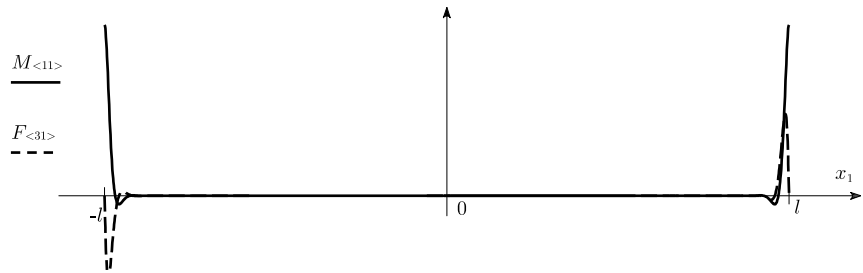}
\caption{Rim Flexure Problem: qualitative diagrams of bending
moment $M{\<{11}}$ and shear force $F{\<{31}}$.} \label{w_flex}
\end{figure}

\begin{figure}[!h]
\centering
\includegraphics[scale=0.7]{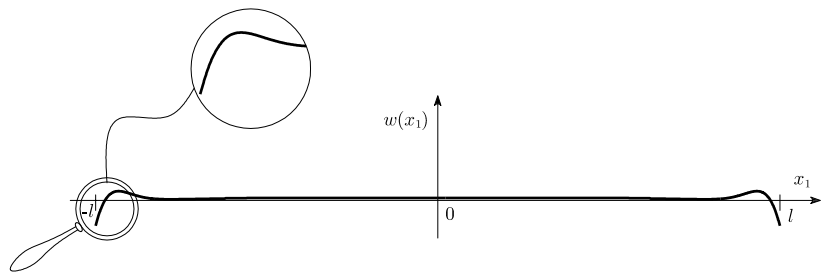}
\caption{Rim Flexure Problem: radial displacement $w$.}
\label{w_flex}
\end{figure}

\begin{figure}[!h]
\centering
\includegraphics[scale=0.4]{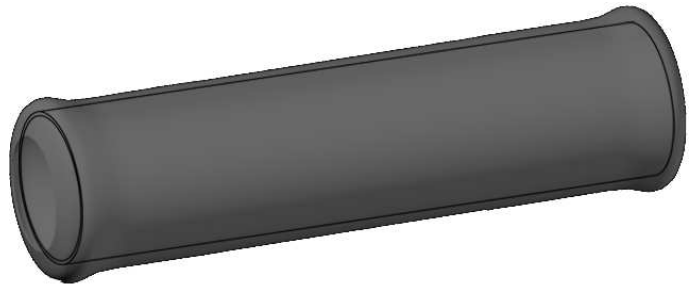}
\caption{Rim Flexure Problem: cartoon visualization of deformed
and undeformed shapes.} \label{flexure}
\end{figure}

\vfill \pagebreak

\section{Cylindrical Shells: Thinness, Slenderness, Contraction Moduli, and Stiffnesses}\label{stif}
As all figures from \eqref{d_trac} to \eqref{flexure} make
evident, a phenomenon of boundary localization takes place,
whatever the shell's thickness, in the boundary-value problems of
traction, pressure and rim-flexure we have solved analytically.
This phenomenon is more and more pronounced as the shell's length
grows ceteris paribus. In our opinion, this fact amply justifies
the use of the much simpler formulas for slender shells that we
derive in this section.

\subsection{Thin shells and slender shells}
%
%As anticipated, once axial twisting is accounted for, we are left
%with a system of three partial differential equations for  the
%fields ${a\<1}$, $w$ and the constant $\gamma$; in the next
%subsection, we shall try and solve this system in the cases of
%pressure, uniform rim flexure and axial traction. It will be
%apparent that noticeable simplifications follow when the
%approximations judged appropriate for thin and slender shells are
%introduced.

We term \emph{thin} a cylindrical shell of diameter
$2(\rho_o+\varepsilon)$ and length $2l$  if $\varepsilon/\rho_o\ll
1$, \emph{slender} if $\rho_o/l\ll 1$; most of times, $\rho_o$ is
indeed smaller than $l$, so that a thin shell is slender as well
(and $\varepsilon/l\ll 1$). For $\rho_o/l$ fixed, a thinner and
thinner shell looses its bending and twisting stiffness more and
more; in the limit for $\varepsilon/\rho_o\rightarrow 0$, it
responds to loads as a tubular membrane.  On the other hand, for
both $\varepsilon$ and $\rho_o$ fixed, longer and longer shells
become slender and slender, without loosing their shell-like
response.

To exemplify the simplifications ensuing from taking
large-thinness limits, we take the integral balance \eqref{integ},
that has the same displacement form \eqref{pregammat} in all three
boundary-value problems of traction, pressure, and rim flexure.
Since
\begin{equation}\label{approx}
\lim_{\varepsilon/\rho_o\rightarrow
0}\frac{1}{2\frac{\varepsilon}{\rho_o}}\log\frac{1+\frac{\varepsilon}{\rho_o}}{1-\frac{\varepsilon}{\rho_o}}=1,
\end{equation}
we see that \eqref{pregammat} reduces to
\[%\label{postgammat}
\mu(1-\nu_{12}\nu_{21})\rho_o\gamma+\eta^{-1}(\nu_{31}+\nu_{21}\nu_{32})\frac{\rho_{o}}{l}a{\<1}(l)
 +(\nu_{32}-\nu_{31}\nu_{12})\frac{1}{2l}\int_{-l}^{+l}w=0,
\]
where of course the values of both $a{\<1}(l)$ and
$\int_{-l}^{+l}w$ are problem-dependent.
% note that the second addendum vanishes in the large-slenderness limit.
\vskip 6pt \remark CNTs, no matter if single- or multi-wall, are
as a rule slender. However, they may be thin or not, in the sense
of the above definition \cite{G}. Of course, all shell theories
concern thin objects, but the thinness notions they are
constructed upon may differ (see the discussions in \cite{PPGB}
and \cite{PPGU}); in particular, those notions need not be
expressed in terms on one purely geometrical aspect ratio. Our
shell theory works whatever that ratio, because its subtler and
more complex notion of thinness is the one typical of the
\emph{method of internal constraints}, a method to derive the
mathematical models of linear structure mechanics firstly sketched
in \cite{PPG69}. \vskip 4pt
\subsection{Axisymmetric equilibria of slender shells}
Hereafter, we display the slenderness approximations of the
solutions to the fundamental problems analyzed in the previous
section, except for the torsion problem, where no such
approximation is in order, because the solution does not depend on
the cylinder's length.

\begin{enumerate}[(i)]
\item {\underline{Traction}}. Looking at \eqref{consttr}, it easy to
conclude that
$$
\lim_{l\rightarrow\infty}c_1=\lim_{l\rightarrow\infty}c_2=0;
$$
then, \eqref{w} takes the form:
$$
w_h(x_1)\equiv 0.
$$
Moreover, \eqref{gammat} yields that:
\begin{equation}\label{gamm}
\gamma=-\bar\gamma\,\frac{\varepsilon^{-1}p}{2E_1}\,,\quad
\bar\gamma:=\frac {1+a_1
\frac{c}{b}}{a_2+a_1\,\frac{d}{b}}\,\Delta\,.
\end{equation}
With this, one finds that
\begin{equation}\label{wok}
w(x_1)\equiv -\bar w_p\,\frac{\varepsilon^{-1}p}{2E_1},\quad \bar
w_p:=\rho_o\left(\frac{c}{b}+\frac {1+a_1
\frac{c}{b}}{a_2+a_1\,\frac{d}{b}}\right)\Delta\,.
\end{equation}
To arrive to a large-slenderness approximation for the function
$x_1\mapsto a{\<1}(x_1)$, the remaining unknown of our problem, we
turn to  \eqref{ali}. Under the present circumstances, that
equation yields the value of the axial strain in a slender shell:
\begin{equation}\label{asstir}
\Lambda:=a{\<1}^\prime(x_1)\equiv \frac{\Delta+\rho_o^{-1} \bar
w_p(\nu_{21}+\nu_{23}\nu_{31})+\bar\gamma(\nu_{31}+\nu_{21}\nu_{32})}{1-\nu_{23}\nu_{32}}\,
\frac{\varepsilon^{-1}p}{2E_1}.
\end{equation}
\item \underline{Pressure}. This time, we look at \eqref{cbarra} to conclude that, in limit for $l\rightarrow \infty$, both constants $\tilde
c_\alpha$ tend to zero. Thus, once again, $w(x_1)\equiv w_p$, and
we find that
\begin{equation}\label{gammm}
\gamma=\tilde\gamma\, \frac{\rho_o\eps^{-1}\varpi}{2E_1} \,,
\qquad \tilde\gamma:=\frac{a_1\frac{\tilde
c}{b}}{a_2+a_1\frac{d}{b}}\,\Delta\, ,
\end{equation}
and
\begin{equation}\label{wokk}
w(x_1)\equiv \tilde w_p\, \frac{\rho_o\eps^{-1}\varpi}{2E_1} ,
\qquad \tilde w_p:=\rho_o\,\frac{a_2\frac{\tilde
c}{b}}{a_2+a_1\frac{d}{b}}\,\Delta\,.
\end{equation}
Finally, a use of \eqref{prali} yields:
\begin{equation}
a{\<{1}}'(x_1)\equiv-\frac{ \rho_o^{-1}\tilde w_p
(\nu_{21}+\nu_{23}\nu_{31})+\tilde \gamma
(\nu_{31}\nu_{21}\nu_{32})}{1-\nu_{23}\nu_{23}}\,\frac{\rho_o\eps^{-1}\varpi}{2E_1}\,.
\end{equation}
\item \underline{Rim Flexure}. Given that $\;\displaystyle \lim_{l\rightarrow\infty}\hat c_1=\lim_{l\rightarrow\infty}\hat c_2=0\,$, it easy to show that the displacement field is everywhere null, in the
large-slenderness limit.

\end{enumerate}

\subsection{Effective contraction moduli}
Let us now regard a slender cylindrical shell as a
three-dimensional rod-like body, in short, a \emph{probe}. On
defining the \emph{cross-section strain measure}:
\begin{equation}\label{sects}
\overline{\Eb}(x_1):=\frac{1}{2\varepsilon}\int_{-\varepsilon}^{+\varepsilon}
\alpha(\zeta)\, \Eb(x_1,\zeta),
\end{equation}
we have from \eqref{strainax} that
\begin{equation}\label{compE}
\overline{E}{\<{11}}=a{\<{1}}',\quad\overline{E}{\<{12}}=\overline{E}{\<{21}}=\frac{1}{2}\,a{\<{2}}',\quad\overline{E}{\<{22}}=\rho_o^{-1}w,\quad\overline{E}{\<{33}}=\gamma.
\end{equation}
Moreover, $\eqref{cilcil}_3$ implies that the deformed external
radius of a shell of undeformed external radius
$r_0^{ext}=\rho_0+\varepsilon$ is
$r^{ext}=\rho_0+w+\varepsilon\gamma$, so that
\[
\Gamma:=\frac{r^{ext}-r^{ext}_0}{r^{ext}_0}=\overline{E}{\<{22}}+O(\varepsilon).
\]
Motivated by this observation, by an \emph{effective contraction
modulus} we mean:
%:
\begin{itemize}
\item in case of axial traction,
\begin{equation}\label{nuca}
\nu_{CA}:=-\frac{{\overline E}{\<{22}}}{{\overline
E}{\<{11}}}=\frac{\rho_o^{-1}\bar
w_p(1-\nu_{23}\nu_{32})}{\Delta+\rho_o^{-1}\bar
w_p(\nu_{21}+\nu_{23}\nu_{31})+\bar\gamma(\nu_{31}+\nu_{21}\nu_{32})
}\,,
\end{equation}
where $\bar w_p$ and $\bar \gamma$ are given by, respectively,
\eqref{wok} and \eqref{gamm} (note that this definition is
directly reminiscent of the laboratory procedure followed to
measure  Poisson's modulus for isotropic materials);
\item in case of uniform inner pressure,
\begin{equation}\label{nucipi}
\nu_{CP}:=-\frac{{\overline E}{\<{11}}}{{\overline
E}{\<{22}}}=\frac{\rho_o^{-1}\tilde
w_p(\nu_{21}+\nu_{23}\nu_{31})+\tilde\gamma(\nu_{31}+\nu_{21}\nu_{32})}{\rho_o^{-1}\tilde
w_p(1-\nu_{23}\nu_{32})}\,,
\end{equation}
where $\tilde w_p$ and $\tilde \gamma$ are given by, respectively,
\eqref{wokk} and \eqref{gammm}.
\end{itemize}
Both $\nu_{CA}$ and $\nu_{CP}$ depend on thickness, through,
respectively, $\bar w_p,\bar\gamma$ and $\tilde w_p,\tilde\gamma$.

\subsection{Effective  stiffnesses}
It remains for us to introduce suitable notions of effective
traction and torsion stiffnesses for a slender cylindrical shell
regarded as a probe. This we do by mimicking the relative familiar
formulas from one-dimensional rod theory.

As to  \emph{effective traction stiffness}, given that the
traction stiffness of a rod is defined to be (axial load)/(axial
strain), we set:
\begin{equation}\label{rigtra}
s_{A}:=\frac{P}{\Lambda},
\end{equation}
whence
\begin{equation}\label{rigtraz}
s_{A}=\frac{(1-\nu_{23}\nu_{32})}{\Delta+\rho_o^{-1} \bar
w_p(\nu_{21}+\nu_{23}\nu_{31})+\bar\gamma(\nu_{31}+\nu_{21}\nu_{32})
}\,E_1 A(\varepsilon)\,,\quad
A(\varepsilon):=4\pi\rho_o\varepsilon,
\end{equation}
where we have made use of \eqref{axtraz},  \eqref{asstir}, and
$\eqref{compE}_1$,  and where $A(\varepsilon)$ is the area of the
shell's cross-section (needless to say, $s_A$ depends on
$\varepsilon$ also through $\bar w_p$ and $\bar\gamma$).

As to  \emph{effective torsion stiffnesses}, we recall that, for a
twisted rod, one takes it to be (torsion moment)/(twist per unit
length); accordingly, we set:
\begin{equation}\label{rigtors1}
s_{T}:=\frac{T}{\Theta},
\end{equation}
so that, on recalling \eqref{tigrande}-\eqref{rotu}, we have that
\begin{equation}\label{rigtors}
s_T=\frac{GJ(\varepsilon)}{\chi(\varepsilon)},\quad
J(\varepsilon):=4\pi\rho_o^3\varepsilon,\quad
\chi(\varepsilon):=\Big(1+\frac{\varepsilon^2}{\rho_o^2}\Big)^{-1},
\end{equation}
where $J(\varepsilon)$ approximates to within $O(\varepsilon^2)$
terms the polar inertia moment of the cross section, and where
$\chi(\varepsilon)= 1+O(\varepsilon^2)$ is a sort of torsion
factor.\footnote{Note that \eqref{rigtors} holds whatever the
slenderness of the shell under consideration.}
%
%
%\vfill
%\pagebreak

%
%\subsection{The effective thickness of carbon nanotubes}
%

%
 \section{Cylindrical Kirchhoff--Love Shells}\label{kls}
 Recall from Section \ref{kin} the Kirchhoff-Love representation \eqref{kl} for the displacement field:
\[
{\ub}_{K\!L}(x,\zeta)=\Ash(x,\zeta){\ab}(x)+w(x)\nb(x)-\zeta\nablas
w(x),\quad \ab(x)\cdot\nb(x)=0,
\]
and compare it with the more general representation
\eqref{KLdispl}: in both cases the unshearability constraint is
imposed implicitly, together with, but only in the first case,
inextensibility of the fibers orthogonal to the middle surface.
Thus, the thickness of these shells does not change whatever the
applied loads, making the simpler Kirchhoff-Love theory suitable,
in our opinion, for application to single-wall CNTs.

Formally, to recover the representation \eqref{kl}, one only has
to take $\gamma=0$ in \eqref{KLdispl}. To adjourn the developments
of Sections 3, 4, and 5, this measure has to be accompanied by a
few other adjustments that we now categorize and  detail.
\begin{enumerate}[(i)]
\item (Balance Assumptions) The variations entering the Principle of Virtual Powers must have the following form, to be compared with \eqref{vars}:
\begin{equation}\label{varsKL}
\begin{aligned}
&\overset{(0)}\vb=v_1\cb_1+v_2\nb'+v_3\nb,\\
&\overset{(1)}\vb=-v_3,_1\cb_1+\rho_o^{-1}(v_2-v_3,_2)\nb',
\end{aligned}
\end{equation}
Consequently, the integral balance equations $\eqref{equicil}_4$
disappears, and the same happens with the last power-conjugate
pair in each of the boundary conditions \eqref{conto1} and
\eqref{conto2}.

\item (Constitutive Assumptions) The constraint space \eqref{constrsp} reduces to
\[
\mathscr{D}=\mathrm{span}(\Wb_i,\, i=1,2,3);
\]
$S{\<{33}}$ is then reactive and only four constitutive moduli
survive. It is not difficult to show that \eqref{consta} and
\eqref{constb} must be replaced, respectively, by
\begin{equation}\label{constakl}
\begin{aligned}
\widetilde\Co&=\Delta^{-1}\Big(E_1\Wb_1\otimes\Wb_1+
E_2\Wb_2\otimes\Wb_2+2\Delta\, G\,
\Wb_3\otimes\Wb_3\\
&+E_1\nu_{21}(\Wb_1\otimes\Wb_2+\Wb_2\otimes\Wb_1)\Big),
\end{aligned}
\end{equation}
and
\begin{equation}\label{constbb}
\Delta:=1-\nu_{12}\,\nu_{21}.
\end{equation}
This result is achieved by the first of various applications to
follow of a procedure  that  consists in taking the limits for
$E_3$, $\lambda$, and $\mu$, tending to infinity, and in setting
$\nu_{\alpha 3}=\nu_{3\alpha}=0\;(\alpha=1,2)$. It is important to
realize that, due to the first of \eqref{moreconst} that we here
repeat for the reader's convenience:
\begin{equation}\label{591}
\frac{E_1}{E_2}=\frac{\nu_{12}}{\nu_{21}},
\end{equation}
\eqref{constakl} and \eqref{constbb} integrate a 4-parameter
constitutive representation.
\item (Torsion Problem) The solution given  in Section \ref{torsio} does not change, because it does not involve any of the constitutive moduli that take singular values in the case of Kirchhoff-Love shells.

\item (Traction Problem)
The two governing equations are, mutatis mutandis, \eqref{ali} and
\eqref{eqdiffi}; specifically, the equation corresponding to
\eqref{ali} is:
\begin{equation}\label{a1}
\begin{aligned}
\rho_o a{\<1}^\prime-\frac{1}{3}\varepsilon^2 w^{\prime\prime}+
\nu_{21}\, w=\rho_o\Delta\, \frac{\varepsilon^{-1}p}{2E_1}\,,
\end{aligned}
\end{equation}
while the one corresponding to \eqref{eqdiffi} reads:
\begin{equation}
\varepsilon^2 \rho_o^2\left( 1-
\frac{1}{3}\frac{\varepsilon^2}{\rho_o^2}\right)\,w{''''}+\varepsilon^2
a\,w''+b\,w+c\,\rho_o\Delta\,\frac{\varepsilon^{-1}p}{2E_1}=0,
\end{equation}
where the constants have the following expressions, that can be
recovered from \eqref{ostanti}:
\begin{equation}\label{klcos}
a:=2\,\nu_{21},\quad
b:=3\eta\left(\frac{1}{2\frac{\varepsilon}{\rho_o}}\log\frac{1+\frac{\varepsilon}{\rho_o}}{1-\frac{\varepsilon}{\rho_o}}-\nu_{12}\nu_{21}\right),\quad
c:=3\,\nu_{21}.
\end{equation}
 As to equation \eqref{freccia}, it is replaced by
\begin{equation}\label{w_tracKL}
w(x_1)=w_h(x_1)+w_{pA},\quad
w_{pA}=-\rho_o\,\nu_{12}\,\delta(\varepsilon)\,\frac{\varepsilon
^{-1}p}{2E_1}\,,
\end{equation}
where
\begin{equation}
 \delta(\varepsilon):=\frac{(1-\nu_{12}\nu_{21})}{\frac{1}{2\frac{\varepsilon}{\rho_o}}\log\frac{1+\frac{\varepsilon}{\rho_o}}{1-\frac{\varepsilon}{\rho_o}}-\nu_{12}\nu_{21}}\quad (\textrm{note for later use that}\;\,\lim_{\varepsilon/\rho_o\rightarrow 0}\delta(\varepsilon)=1).
\end{equation}
%
%note for later use that $\displaystyle\lim_{\varepsilon/\rho_o\rightarrow 0}\delta(\varepsilon)=1$.
For slender shells, $w(x_1)=w_p$ and \eqref{a1} allows to conclude
that
\begin{equation}\label{a_tracKL}
a{\<{1}}(x_1)= \big(1-\nu_{12}\nu_{21}\left(1-
\delta(\varepsilon)\right)\big)\frac{\varepsilon^{-1}p}{2E_1}\,x_1.
\end{equation}

\item (Pressure Problem) The two relevant equations,
\eqref{prali} and \eqref{preqdiff}, become, respectively:
\begin{equation}\label{pra1}
\begin{aligned}
\rho_o a{\<1}^\prime-\frac{1}{3}\varepsilon^2
w^{\prime\prime}=-\nu_{21}w,
\end{aligned}
\end{equation}
and
\begin{equation}
\varepsilon^2 \rho_o^2\left( 1-
\frac{1}{3}\frac{\varepsilon^2}{\rho_o^2}\right)\,w{''''}+\varepsilon^2
a\,w''+b\,w-3\rho_o^2\Delta\frac{\varepsilon^{-1}\varpi}{2E_1}=0,
\end{equation}
where the constants $a$ and $b$ are the same as in equation
\eqref{klcos}. Equation \eqref{prfreccia} is replaced by
\begin{equation}\label{w_presKL}
w(x_1)=w_h(x_1)+w_{pP}, \quad w_{pP}=
\rho_o^2\,\delta(\varepsilon) \frac{\varepsilon^{-1}\varpi}{2E_2}.
\end{equation}
By using \eqref{pra1}, we conclude that, for slender shells,
\begin{equation}\label{a_presKL}
a{\<{1}}=-\nu_{21}\,\delta(\varepsilon)\,\frac{\rho_o\eps^{-1}\varpi}{2E_2}\,x_1.
\end{equation}
\item (Effective Contraction Moduli. Effective Stiffnesses)
As to contraction moduli, formulas \eqref{nuca} and \eqref{nucipi}
become:
\begin{equation}\label{nusimp}
\nu_{CA}=\frac{\delta(\varepsilon)}{1-\nu_{12}\nu_{21}\left(1-
 \delta(\varepsilon)\right)}\,\nu_{12}\quad\textrm{and}\quad \nu_{CP}=\nu_{21}.
\end{equation}
As to traction stiffness, \eqref{rigtraz} reduces to
\begin{equation}\label{ssimp}
 s_{A}=\frac{1}{1-\nu_{12}\nu_{21}\left(1-
 \delta(\varepsilon)\right)}\,E_1A(\eps);
 \end{equation}
 torsion stiffness continues to be given by formula \eqref{rigtors}.
 \item (Determination of Effective Constitutive Moduli)
In principle, a set of simple torsion, traction, and pressure,
experiments or simulations allows to deduce the values of
$s_T,s_A,\nu_{CA}$, and $\nu_{CP}$, from the prescribed values of
applied torque, axial load, and inner pressure and the measured or
computed values of $\Theta,\Lambda$, and $\Gamma$. In the
large-thinness limit, one would have:
\[
GJ(\varepsilon)=s_T,\quad E_1 A(\varepsilon)=s_A,\quad
\nu_{12}=\nu_{CA},\quad \nu_{21}=\nu_{CP},
\]
so that the values of the four constitutive moduli
$G,E_1,\nu_{12}$, and $\nu_{21}$ would follow, provided one could
measure or evaluate the values of the geometrical parameters
$\rho_o$ and $\varepsilon$. Now, given the well-known
difficulties, commonly referred to as the \emph{Yakobson's
Paradox} \cite{Sh},  in choosing  a representative value for the
wall thickness of a SWCNT, we think it best to characterize the
mechanical response of a Kirchhoff-Love shell by the contraction
moduli $\nu_{12}, \nu_{21}$ and the \emph{effective constitutive
moduli} $\widetilde G:=G\varepsilon$, $\widetilde
E_\alpha:=E_\alpha\varepsilon,\;(\alpha=1,2)$.

\remark We find it appropriate to expand a little on this last
point. For slender and thin cylindrical shells, we find that:
\begin{itemize}
\item [-] when subject to end traction,
\begin{equation}\label{constpar}
w_T=-\rho_o\nu_{12}\frac{p}{2\widetilde E_1}, \qquad
a{\<{1}}'_T=\frac{p}{2\widetilde E_1},
\end{equation}
 by way of equations \eqref{w_tracKL} and
\eqref{a_tracKL};
\item [-] when subject to inner pressure,
\begin{equation}\label{constpar1}
w_P=\rho_o^2\,\frac{\varpi}{2\widetilde E_2}, \qquad
a{\<{1}}'_P=-\nu_{21}\frac{\rho_o\varpi}{2\widetilde E_2},
\end{equation}
by way of equations \eqref{w_presKL} and \eqref{a_presKL}.
\end{itemize}
Equations \eqref{constpar} and \eqref{constpar1} can be regarded
as a system of four equations in the unknowns $\widetilde E_1$,
$\widetilde E_2$, $\nu_{12}$, $\nu_{21}$, whose solution is:
\begin{equation}\label{constpar2}
\begin{aligned}
\widetilde E_1=\frac{p}{2a{\<{1}}'_T}, \quad \widetilde
E_2=\rho_o^2\frac{\varpi}{2w_P}, \quad \nu_{12}=-\frac{w_T}{\rho_o
a{\<{1}}'_T}, \quad \nu_{21}=-\rho_o\frac{a{\<{1}}'_P}{w_P}.
\end{aligned}
\end{equation}
Interestingly, the consistency condition $\eqref{moreconst}_1$,
which can now be written as
\[
\frac{\widetilde E_1}{\widetilde E_2}=\frac{\nu_{12}}{\nu_{21}},
\]
implies that
\[
 \frac{p}{w_T}=
\frac{\varpi}{a{\<{1}}'_P},
\]
%
%\[
%\frac{p\,w_P}{\rho_o^2\,a{\<{1}}'_T\,\varpi}=\frac{\widetilde E_1}{\widetilde E_2}=\frac{\nu_{12}}{\nu_{21}}=
%\frac{w_Tw_P}{\rho_o^2\,a{\<{1}}'_T\,a{\<{1}}'_P}\quad\Rightarrow\quad \frac{p}{w_T}=
%\frac{\varpi}{a{\<{1}}'_P},
%\]
%
a relation that can be used to check whether the present
simplified form of our shell theory is applicable.
 \end{enumerate}
\section{Conclusions and further developments}
We have given a detailed presentation of a theory of linearly
elastic orthotropic shells with potential application to the
continuous modeling of carbon nanotubes. The novelty of this
theory resides in two features: (1) the type of \emph{orthotropic
response} we have selected seems suitable, with minimal tuning, to
capture chirality, not only in the extreme cases of zig-zag and
armchair SWCNTs, but also when it varies in an essentially
undetectable manner from wall to wall of a MWCNT; (2) the
possibility of accounting for overall \emph{thickness changes},
that should be almost exclusively due to changes in inter-wall
separation. As a matter of fact, the referential thickness of an
ideal MWCNT is dictated by inter-wall forces of van der Waals
type, whose action may be modeled essentially in two ways: either
they can be regarded small with respect to applied loads, and
ignored altogether; or they can be thought of as inducing a
referential equilibrated stress state that should be taken into
account when studying the effects of boundary conditions, no
matter if hard or soft: our sounding of this latter approach is
encouraging.

In addition, we have proposed a simpler version of the theory, in
which orthotropy is preserved but thickness changes are excluded
in all admissible deformational vicissitudes; we believe this
simpler theory to  fit SWCNTs, whose effective thickness when
regarded as cylindrical shells should not change appreciably when
loaded no matter how evaluated.

Presuming a rather complex material response requires
specification of a number of constitutive parameters -- seven when
thickness changes are allowed, four when they are not. Luckily,
another feature of our present theory is that, in both its
versions, it leads to a number of significant boundary-value
problems that can be solved explicitly in closed form. These
problems are: torsion, axial traction, uniform inner pressure, and
rim flexure; were their solutions coupled with the corresponding
measurements and/or simulation results, applicability of our
theory could be unequivocally assessed and all constitutive
parameters in it uniquely determined. It is not difficult to paste
explicit solutions of the type we here derived for two or more
coaxial shells, both in statical and dynamical situations; we are
currently developing this line of research with a view to a better
understanding of van der Waals forces and the role and evolution
of defects \cite{DCFPG}.

\vspace{0.5cm} \noindent \textbf{Acknowledgements}\\ \noindent We
gratefully acknowledge a useful discussion with Professor M.
Bertsch.


\begin{thebibliography}{}

\bibitem{CFPG} C. Bajaj, A. Favata, P. Podio--Guidugli, On a Scale-Bridging Mechanical Model of Carbon
Nanotubes, Forthcoming.

\bibitem{DCFPG} A. Di Carlo, A. Favata, P. Podio--Guidugli, Modeling Multi-Wall Carbon Nanotubes as Elastic Multi-Shells, 2011 (Forthcoming).

\bibitem{DoC} M.P. Do Carmo, Differential Geometry of Curves and Surfaces. Prentice Hall, New Jersey (1976).

\bibitem{FPG} A. Favata, P. Podio-Guidugli, What shell theory fits carbon nanotubes? To appear in Proceedings of Euromech 527, Lutherstadt Wittenberg, Aug 22-26 2011.

\bibitem{G} R.V. Goldstein, V.A. Gorodtsov, A.V. Chentsov, S.V. Starikov, V.V. Stegailov, G.E. Norman, {To description of mechanical properties of nanotubes. Tube wall thickness problem. Size effect}, Russian Academy of Sciences, A.Yu. Ishlinsky Institute for Problems in Mechanics, Preprint  937 (2010).

\bibitem{Gu} M.E. Gurtin, The Linear Theory of Elasticity. Pp. 1-295 of \emph{Handbuch der Physik
VIa/2}, Springer (1972).

\bibitem{LPG1} M. Lembo, P. Podio--Guidugli, Internal constraints, reactive stresses,
and the Timoshenko beam theory. J. Elasticity \textbf{65} (2001)
131-148.

\bibitem{LPG2} M. Lembo, P. Podio--Guidugli,  How to use reactive stresses to improve plate-theory approximations of the stress field
in a linearly elastic plate-like body. Int. J. Sol. Structures
\textbf{44} (2007) 1337-1369.

\bibitem{PPG69} P. Podio-Guidugli, An exact derivation of the thin plate equation. J. Elasticity \textbf{22} (1989) 121-133.

\bibitem{PPG1} P. Podio--Guidugli, {\it Lezioni sulla teoria lineare dei gusci elastici
sottili}, Masson, Milano (1991).

\bibitem{PPG}P. Podio--Guidugli, { A Primer in Elasticity},
Kluwer (2000).
%Springer, 2nd Ed. (forthcoming).

\bibitem{PPGB} P. Podio--Guidugli, On structure thinness, mechanical and variational, pp. 227-242 of \emph{Variational Formulations in Mechanics: Theory and Applications}, E. Taroco, E.A. de Souza Neto, and A.A. Novotny (Ed.s), CIMNE, 2007

\bibitem{PPGU} P. Podio--Guidugli, Concepts in the mechanics of thin structures. Pp. 77-110 of CISM Vol. 503,
A. Morassi and R. Paroni (Eds.), Springer (2008).

\bibitem{PGV} P. Podio--Guidugli, M.
Vianello, The representation problem of constrained linear
elasticity. J. Elasticity \textbf{28} (1992) 271-276.

\bibitem{PGV2} P. Podio--Guidugli, M.
Vianello,  Hypertractions and hyperstresses convey the same
mechanical information,  Cont. Mech. Thermodyn. \textbf{22} (2010)
163-176.

\bibitem{Sh} O.A. Shenderova, V.V. Zhirnov, D.W. Brenner, Carbon
Nanostructures. Crit. Rev. Solid State Mater. Sci. 27, 227–356
(2002).


\end{thebibliography}


%

\end{document}